\documentclass[
    reprint,                
    superscriptaddress,      
    aps,                     
    prd,                     
    floatfix,                
    onecolumn,
    11pt
]{revtex4-2}

\input{packages.sty}

\input{macros.sty}

\begin{document}



\title{Towards a Self-Driving Trigger at the LHC: Adaptive Response in Real Time}

\author{Shaghayegh~Emami}
\affiliation{University of Michigan, Ann Arbor, Michigan 48109, USA}

\author{Cecilia~Tosciri}
\affiliation{Enrico Fermi Institute, University of Chicago, Chicago, IL 60637, USA}

\author{Giovanna~Salvi}
\affiliation{University of Michigan, Ann Arbor, Michigan 48109, USA}

\author{Zixin~Ding}
\affiliation{Department of Computer Science, University of Chicago, Chicago, IL 60637, USA}

\author{Yuxin~Chen}
\affiliation{Department of Computer Science, University of Chicago, Chicago, IL 60637, USA}

\author{Abhijith~Gandrakota}
\affiliation{Fermi National Accelerator Laboratory, Batavia, IL 60510, USA}

\author{Christian~Herwig}
\affiliation{University of Michigan, Ann Arbor, Michigan 48109, USA}

\author{David~W.~Miller}
\affiliation{Enrico Fermi Institute, University of Chicago, Chicago, IL 60637, USA}
\affiliation{Department of Physics, Kavli Institute for Cosmological Physics, University of Chicago, Chicago, IL 60637, USA}

\author{Jennifer~Ngadiuba}
\affiliation{Fermi National Accelerator Laboratory, Batavia, IL 60510, USA}

\author{Nhan~Tran}
\affiliation{Fermi National Accelerator Laboratory, Batavia, IL 60510, USA}

\begin{abstract}

Real-time data filtering and selection -- or \textit{trigger} -- systems at high-throughput scientific facilities such as the experiments at the Large Hadron Collider (LHC) must process extremely high-rate data streams under stringent bandwidth, latency, and storage constraints. Yet these systems are typically designed as static, hand-tuned menus of selection criteria grounded in prior knowledge and simulation. In this work, we further explore the concept of a \textit{self-driving trigger}, an autonomous data-filtering framework that reallocates resources and adjusts thresholds dynamically in real-time to optimize signal efficiency, rate stability, and computational cost as instrumentation and environmental conditions evolve. We introduce a benchmark ecosystem to emulate realistic collider scenarios and demonstrate real-time optimization of a menu including canonical energy sum triggers as well as modern anomaly-detection algorithms that target non-standard event topologies using machine learning.
Using simulated data streams and publicly available collision data from the Compact Muon Solenoid (CMS) experiment, we demonstrate the capability to dynamically and automatically optimize trigger performance under specific cost objectives without manual retuning.  
Our adaptive strategy shifts trigger design from static menus with heuristic tuning to intelligent, automated, data-driven control, unlocking greater flexibility and discovery potential in future high-energy physics analyses.

\end{abstract}


\maketitle
\newpage
\tableofcontents

\section{Introduction}
\label{sec:intro}

In scientific domains characterized by high data throughput, algorithms tasked with filtering and selecting data for storage and analysis -- frequently referred to as \textit{trigger algorithms} -- must operate under extreme bandwidth, computation, and storage capacity requirements: data rates of $10^{6}$--$10^{9}$ per second (MHz--GHz), bandwidths of $10^{12}$--$10^{15}$ bits per second (Tb/s--Pb/s), and storage volumes of $10^{15}$--$10^{18}$ bytes per year (PB/y--EB/y). To meet these demands, data filtering algorithms must be highly selective, often operating with selection rates at the level of 1 part in $10^5$. With sensors numbering from the millions to the billions, these data processing tasks are also moving ever closer to the edge (i.e., closer to the sensor itself), further stratifying and complexifying the data filtering process. 

Despite these complex environments, many data filtering paradigms for discovery science domains, such as at high-energy particle colliders like the Large Hadron Collider (LHC), rely heavily on detailed prior knowledge of the feature space being probed, precise simulations of the system and instrumentation, and highly modularized approaches to optimization. Consequently, the design and operation of these experiments require enormous manual effort and human expertise, and also introduce inherent inefficiencies such as redundant feature labeling schemes and cost-ineffective algorithm execution. Furthermore, both known and unknown operator biases present in these algorithms raises the fundamental concern of whether rare or new phenomena are being missed entirely. 

In this work, we further explore the concept of a \textit{self-driving trigger system}: a hardware-aware autonomous data processing and filtering system~\cite{Mahesh:2021iph}. The core objective of this work is to establish systems capable of continuously and autonomously learning what data to collect and record, and identifying and responding dynamically to changing conditions and new observations. Such a system aims to provide a \textit{cost-effective explanation} for why data are saved or discarded, subsequently updating its decision framework in a hardware- and resource-aware manner. This effort builds on advances in reinforcement learning, explainable artificial intelligence (AI), and control optimization to autonomously adapt and refine selection policies.

Recent studies~\cite{Mahesh:2021iph,Evans:2025jiw} have proposed the integration of adaptive algorithms into the trigger decision chain to dynamically reallocate bandwidth and adjust thresholds in response to real-time changes in experimental conditions. Inspired by these efforts, we introduce a benchmark framework to evaluate autonomous trigger control strategies under realistic scenarios at collider experiments. Specifically, we focus on the sets of data selection rules, so-called \textit{trigger menus}, that encode the selection criteria and constraints. These menus are often designed based on predefined science goals and expected detector performance. While robust under stable conditions, such menus are effectively static and are inherently inflexible: they do not automatically adapt to time-dependent variations in the operation of the instrumentation, the conditions that evolve throughout a data taking period, or even in the potential appearance of new features in the data that were previously absent or not statistically significant. Consequently, the system may fail to identify rare and potentially interesting data or misinterpret otherwise uninteresting data. In the context of large scientific facilities, updates and adjustments to these complex filtering rules to account for these issues typically necessitate extensive human intervention and can take weeks or even months to assess and implement. This motivates the development of smarter, more autonomous systems that can dynamically respond to real-time changes in experimental conditions and adapt to minimize biases and maximize the effectiveness of the data collection and discovery potential.

\begin{figure}[htbp]
\centering
\includegraphics[width=0.6\textwidth]{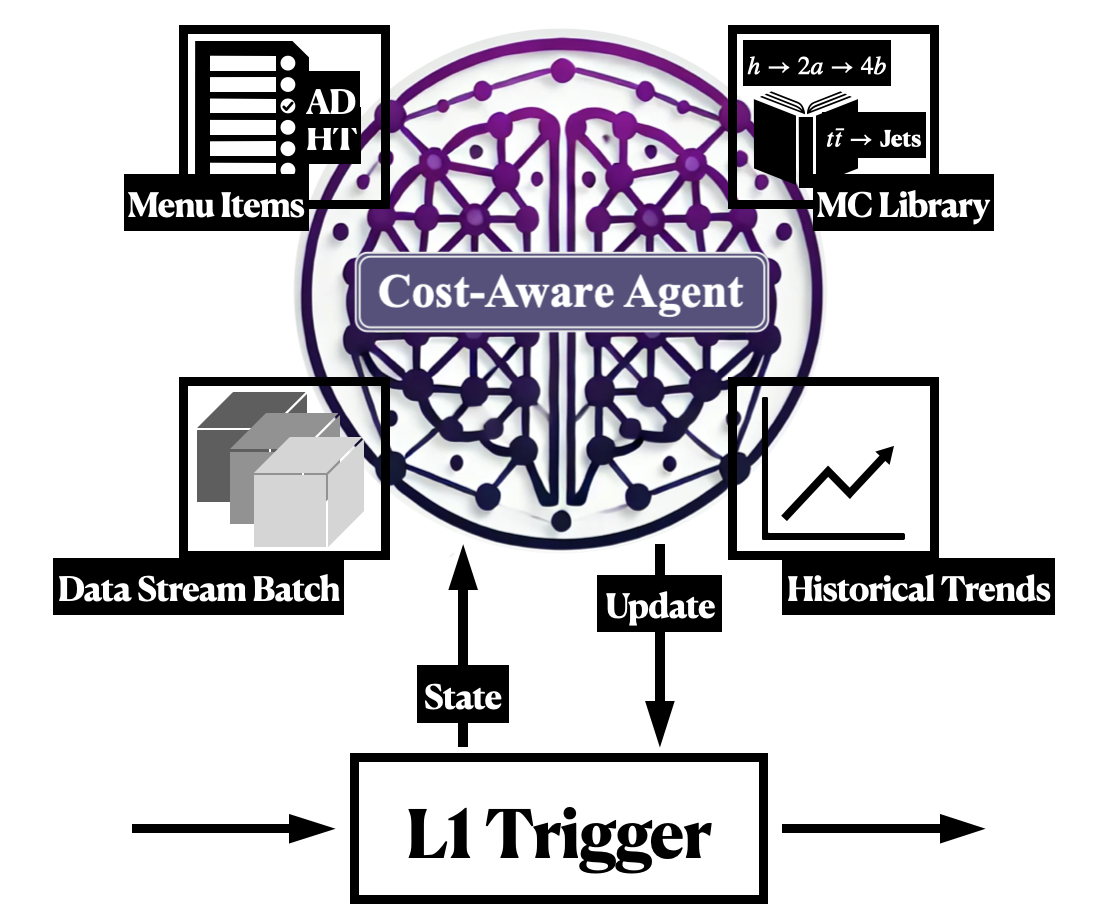}
\caption{Schematic overview of the adaptive trigger control system. A cost-aware agent receives batches of collision data, evaluates trigger performance and computational costs, and updates L1 trigger thresholds and bandwidth allocations in real time through a feedback-driven control loop.}
\label{fig:big-picture}
\end{figure}

We envision the self-driving trigger as an \textit{agent} that interacts directly with the \textit{environment} of the hardware-based Level-1 (L1) trigger system, as sketched in Fig.~\ref{fig:big-picture}.
At regular time intervals the agent and environment interact: the L1 trigger transmits a record of recent trigger decisions to the control agent, and the agent prescribes updated selection rules to the L1 trigger that will be used to collect the next batch of events.
The agent's prescription may incorporate many factors, including short- and long-term trends in the selection rates observed in data, correlations among the decisions of the active trigger algorithms, and auxiliary factors such as the computation cost of processing the selected events.
In addition to these purely data-driven considerations, we propose to make use of labeled simulation by furnishing the agent with a modest library of Monte Carlo (MC) events that encompasses several physical processes across a range of expected experimental conditions.
Access to simulation would allow the agent to estimate the trigger efficiency of a candidate menu for a few key signatures, based on events with similar conditions to those currently seen in data.
A flexible reward mechanism allows each of these considerations to factors into the agent's real-time prescriptions.

We test this approach using CMS Open Data and corresponding Monte Carlo (MC) simulations~\cite{CMSOpenData}, focusing on the task of selecting events with interesting jet activity.
A simplified trigger menu includes a traditional algorithm to select the most energetic events, as well as a modern approach for unsupervised anomaly detection (AD) with neural networks.
Our results establish a proof-of-concept that an adaptive trigger systems present a powerful and robust innovation over the status quo, which can be easily tailored to prioritize a range of different operational objectives including the efficiency for known or expected signals, sensitivity to anomalies, and expected burden on computational resources.

The remainder of this paper is organized as follows:
Section~\ref{sec:info_MC_data}  introduces the datasets, simulation framework, and benchmark tasks. Section~\ref{sec:trigger_strategies} describes the baseline \HT and anomaly detection trigger algorithms. In Section~\ref{sec:perf_fixed_menu}, we characterize the performance and limitations of a fixed trigger menu in simulated event streams. Sections~\ref{sec:real_time_control}--\ref{sec:local_controller} develop and compare real-time control strategies, starting from the independent adjustment of each algorithm and extending to a single controller that allocates bandwidth across several triggers based on different cost and physics objectives. Section~\ref{sec:controller_perf_data} then demonstrates the behavior of these controllers on publicly available collision data from the CMS experiment at the LHC. Section~\ref{sec:summary_plots} presents a direct comparison between the Monte Carlo simulation and the real data's approach, assessing the robustness and transferability of the proposed control strategies. Finally, Section~\ref{sec:outlook} summarizes the main conclusions of this work and outlines the possible directions for future developments.

\section{Datasets and Simulation Framework}
\label{sec:info_MC_data}

The CMS~\cite{CMSOpenData} and ATLAS~\cite{ATLASOpenData2015-2016} Collaborations have each released datasets of reconstructed collision events that are publicly accessible through the CERN Open Data portal~\cite{CERNOpenDataPortal}. The availability of the data is essential for the study of real-time detector control described in this manuscript, as it relies heavily on variations in detector conditions that may only be present in the real data. Monte Carlo (MC) simulations of various physical processes are also provided to facilitate the interpretation of this dataset and are generated to reflect, on average, the experimental conditions in which the collision data were recorded.

Our studies make use of the CMS Open Data dataset corresponding to proton-proton collisions recorded at a center-of-mass energy $\sqrt{s}=13$~TeV in 2016. Specifically, we use data collected during periods G and H of Run~2, amounting to a total of 16.4\ifb~\cite{dataG,dataH}.
To date, this is the most recent major dataset that has been released by either of the experimental collaborations. The collision datasets and MC simulation samples are processed with detector simulation and reconstruction algorithms matching the 2016 data taking conditions, including the pileup profile, beam energy, and detector calibration parameters.

In this benchmark scenario, we implement and test our approach using two representative trigger paths:
\begin{itemize}
    \item \HT trigger, collecting events with large scalar sums of jet transverse momenta (\pt),\\ $
H_T = \sum_{j \in \mathcal{J}} p_T^{\,j},
\,
\mathcal{J} = \left\{ j \,\middle|\, p_T^{\,j} > 20~\mathrm{GeV},\ |\eta^{\,j}| < 2.5 \right\}.$
    \item Anomaly detection trigger, AD, trained to identify events that deviate from known background topologies.
\end{itemize}

\noindent Together, these triggers capture the dual challenge of selecting rare signals and carrying out this task based on a set of rich but inherently shifting (pileup sensitive) observables. While the \HT trigger follows the classic heuristic of controlling rates by selecting the most energetic events, the  AD  trigger serves as an alternate strategy that may be particularly promising in selecting novel or poorly modeled processes. 
A detailed description of each algorithm is presented in Section~\ref{sec:trigger_strategies}.

We validate our strategy using a combination of CMS data and simulation and provide a reproducible framework for evaluating adaptive trigger control strategies. Data are selected from an unbiased event stream (\texttt{ZeroBias}), recorded by random triggers that only require the presence of a filled LHC bunch.
In the following, we refer to the ZeroBias sample as the ``background data" sample.
Such triggers are heavily `pre-scaled', whereby only a small fraction of events passing the trigger logic are actually recorded.
Because this stream is typically used for experimental diagnostics rather than physical measurements, the selected events make up only a small fraction of the total trigger rate.

At the LHC, a fill refers to a period of time during which beams are injected, accelerated, and kept in stable collisions, typically lasting several hours. Within each fill, data are collected in multiple runs, which correspond to contiguous periods of data taking defined by stable detector and trigger configurations.
Considering sequential data from a single run within a fill allows us to track the gradual evolution of beam conditions and pileup over time with relatively consistent settings.
Thus, in order to mimic a system capable of adjusting trigger rules in real-time, we primarily consider the largest run contained in the dataset (Run 283408), which includes 1.99 million events in the background data stream. 

Monte Carlo samples of events corresponding to  Standard Model (SM) and Beyond Standard Model (BSM) processes were also used to design the real-time control strategy and assess its performance.
Specifically, the following processes have been considered:
\begin{itemize}
    \item Simulated minimum bias collision events selected using loose trigger requirements, with pileup and conditions matching the 2016 data taking period~\cite{minbiasSample}, simulated with Pythia8~\cite{pythia8}. 
    \item Production of top quark pairs (\ttbar), as a representative Standard Model (SM) process with fully hadronic final states and a wide variety of kinematic configurations~\cite{ttbarSample}. The hard-scatter process is simulated at next-to-leading order in the strong coupling constant using Powheg~\cite{powheg}, interfaced with Pythia8 for parton showering and hadronization.

    \item Exotic decays of the Standard Model Higgs boson ($h$) to new pseudo-scalar particles ($a$), simulated using MadGraph~\cite{madgraph} interfaced with Pythia8, for a range of $a$ masses between 12 and 60 GeV. Both pseudo-scalars are assumed to decay promptly to $b$-quark pairs with a 100\% branching ratio, resulting in a $h \to aa \to b\bar b b \bar b$ (simply referred to as \haaFourB\ in the following sections) final state~\cite{exoSample}.
\end{itemize}

To design and evaluate the real-time control strategy, it is essential to model how the collision environment evolves during a typical LHC fill.
In real data, the number of primary vertices (\NPV) typically decreases over time, as the number of protons in the beam drops due to inelastic collisions and beam losses. This reduction in luminosity leads to lower pileup and fewer reconstructed vertices per event. Figure~\ref{fig:npv_vs_time} illustrates this behavior in CMS Run 283408 by showing the average \NPV\ reconstructed as a function of time, shown as the fraction of the run that has elapsed.
\begin{figure}[htbp]
\centering
\includegraphics[width=0.45\textwidth]{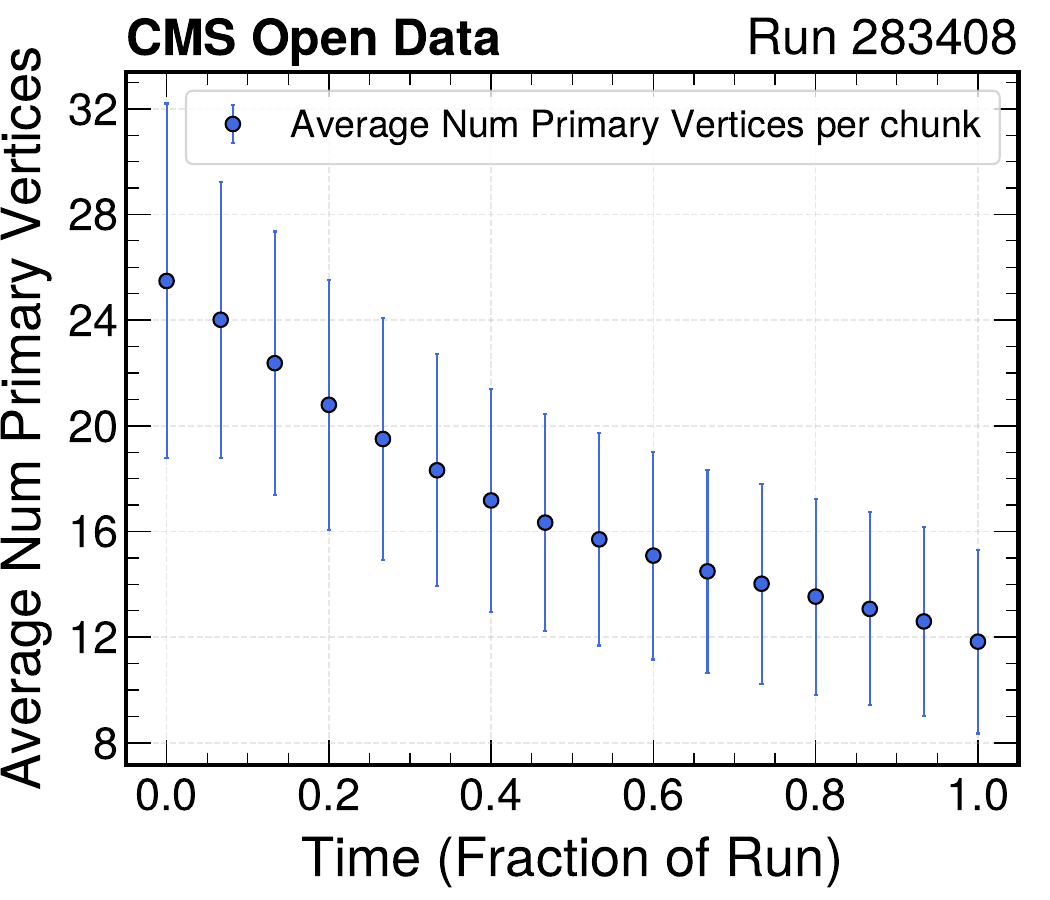}
\caption{
    The average number of primary vertices (\NPV) reconstructed at various points throughout Run 283408 collected by the CMS Experiment in 2016.
    Time is measured as the fraction of the run that has elapsed, with the gradual decrease of \NPV reflecting the gradual drop in luminosity and pileup as the fill progresses.}    
\label{fig:npv_vs_time}
\end{figure}

A similar time evolution pattern is observed in the variable $H_T$, the scalar sum of trigger jet \pt, defined in detail in Section~\ref{sec:trigger_strategies}. 
As shown in Fig.~\ref{fig:ht_evolution-a}, the energy distribution gradually shifts toward lower values as the fill progresses. 
To visualize this trend more clearly, Fig.~\ref{fig:ht_evolution-b} shows the same distributions as a series of violin plots, highlighting the evolution of the median and quartiles of \HT as a function of the run-time fraction. 
These distributions convey the extent to which the experimental trigger data undergo significant changes (in both scale and shape) over the course of a single run, and the inherent time-dependence of an \HT trigger algorithm.

\begin{figure}[htbp]
    \centering
    \begin{subfigure}[t]{0.47\textwidth}
        \vspace{0pt}
        \centering
        \includegraphics[width=\linewidth]{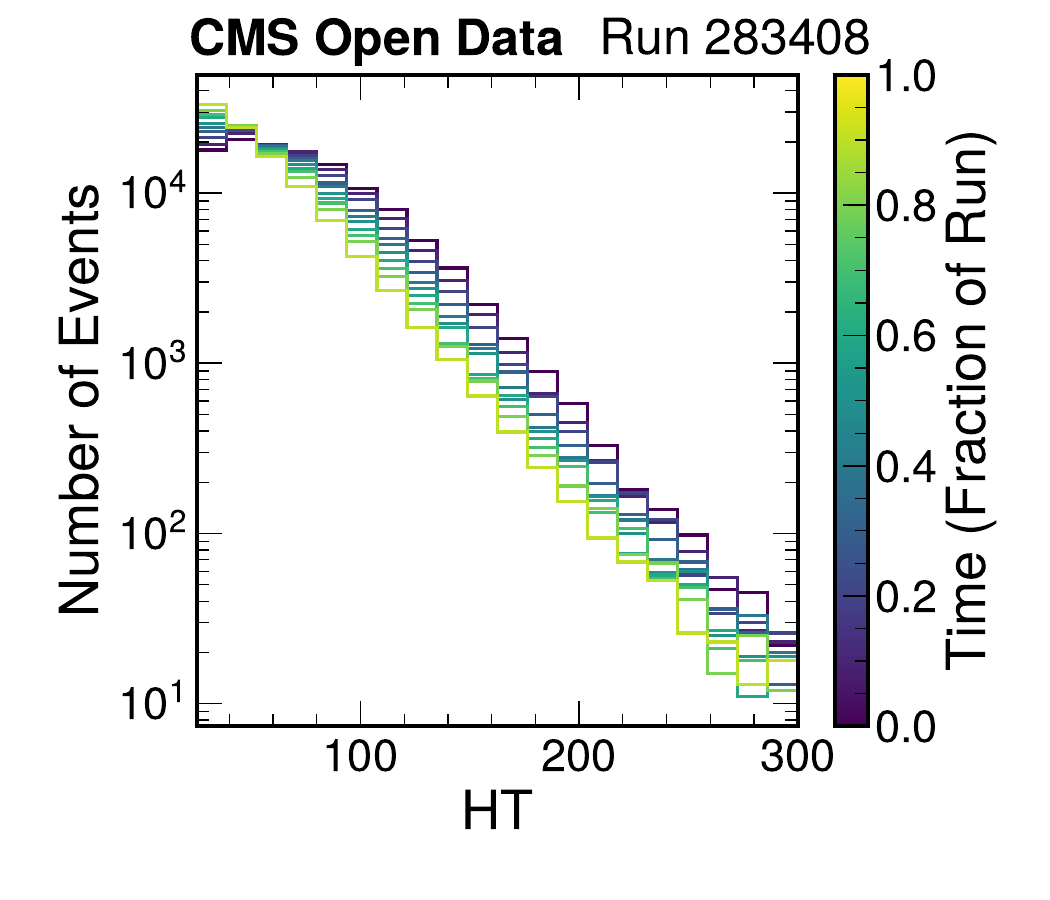}
        \caption{}
        \label{fig:ht_evolution-a}
    \end{subfigure}
    \hspace{0.02\textwidth}
    \begin{subfigure}[t]{0.47\textwidth}
        \vspace{0pt}
        \centering
        \includegraphics[width=\linewidth]{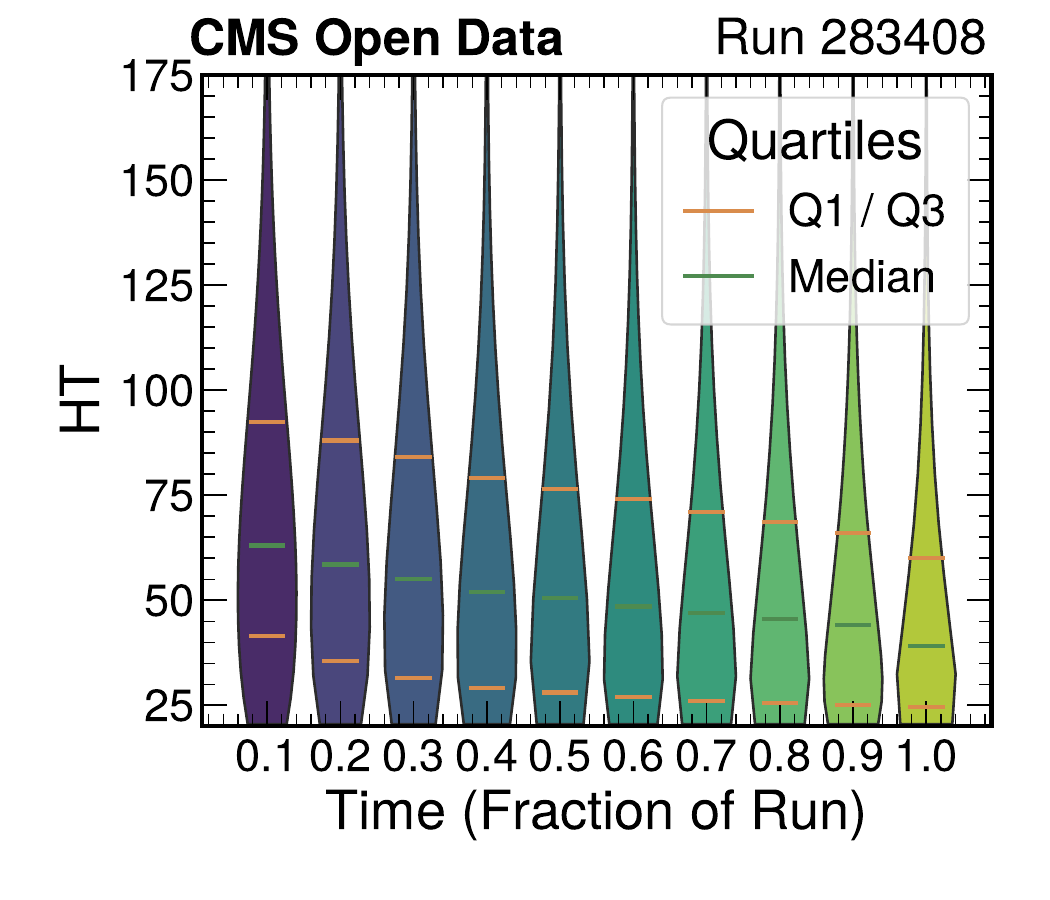}
        \caption{}
        \label{fig:ht_evolution-b}
    \end{subfigure}

    \caption{Evolution of \HT during a 2016 LHC run recorded by CMS (Run 283408).
    (a) Histograms show the progressive shift of the \HT distribution toward lower values over the course of the fill.
    (b) Violin plots display the evolution of the median and quartiles of \HT as a function of the run-time fraction,
    highlighting the correlation between $H_T$, time, and \NPV\ through variations in the pileup energy density.
    }
    \label{fig:ht_evolution}
\end{figure}

Events produced with MC simulation are inherently unordered, with each event's value of \NPV\ having been randomly sampled from the aggregate distribution measured over the full 2016 data sample.
To develop and test a real-time control strategy, an ad hoc ordering must be enforced on the simulated datasets that captures the general features observed in collision data.
To this end, we emulate the gradual decrease in the average number of pileup collisions over time within a fill by enforcing an approximate \NPV-sorting of the simulated events.
However, because the true value of \NPV\ follows a Poisson distribution and therefore fluctuates on an event-by-event basis, the same effect is naturally imposed in our simulation samples.

Arbitrary levels of variation can be studied in simulation with this approach. Figures~\ref{fig:npv_time-mc} and ~\ref{fig:ht_evolution_mc} demonstrate that the time variable induces time-dependent trends in simulation samples similar to those observed in data.
\begin{figure}[htbp]
    \centering
    \begin{subfigure}[t]{0.465\textwidth}
     \vspace{0pt}
        \centering
        \includegraphics[width=\linewidth]{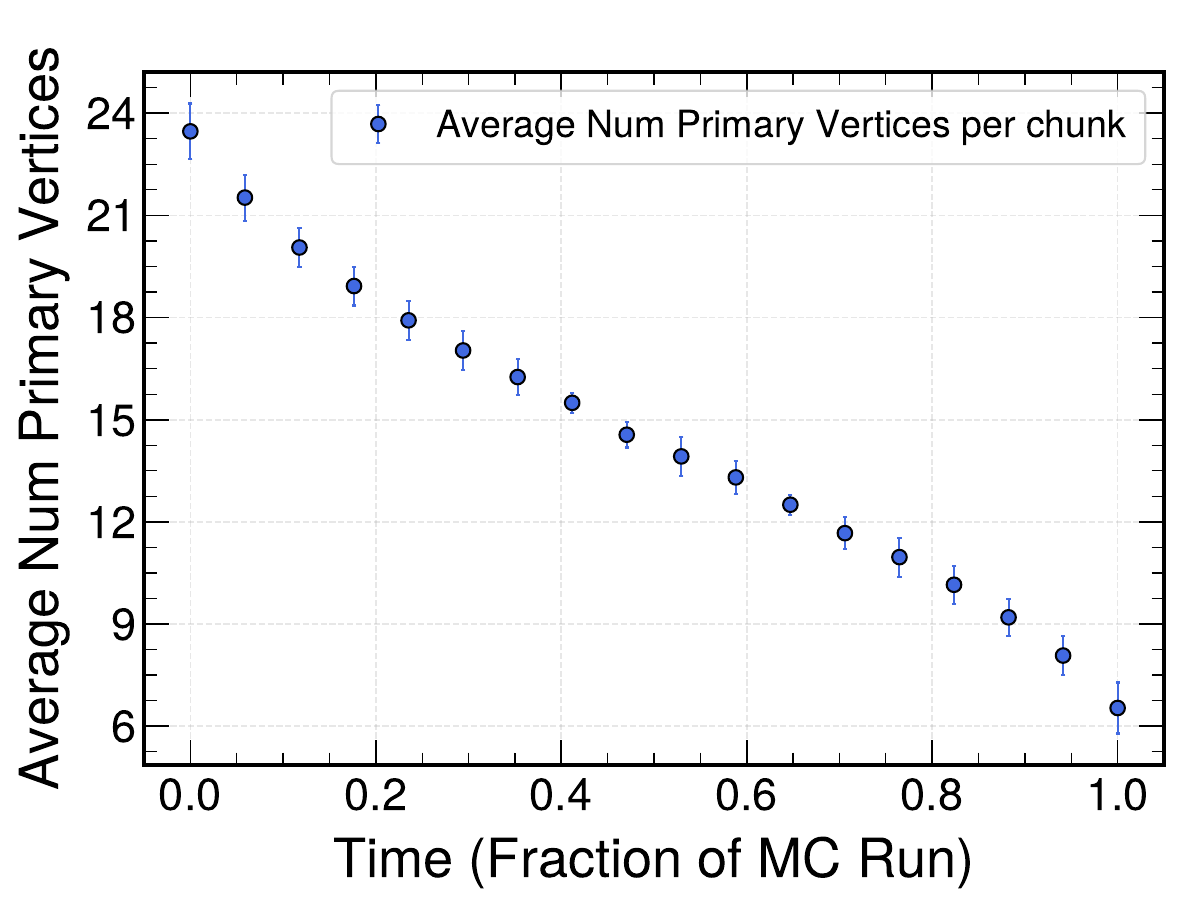}
        \caption{}
            \label{fig:npv_time-mc}
    \end{subfigure}    
    \hspace{0.02\textwidth}
    \begin{subfigure}[t]{0.49\textwidth}
        \vspace{0pt}
        \centering
        \includegraphics[width=\linewidth]{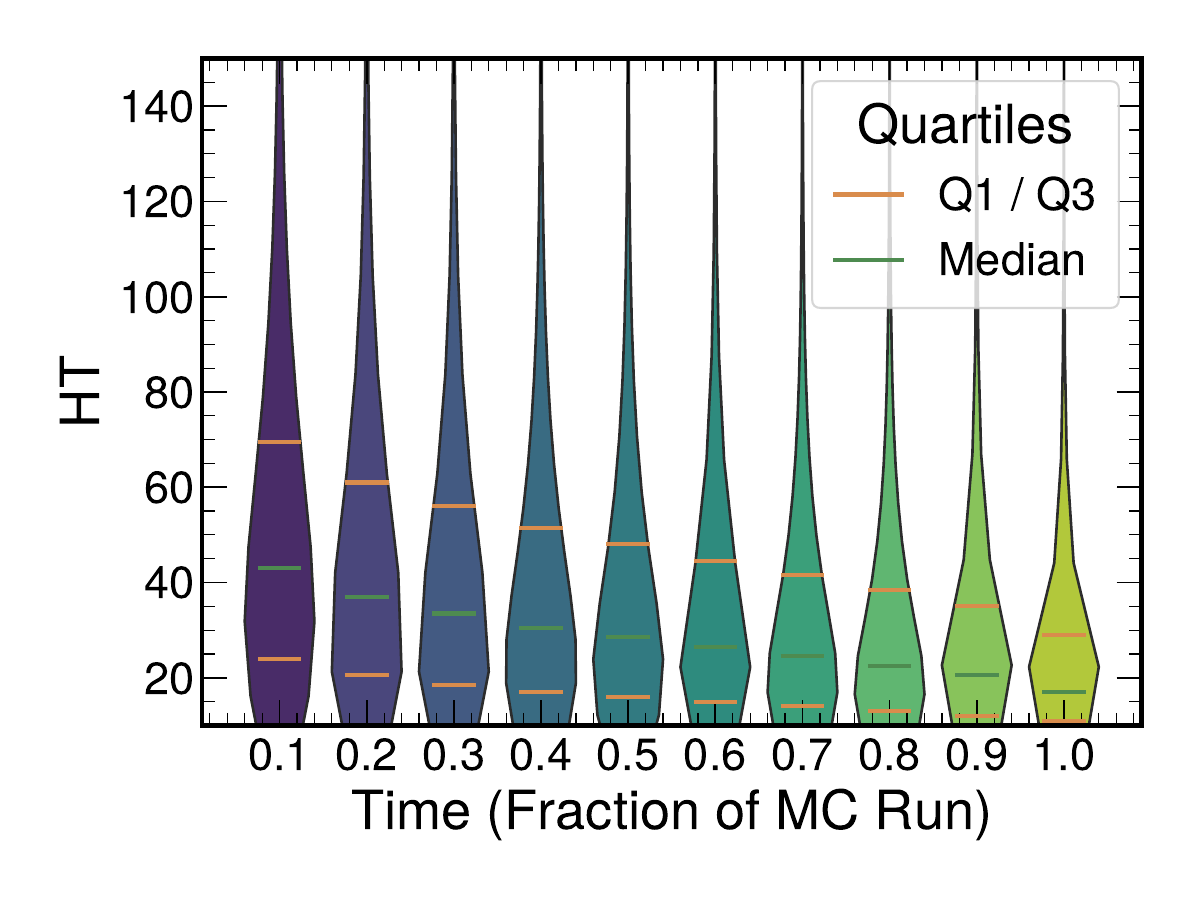}
        \caption{}    
            \label{fig:ht_evolution_mc}
    \end{subfigure}
    \caption{
    Introducing a time index in simulated events by ordering them according to the \NPV distribution, smeared with Poisson noise to mimic realistic fluctuations. 
    (a) \NPV versus time, illustrating how the constructed index reproduces the characteristic pileup evolution seen during an LHC fill. 
    (b) Violin plots of \HT versus time, showing the expected decrease of hadronic activity over the fill as the pileup drops. 
    }
    \label{fig:evolution_mc}
\end{figure}
To summarize our benchmark strategy, the input to the hardware-based L1 trigger logic is treated as background, and the trigger menu is configured to pass 99.75 percentile of events using thresholds on HT and AD menu items. These thresholds are dynamically adjusted by a controller layer operating on top of the L1 logic, which autonomously adapts the menu as a function of time. The whole setup focuses on the hadronic aspect of the events since it is the most challenging part to trigger, but it is easily extendable to other features in the experiment. Two types of background are considered: simulated minimum bias events and real \texttt{ZeroBias} data. While real data naturally carries time information, simulated events do not; therefore, we exploit the time evolution using the number of primary vertices, \NPV, as a proxy for time. Simulated background events are therefore sorted and smeared according to \NPV, producing a time-indexed background stream that mimics realistic running conditions. 

The controller additionally has access to a simulated signal library, consisting of \ttbar\ and \haaFourB\ samples, which is used exclusively to evaluate the performance of the evolving trigger menu. The treatment of signal samples depends on the choice of background. When using a simulated background, signal events are ordered using the same \NPV -based time indexing, and for the case of real data as background, each data event is randomly matched to signal events with the same \NPV. Our benchmark, therefore, begins with studying the simulated background to develop and validate the framework and controller strategies, and subsequently transitions to the data background with matched signal samples to evaluate methods under real experiment conditions.

\section{Features of the Trigger Data and Algorithms} \label{sec:trigger_strategies}

We employ two representative trigger algorithms that represent distinct operational challenges to benchmark the behavior of static versus adaptive strategies. The first is the \HT trigger, a standard algorithm to collect events with significant hadronic activity in the ATLAS and CMS experiments, which generally suffers from sizable pileup dependence.
The second is an anomaly detection (AD) trigger, based on a machine learning approach designed to capture rare or \textit{a priori} unknown features in collision events with minimal bias toward any specific new-physics model. These triggers reflect complementary experimental considerations and physics priorities: 
one is based on known signatures and relative operational stability, while the other aims to capture new signatures even if an end-to-end analysis workflow with a fully optimized trigger strategy has not yet been fully defined.

In the current ATLAS and CMS experiments, hardware-based L1 trigger systems perform the initial stage of real-time event reconstruction and selection. The L1 trigger processes data with latencies of a few microseconds and reduces the raw 40 MHz collision rate to around 100 kHz. This drastic reduction is achieved through coarse event reconstruction based on calorimetric and muon data. 
For hadronic events, L1 jets are reconstructed from regional energy deposits using simplified clustering algorithms. Due to the presence of overlapping $pp$ collision byproducts, raw jet energies must be corrected, typically by estimating and subtracting pileup contributions on an event-by-event basis. Further technical details on jet reconstruction and pileup mitigation at L1 can be found in Refs.~\cite{ATLAS:2021tnq,CMS:2020cmk}.

Building on the datasets described in the previous section, comprising both MC simulated background and signal samples, as well as real collision data, we extract the event-level jet information. This includes up to eight L1 jets, with per-jet features encoded in an array 
encompassing detector pseudorapidity ($\eta$), azimuthal angle ($\phi$), and transverse momentum ($p_T$), together with the number of primary vertices associated with the event \NPV.

These features serve as inputs for the two trigger algorithms, and enable a direct comparison between a fully-supervised trigger and a data-driven approach. Figs.~\ref{fig:jet_features-a}, ~\ref{fig:jet_features-b}, ~\ref{fig:jet_features-c}, ~\ref{fig:jet_features-d},
show the distributions of key observables for different datasets, including $H_T$, jet multiplicity, jet \pt\, and missing transverse hadronic energy $H_T^{\text{miss}}$, defined as the magnitude of the vector sum of jet momenta in the $(x, y)$ plane: 
\begin{align}
H_{\mathrm{T}}^{\text{miss}}=\lVert\vec p_{\text{miss}}^{\,\mathrm{T}}\rVert
&=\sqrt{\Bigl(\sum_{j}p_{T,j}\cos\phi_j\Bigr)^{2}
      +\Bigl(\sum_{j}p_{T,j}\sin\phi_j\Bigr)^{2}},  \nonumber \\ 
\vec p_{\text{miss}}^{\,\mathrm{T}}
   &= \sum_{j} p_{T,j}\bigl(\cos\phi_j\,\hat x+\sin\phi_j\,\hat y\bigr). \nonumber
\end{align}

The two triggers allow us to isolate the effects of different trigger strategies. In particular, while the \HT trigger prioritizes highly energetic events with a simple energy threshold, the AD trigger is more flexible, selecting events with the most anomalous scores as evaluated by a custom machine learning model.

\begin{figure}[htp]
    \centering
    \begin{subfigure}{0.4\textwidth}
        \centering
        \includegraphics[width=\linewidth]{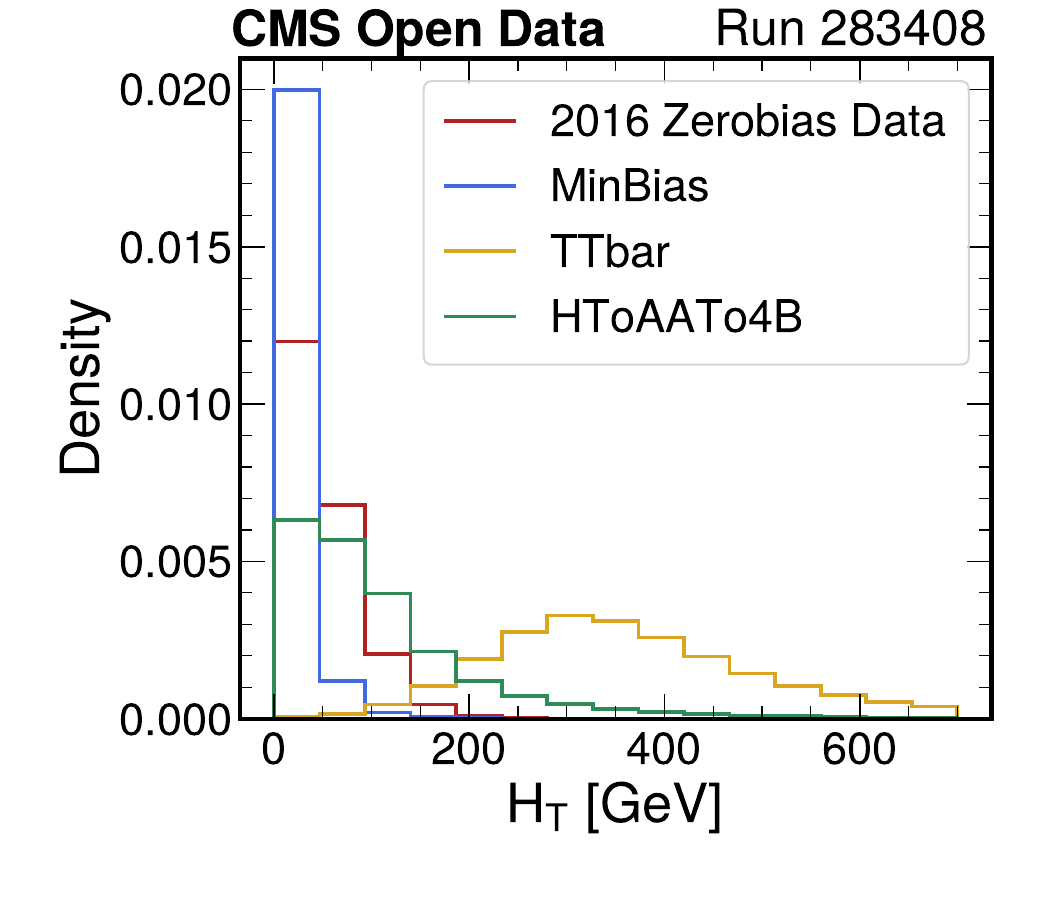}
         \caption{}\label{fig:jet_features-a}
    \end{subfigure}
    \begin{subfigure}{0.4\textwidth}
        \centering
        \includegraphics[width=\linewidth]{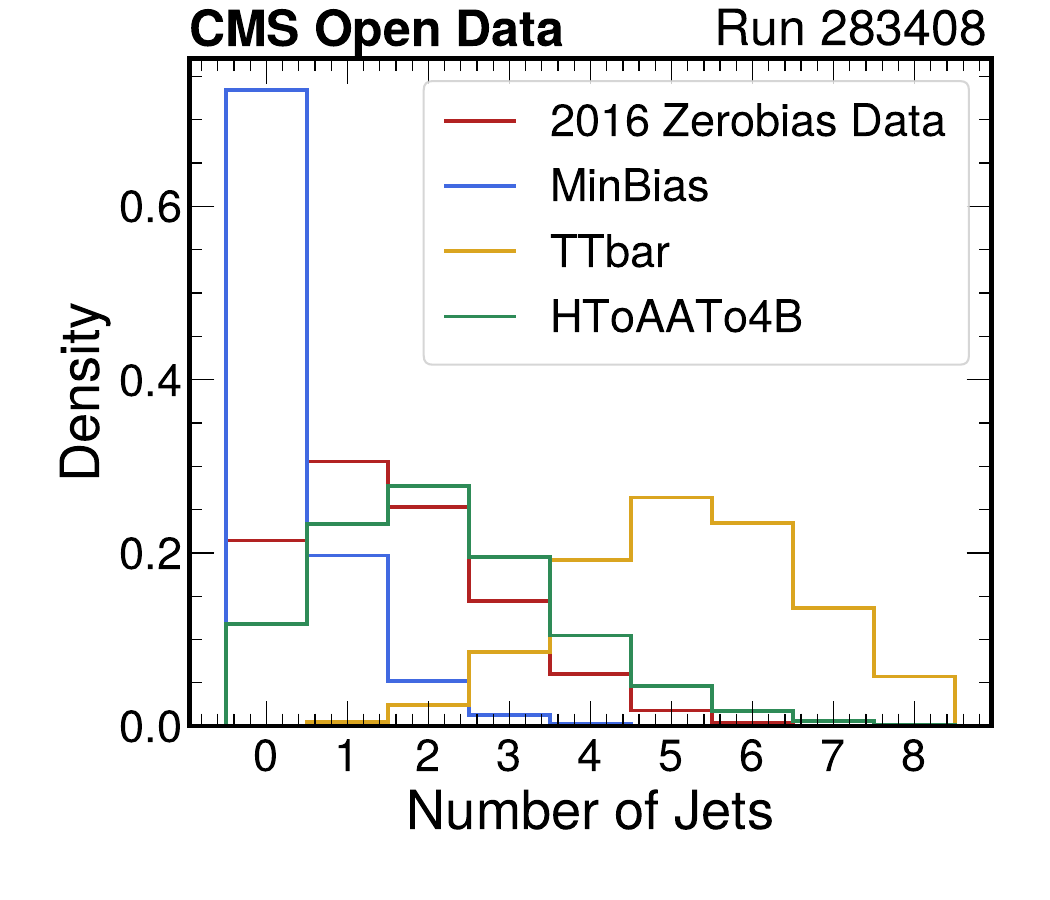}
        \caption{}\label{fig:jet_features-b}
    \end{subfigure}
    
    \vspace{2ex}
    \begin{subfigure}{0.4\textwidth}
        \centering
        \includegraphics[width=\linewidth]{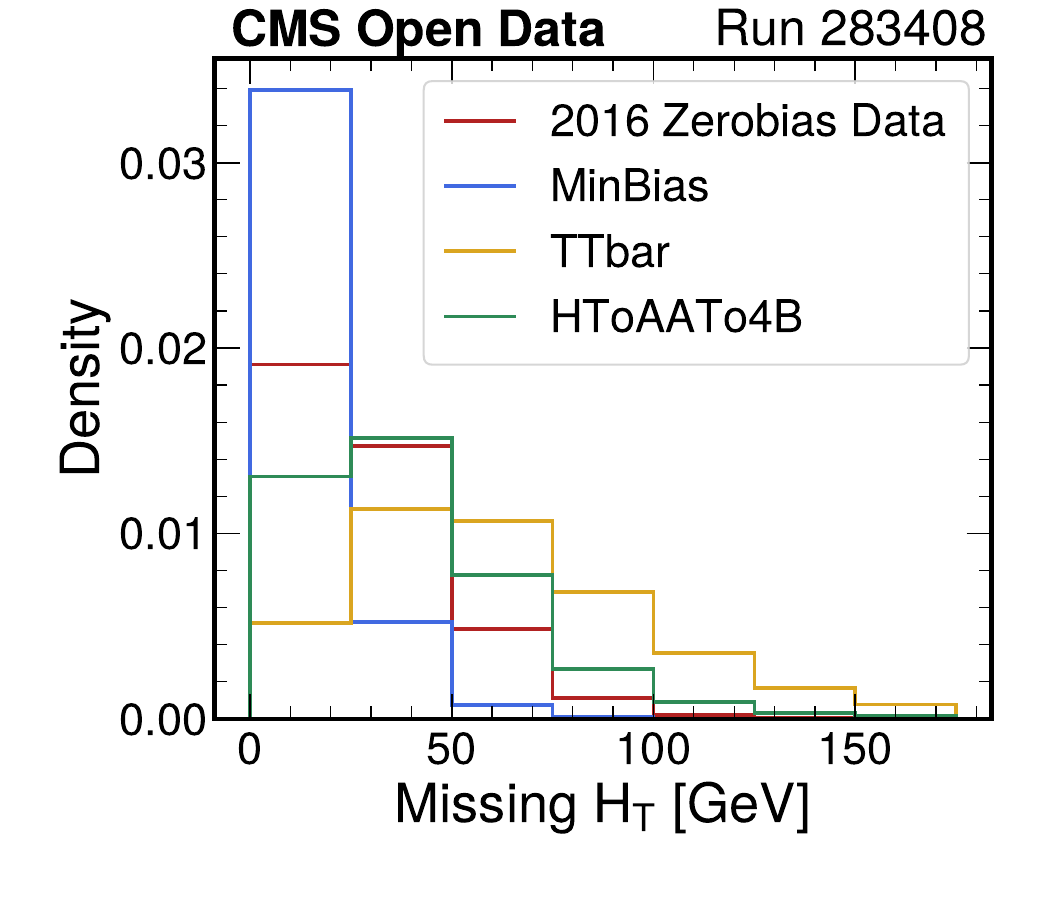}
            \caption{}\label{fig:jet_features-c}
    \end{subfigure}
    \begin{subfigure}{0.4\textwidth}
        \centering
        \includegraphics[width=\linewidth]{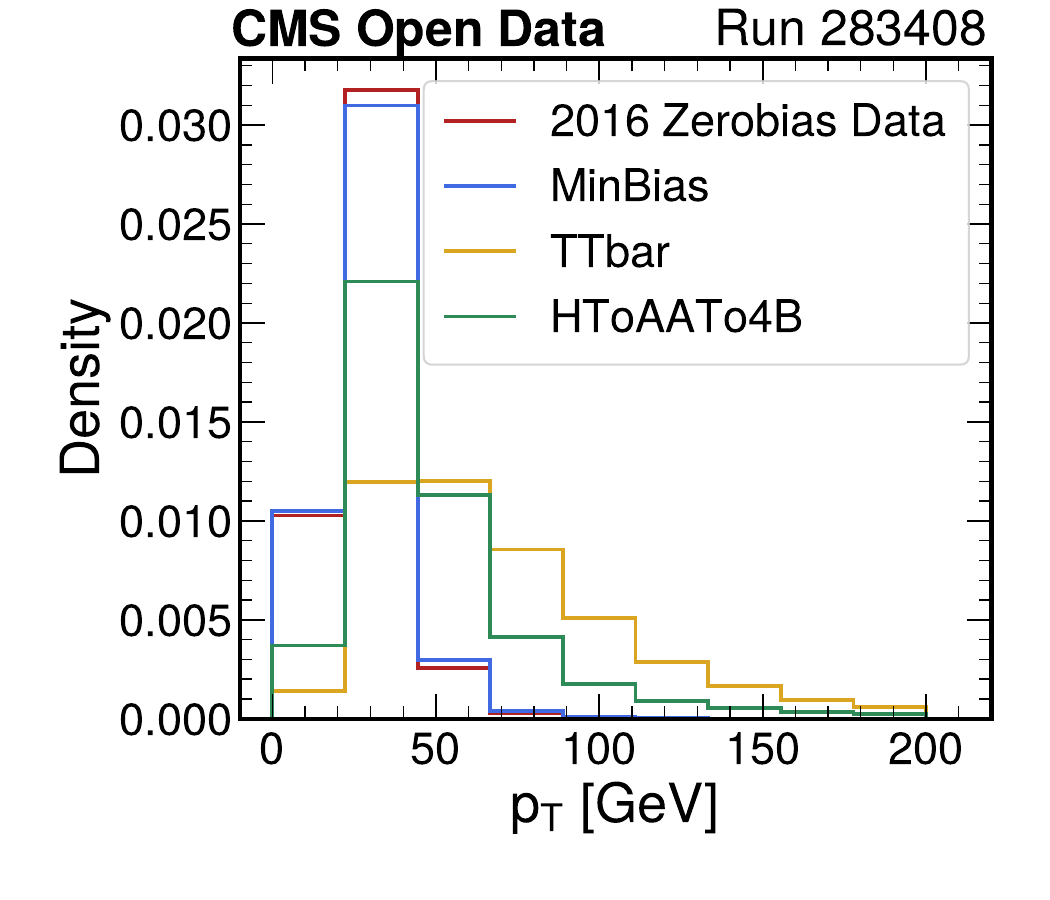}
            \caption{}\label{fig:jet_features-d}
    \end{subfigure}
    
    \caption{Comparison of key per-event features across different samples: real collision data, minimum bias simulation, SM \ttbar\ events, and an exotic benchmark process \haaFourB. 
    These distributions provide insight into the kinematic and topological differences that the trigger strategies aim to capture.}
    \label{fig:jet_features} 
\end{figure}

\subsection{Anomaly Detection Algorithm}\label{anomaly_detection}

We use an autoencoder (AE) model to compute the anomaly score for each event. As the core component of the AD trigger, the AE is essential for assessing the trigger's sensitivity to rare physics signals and its robustness under varying experimental conditions. Our approach to anomaly detection at the L1 is inspired by state-of-the-art methods employed by CMS \cite{CMS-AD} and ATLAS \cite{GELATO} that leverage AE models. 

Autoencoders \cite{hinton2006reducing,an2015variational} are self-supervised neural networks trained to reproduce their inputs at the output layer. They consist of an encoder $f_{\theta}$ that maps an input feature vector $x \in \mathbb{R}^{D}$ to a lower-dimensional latent representation $z = f_{\theta}(x) \in \mathbb{R}^{d}$ with $d \ll D$, and a decoder $g_{\phi}$ that maps back to a reconstruction $\hat{x} = g_{\phi}(z)$. The parameters $(\theta, \phi)$ are optimized to minimize a reconstruction loss over background events, such that the latent space captures the structure of the most frequent (background) patterns. At test time, events that are well represented by this learned background manifold are reconstructed with small error, while rare (i.e., anomalies) yield larger reconstruction error. We use the reconstruction error as the anomaly score, defined as the mean squared error (MSE) between each input array and its reconstruction. 

We adopt a fully connected multi-layer perceptron with a single encoding and decoding layer, consistent with prior work and compatible with a compact hardware implementation (e.g., with \texttt{hls4ml}~\cite{Duarte:2018ite}) for trigger systems across a variety of scales. In our implementation, the AD model takes a 25-dimensional input representation of each event, compresses it to a $d$-dimensional latent space, and then reconstructs it back to 25 dimensions. The model can be trained with either real collision data or MC simulation of minimum bias events to learn the typical background features. Events that deviate from these learned features are flagged as anomalous from the L1-trigger perspective and provide promising candidates for offline follow-up studies, where additional information and more powerful analysis techniques are available.

A key aspect in our input representation is the choice to provide the model with a proxy for the time variable, in this case \NPV. This helps the model learn the detector response to pileup variations more effectively. This is a natural choice, as all standard hadronic trigger algorithms must attempt to `calibrate away' pileup dependence (through energy density corrections, for example), even though ultimately some residual dependence almost always remains. A schematic diagram of the AD model is presented in Fig~\ref{fig:ae_model}. Each event is represented by up to eight jets, characterized by their kinematic properties ($\eta$, $\phi$, and $p_T$), together with the event's \NPV, resulting in a total of 25 input features.

\begin{figure}[htbp]
    \centering
    \includegraphics[width=0.55\textwidth]{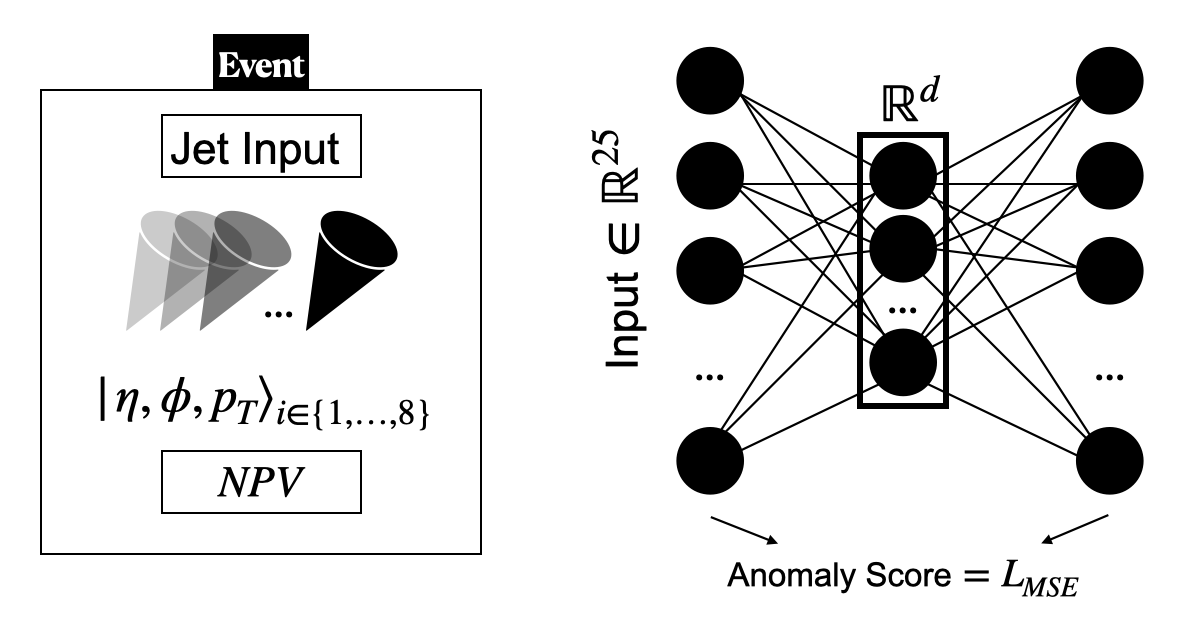}
    \caption{Schematic representation of the autoencoder model used for anomaly detection.}
    \label{fig:ae_model}
\end{figure}

A guiding consideration for this study is to develop an AD model that is comparable in size and scope to the actual models that have been implemented so far in the trigger hardware of the ATLAS and CMS experiments.
We thus seek to construct a compact and efficient model that can effectively perform AD tasks on the datasets described above, motivating a simple architecture that still satisfies the fundamental requirements of the project, and has a sufficient amount of data to be trained. 

Subject to these conditions, the anomaly detection model used in the trigger menu is trained on background samples. In simulation studies, the model is trained on an independent simulated background sample, while the data model is trained using the second longest run available for that year (Run 283876~\cite{dataG}). This is to ensure sufficient statistics while avoiding overlap with the evaluation sample for control studies~\cite{dataH}.

To this end, we trained a suite of models with varying latent dimensions and determined an AD score threshold for each that would achieve a rejection factor of 400 the test background sample, commensurate with typical L1 trigger requirements.
The efficiency for signals to exceed these thresholds is shown 
in Figs.~\ref{fig:signal_eff_vs_latent_data} and ~\ref{fig:signal_eff_vs_latent_MC} for both \ttbar\ and \haaFourB\ samples. These figures also show the fraction of signal events selected exclusively by the AD path by applying the same background rejection threshold to both \HT and AD. 

\begin{figure}[htbp]
    \centering
    \begin{subfigure}[t]{0.47\textwidth}
        \vspace{0pt}
        \centering
        \includegraphics[width=\linewidth,trim={0 0 5mm 0},clip]{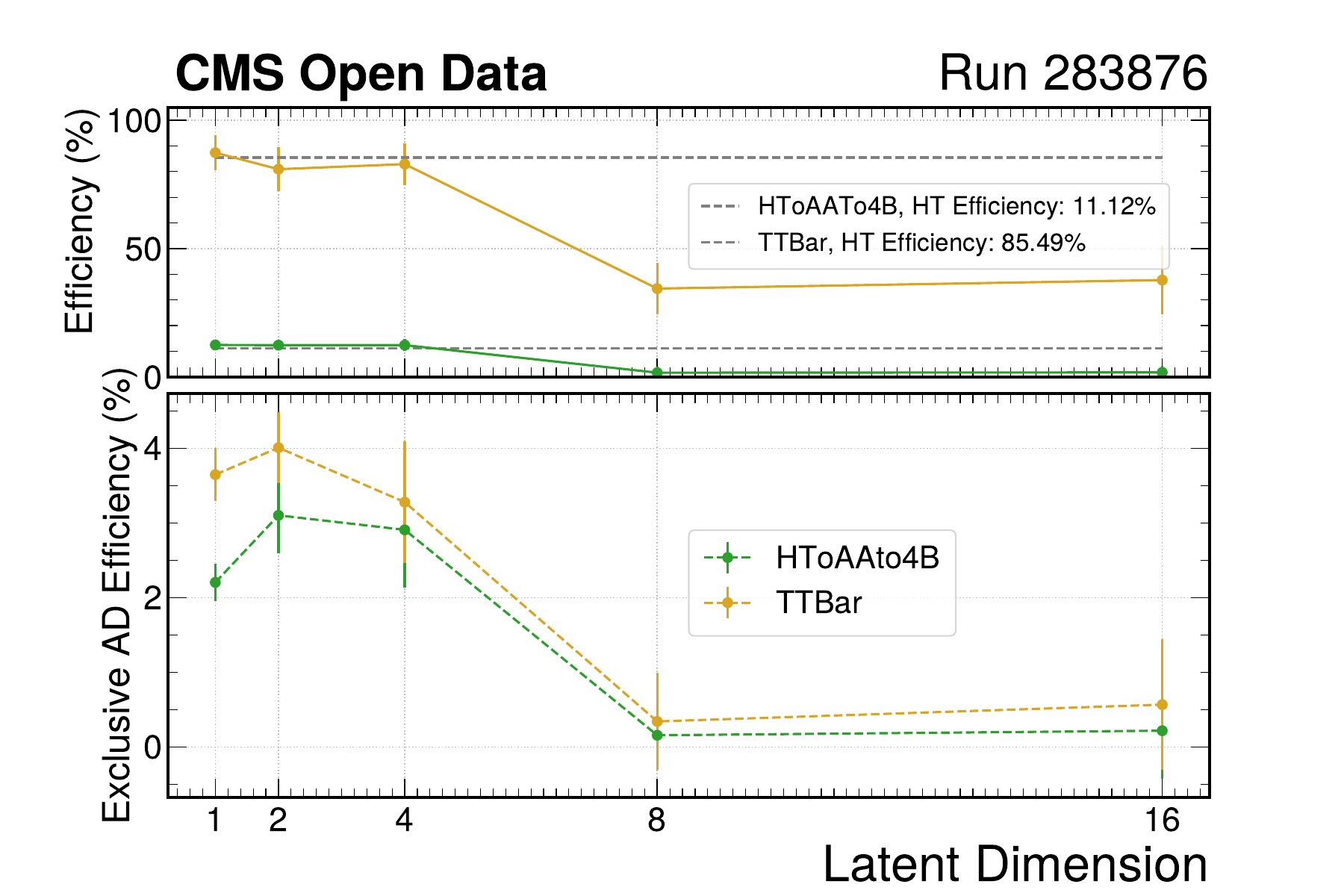}
        \caption{AD model trained on data background.}
        \label{fig:signal_eff_vs_latent_data}
    \end{subfigure}
    \hspace{0.02\textwidth}
    \begin{subfigure}[t]{0.47\textwidth}
        \vspace{0pt}
        \centering
        \includegraphics[width=\linewidth,trim={0 0 5mm 0},clip]{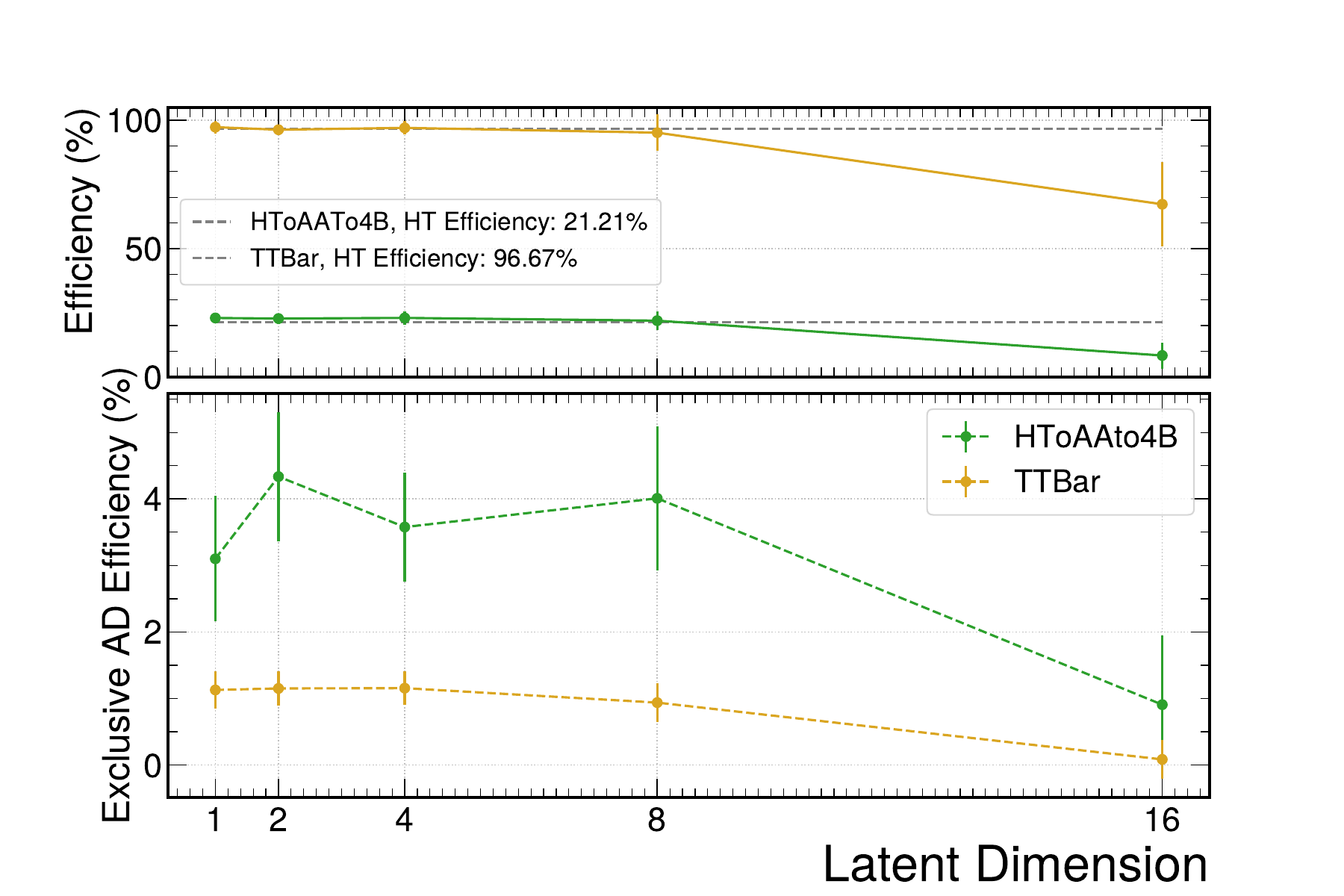}
        \caption{AD model trained on simulation background.}
        \label{fig:signal_eff_vs_latent_MC}
    \end{subfigure}

    \caption{
    Signal efficiency as a function of latent dimension for \ttbar\ (yellow) and \haaFourB\ (green) simulated signals.
    }
    \label{fig:signal_eff_vs_latent}
\end{figure}

It is observed that models with smaller dimensions show higher signal efficiency and larger exclusive contributions relative to the \HT trigger. 

While the available training statistics can influence the performance of higher-dimensional models, the presence of detector noise and data-driven fluctuations means that learning detailed representations does not necessarily improve discrimination. In practice, more compact latent spaces provide a more stable and robust separation between signal and background in data. Therefore, given the consistent performance, lower resource requirements, and resilience to noise, for the purposes of this benchmark study, the configuration with $d=2$ is adopted as the baseline.

Figure~\ref{fig:eff_ht} compares the signal efficiency obtained using an \HT trigger versus an AD trigger on simulated or real data events. The results show that for an \HT threshold and an AD score cut chosen to yield a consistent L1 rate reduction, the AD trigger recovers signal efficiency in the region below the \HT threshold, where the \HT trigger rejects essentially all signal events. When operated in parallel with \HT, the AD trigger therefore retains additional sensitivity at lower energy scales. This illustrates the complementarity of the AD  approach compared to energy sums alone, providing additional sensitivity to processes with lower energy scales for both the \ttbar\ and \haaFourB\ event topologies.

\begin{figure}[htbp]
    \centering
    \begin{subfigure}{0.45\textwidth}
        \centering
        \includegraphics[width=\linewidth,trim={0 0 5mm 0},clip]{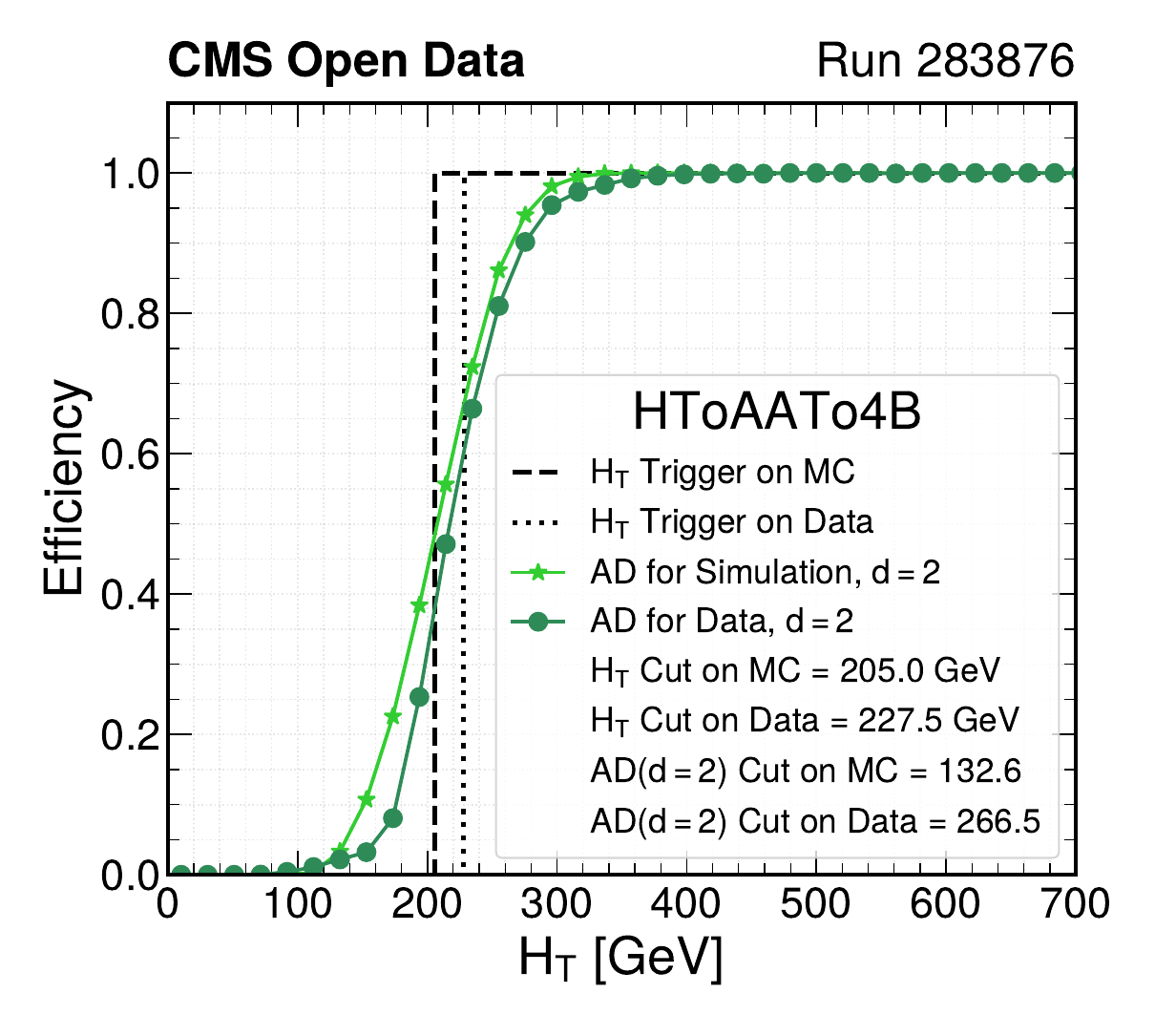}
        \caption{\haaFourB}
        \label{fig:eff_ht_a}
    \end{subfigure}\qquad
    \begin{subfigure}{0.41\textwidth}
        \centering
        \includegraphics[width=\linewidth,trim={5mm 0 0 0},clip]{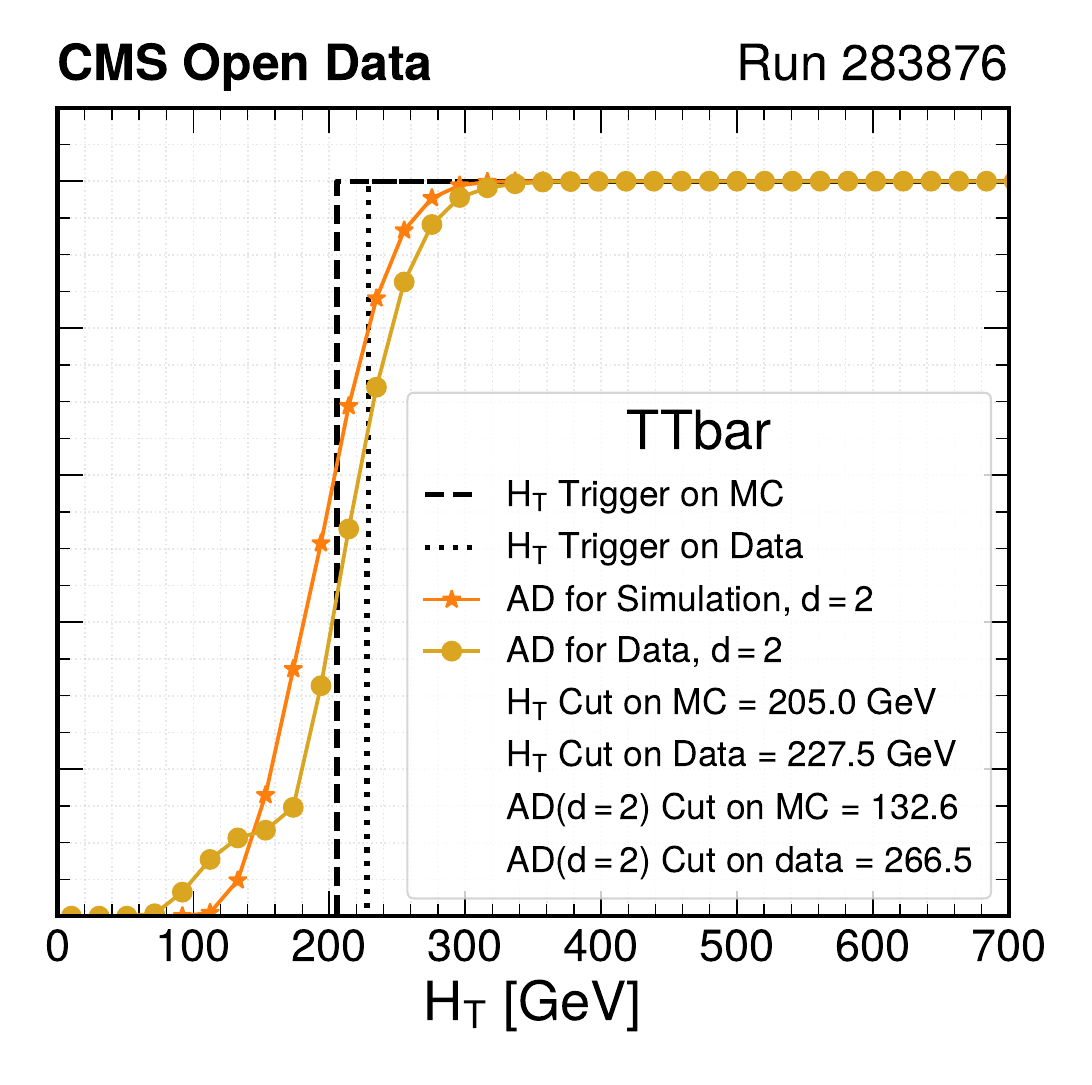}
        \caption{\ttbar}
        \label{fig:eff_ht_b}
    \end{subfigure}
    \caption{
    Efficiency as a function of \HT for (a) \haaFourB\ and (b) \ttbar\ simulated events.  
    This is a demonstration of signal efficiency improvement in the lower $p_T$ kinematic region when using anomaly detection. 
    }
    \label{fig:eff_ht}
\end{figure}

Finally, Fig.~\ref{fig:anomaly_score_hist} shows the anomaly score distributions for the models trained on simulation and real data. The high-score tail is most relevant for signal efficiency: improved separation between signal and background in this region translates directly into higher signal acceptance.

\begin{figure}[htbp]
    \centering
    \begin{subfigure}{0.44\textwidth}
        \centering
        \includegraphics[width=\linewidth,trim={0 0 5mm 0},clip]{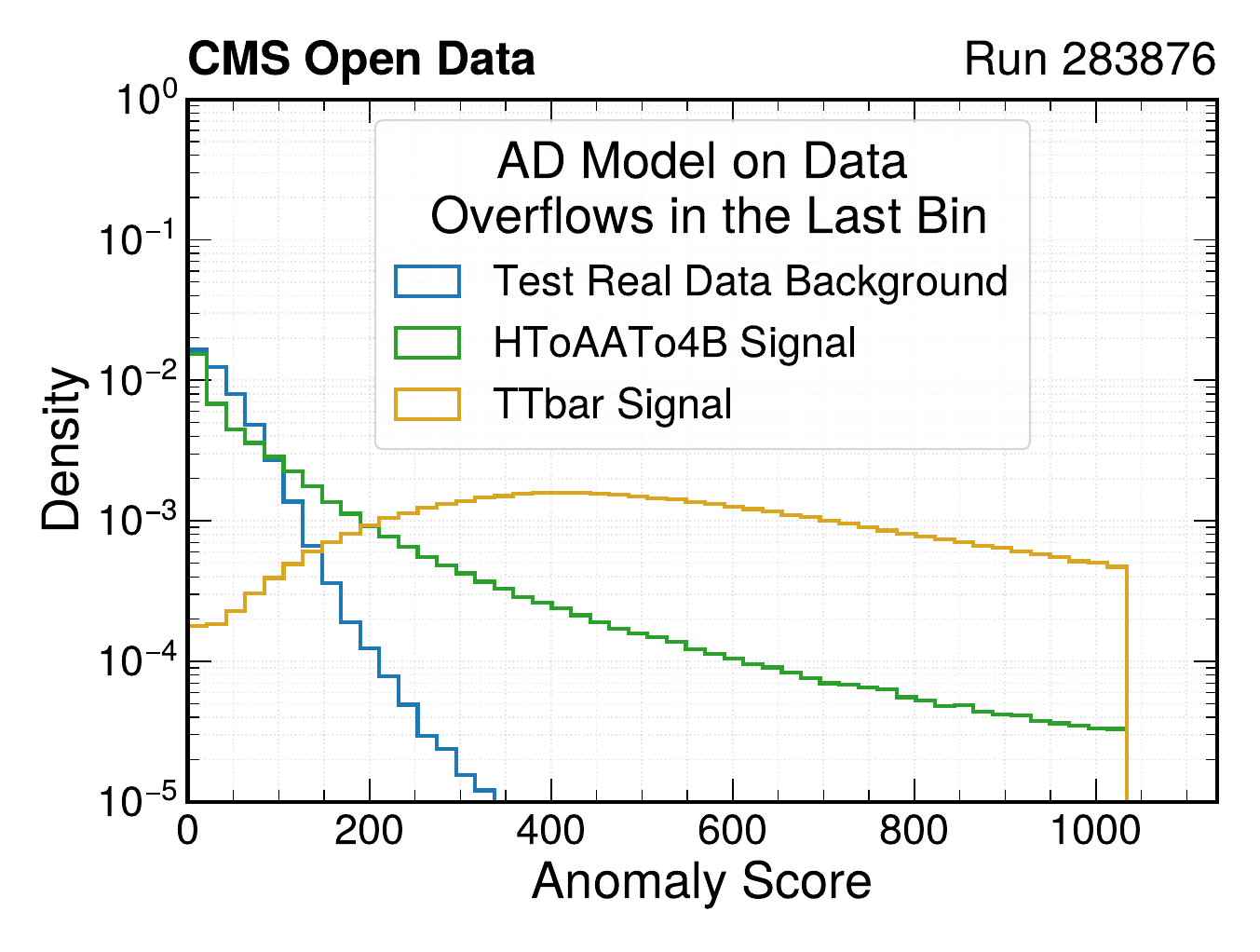}
        \caption{Data background.}
        \label{fig:AS_hist_a}
    \end{subfigure}\qquad
    \begin{subfigure}{0.46\textwidth}
        \centering
        \includegraphics[width=\linewidth]{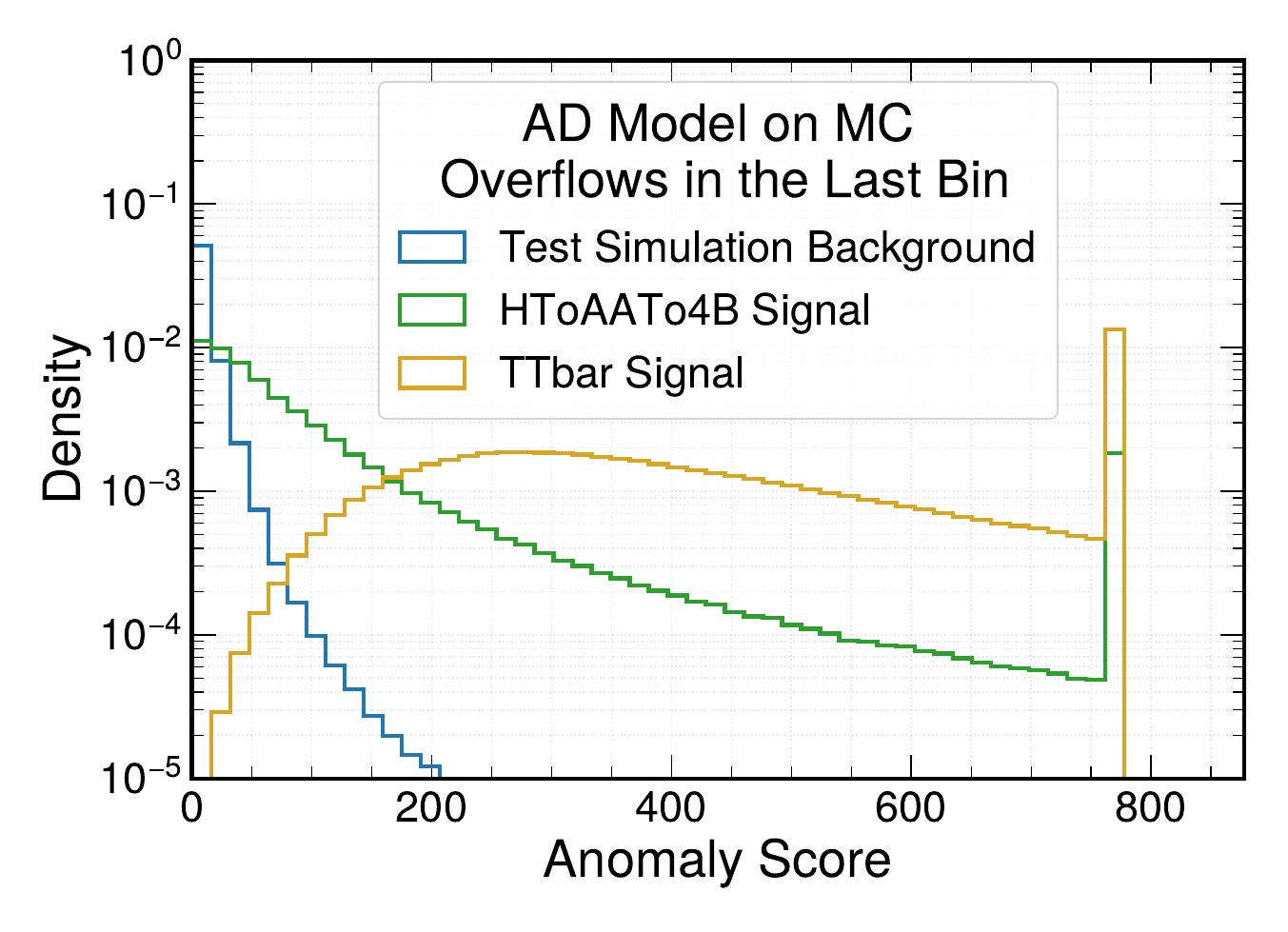}
        \caption{Simulation background.}
        \label{fig:AS_hist_b}
    \end{subfigure}

    \caption{
    In both cases, there is a clear separation between the background and the two signal samples: \ttbar\ and \haaFourB.
    }
    \label{fig:anomaly_score_hist}
\end{figure}

\section{Performance of a Fixed Trigger Menu in Simulated Events}
\label{sec:perf_fixed_menu}

In this section, we analyze the behavior of a fixed trigger menu on the simulation background without any control mechanisms. The study is done by allocating all of the bandwidth to only one item, which is either the \HT or the AD. The goal is to investigate how their distributions evolve over time and how the trigger rates and signal efficiencies are affected by evolving pileup conditions in our MC samples. This is the first step to formulate the problem before introducing any solutions.
The trigger menu encompasses a set of rules that determine whether an event is saved or rejected, which are determined in our setup by thresholds on an AD score or \HT. To begin, we consider a constant trigger menu, where these thresholds do not vary with time.

\begin{figure}[htbp]
    \centering
    \begin{subfigure}{0.45\textwidth}
        \includegraphics[width=\linewidth]{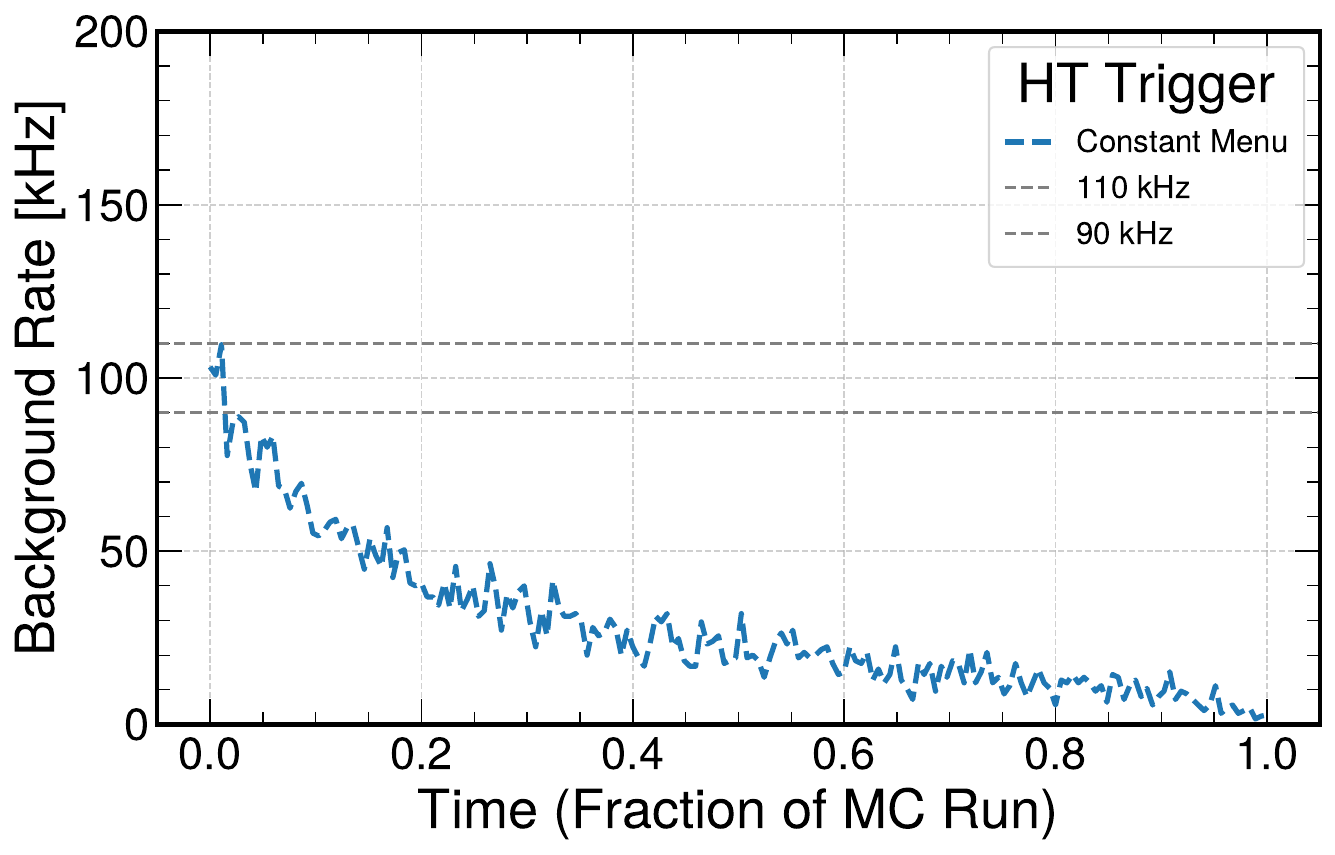}
        \caption{\HT trigger rate.}
        \label{fig:bkg_rate_a}
    \end{subfigure}
    \qquad
    \begin{subfigure}{0.45\textwidth}
        \includegraphics[width=\linewidth]{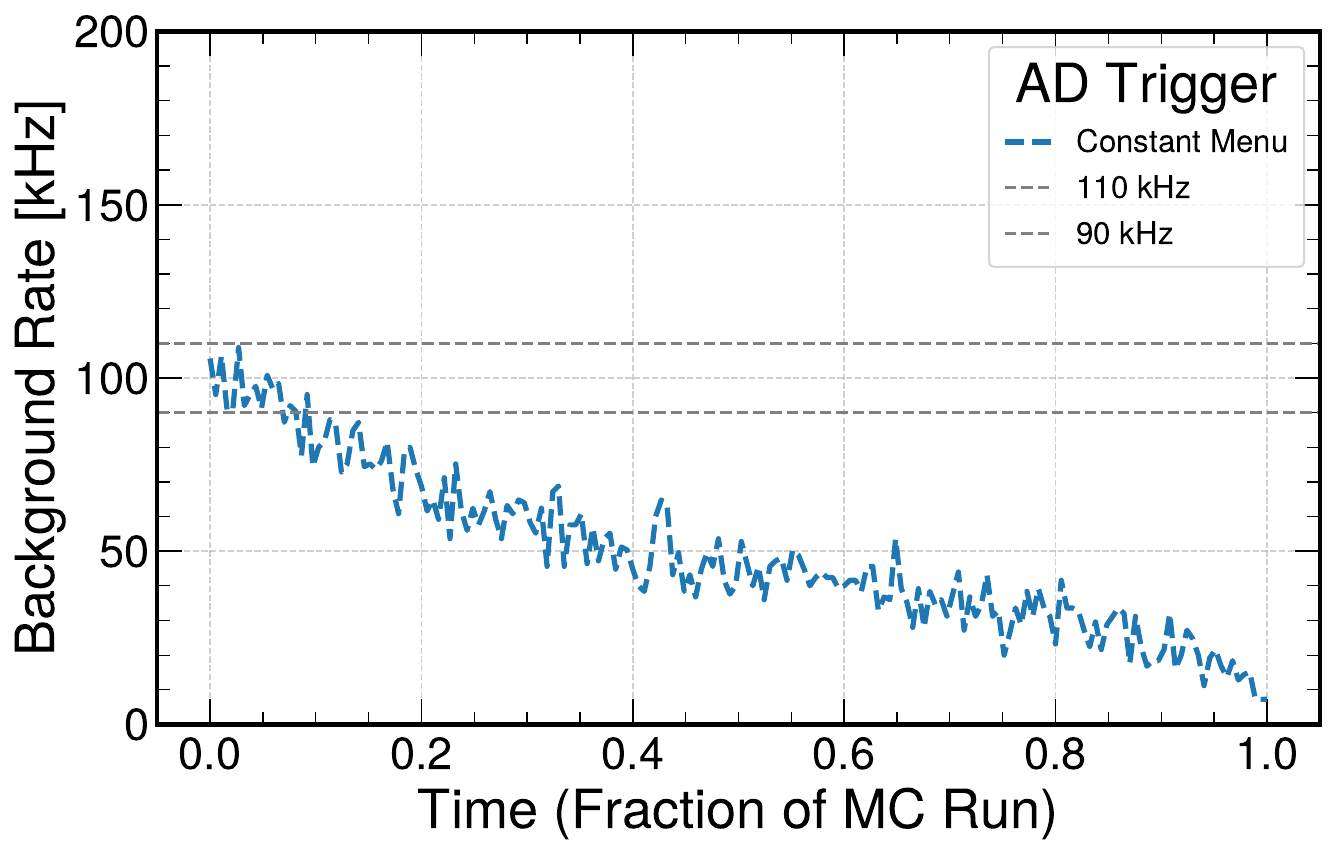}
        \caption{ AD  trigger rates.}
        \label{fig:bkg_rate_b}
    \end{subfigure}

    \caption{
        Evolution of background trigger rates under a constant threshold strategy on MC background sample. 
        (a) \HT trigger. 
        (b) AD trigger.
        Dashed lines provide visual guides.
        Significant rate drifts are observed due to evolving detector conditions. 
    }
    \label{fig:bkg_rate}
\end{figure}

In Fig.~\ref{fig:bkg_rate}, the fixed menu thresholds are determined by requiring a 400x rejection factor for events in the first few batches of data. In this work, a \textit{batch} refers to a fixed number of events from the data stream, processed sequentially. 
As the focus of this study is L1 triggers, 
 we target an acceptance rate of maximally  100~kHz from the background events, assuming 40~MHz collisions. 
 Assuming that this is the maximum tolerable rate given external constraints on the system (bandwidth and storage, for example) and our available resources, then ensuring it does not exceed the limit in the first batch serves as a reasonable safeguard that it will remain within bounds for the rest of the run.

For the \HT trigger, we observe a significant rate decrease as the \NPV decreases over time. This behavior reflects the expected sensitivity of \HT to pileup conditions: fewer interactions per bunch crossing lead to reduced jet activity and lower \HT values, an increasing number of which ultimately fall below the fixed threshold. This familiar decrease in trigger rates over the course of a fill is a well-documented phenomenon in LHC operations~\cite{ATLAS-Run3-Trig, CMS-HLT-Run2}. A similar trend is observed for the  AD  trigger. This suggests that its anomaly score inherits a conventional pileup dependence, analogous to traditional trigger items.

Figure~\ref{fig:sig_rate} shows the relative change in signal efficiency as a function of time, $\frac{\epsilon(t)}{\epsilon(t=0)}$, computed per batch of 50,000 events. This quantity reflects how the instantaneous trigger efficiency evolves, which is highly sensitive to changes in event characteristics such as pileup conditions. 
Figure~\ref{fig:G_sig_rate} complements this view by showing the relative change in cumulative signal efficiency instead, i.e., the cumulative fraction of signal events passing the trigger up to a given point in the run. 
This metric smooths out short-term fluctuations and highlights longer-term trends in trigger performance. These figures show how the ability of \HT and  AD triggers to capture signal events degrades over time due to pileup evolution.

\begin{figure}[htbp]
    \centering
    \begin{subfigure}{0.45\textwidth}
        \includegraphics[width=\linewidth]{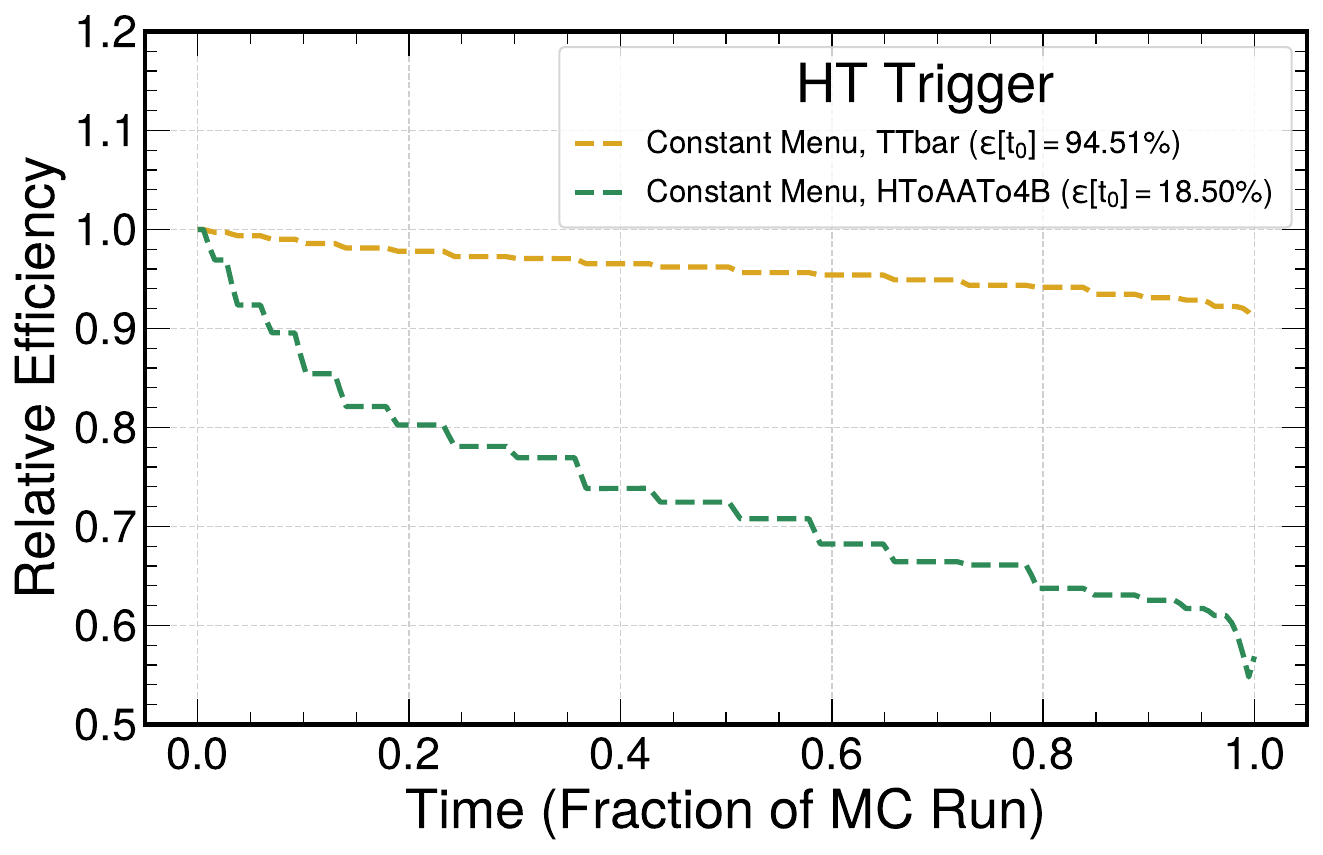}
        \caption{\HT trigger.}
        \label{fig:sig_rate_a}
    \end{subfigure}
    \qquad
    \begin{subfigure}{0.45\textwidth}
        \includegraphics[width=\linewidth]{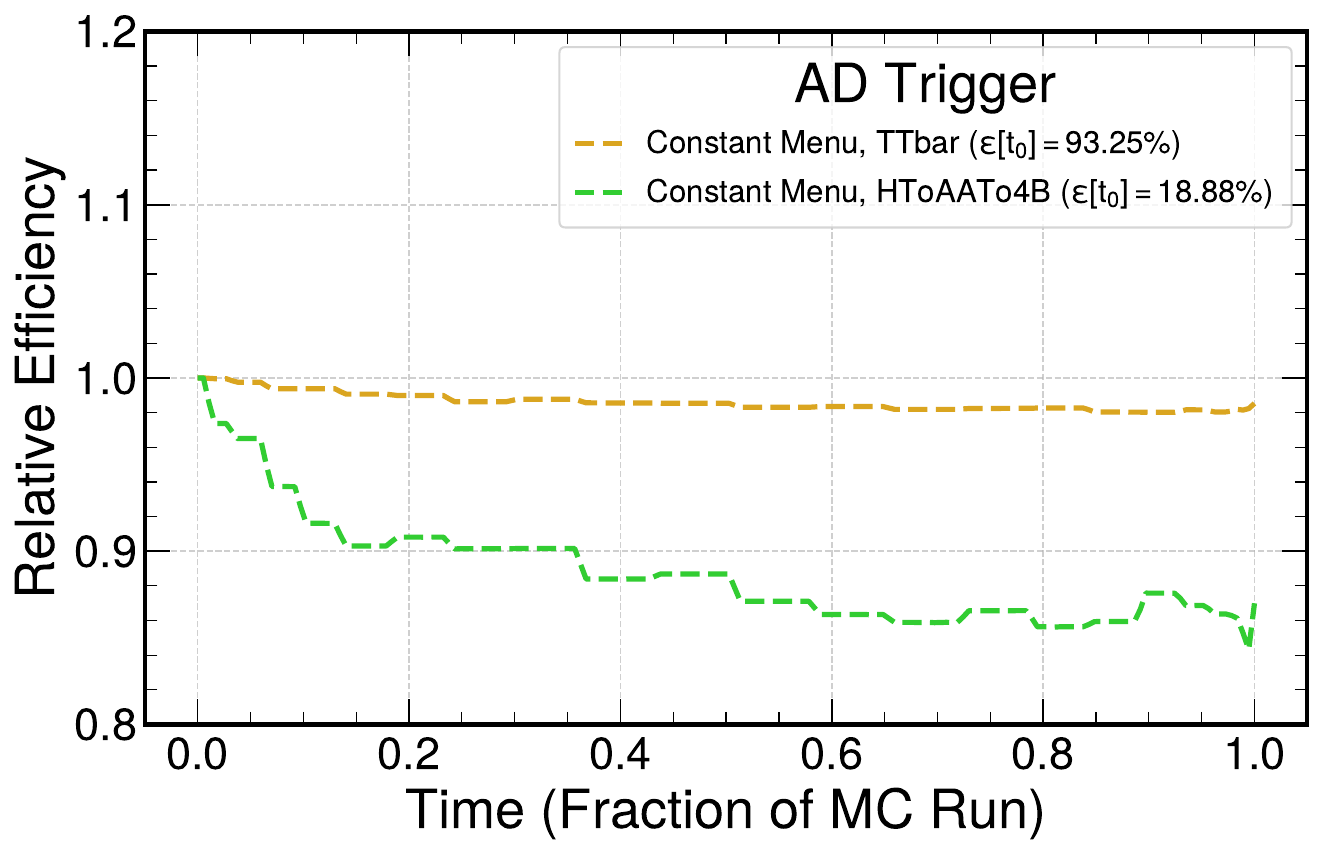}
        \caption{ AD   trigger.}
        \label{fig:sig_rate_b}
    \end{subfigure}

    \caption{
        Relative signal efficiency  
        (a) \HT trigger,
        (b)  AD   trigger.
        This quantity reflects the instantaneous trigger efficiency and is highly sensitive to changes in event characteristics such as pileup conditions.
    }
    \label{fig:sig_rate}
\end{figure}

\begin{figure}[htbp]
    \centering
    \begin{subfigure}{0.45\textwidth}
        \includegraphics[width=\linewidth]{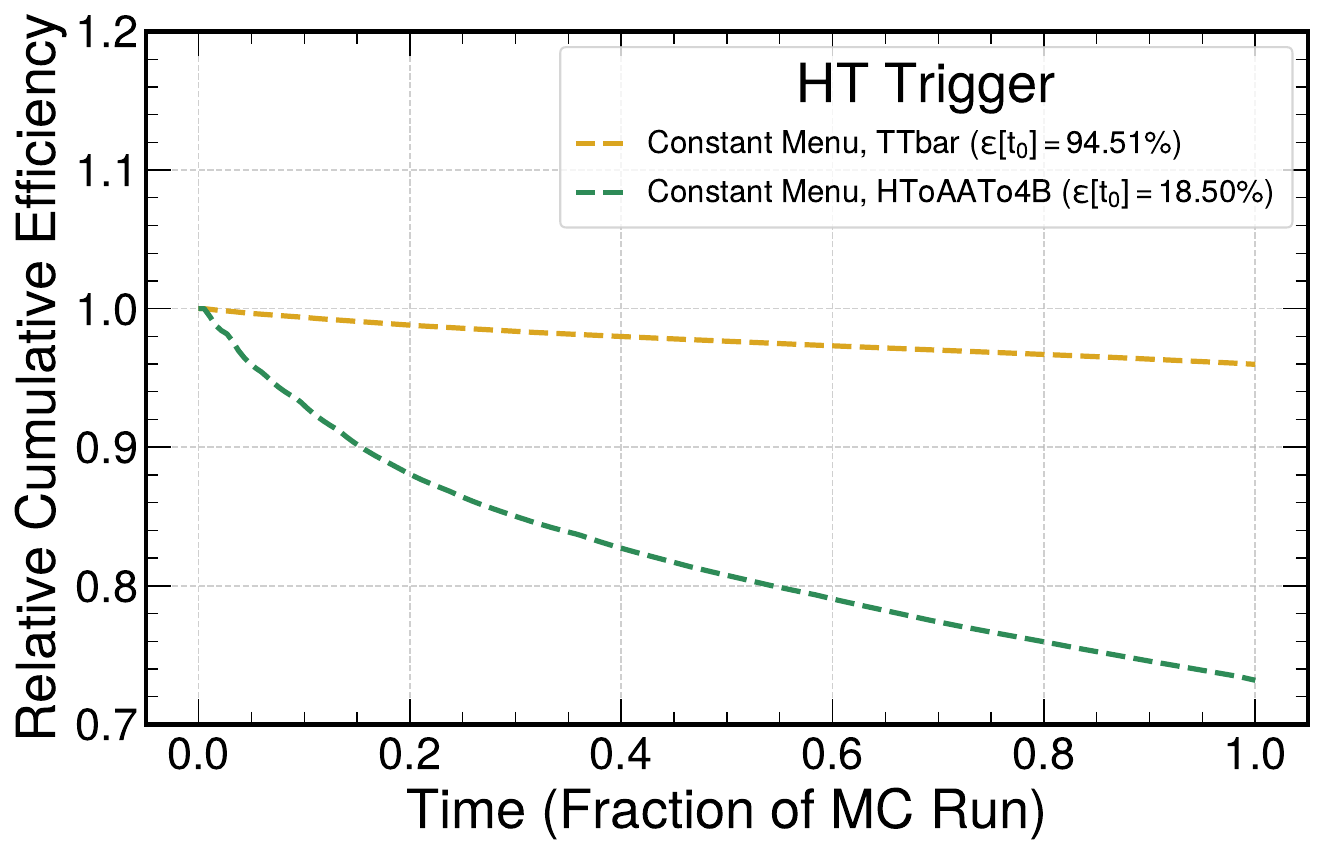}
        \caption{\HT trigger.}
        \label{fig:G_sig_rate_a}
    \end{subfigure}
    \qquad
    \begin{subfigure}{0.45\textwidth}
        \includegraphics[width=\linewidth]{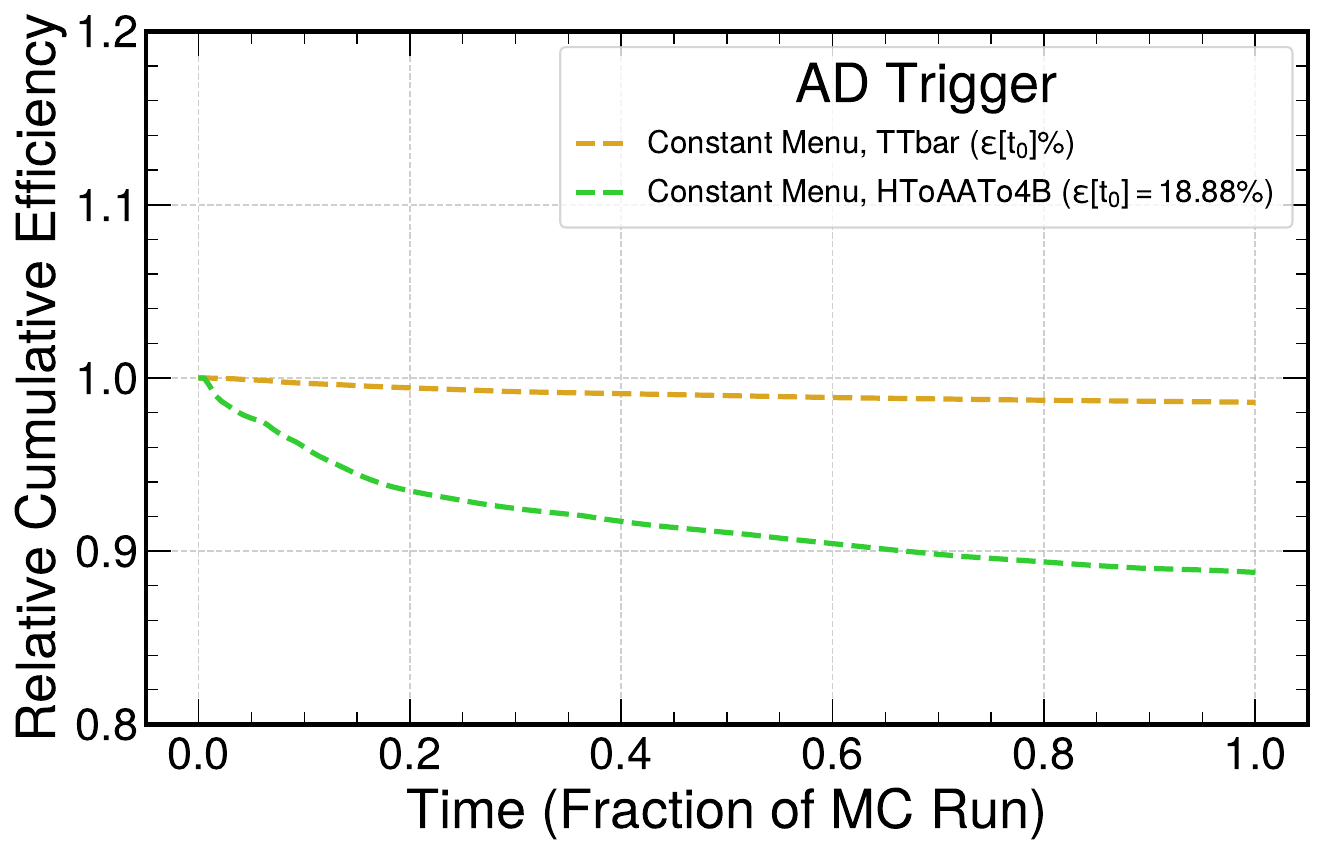}
        \caption{ AD   trigger.}
        \label{fig:G_sig_rate_b}
    \end{subfigure}

    \caption{
        Relative cumulative signal efficiencies as a function of time
        (a) \HT trigger,
        (b)  AD   trigger. 
        The plots show how static thresholds lead to degraded performance as pileup decreases.
    }
    \label{fig:G_sig_rate}
\end{figure}

Ultimately, the loss of efficiency highlights a key limitation of static trigger menus: without adaptation, valuable signal events are increasingly missed as detector and beam conditions evolve.

\section{Real-time Control for Individual Trigger Rules}\label{sec:real_time_control}

The primary objective for a trigger controller is to achieve a stable data collection rate over time (with varying detector conditions) through the dynamic adjustment of trigger rules.
Without a perfect model for all aspects of the changing experimental context (i.e., the environment with which the controller interacts), we invoke the classic solution from control theory of a linear feedback mechanism.
Here, observed deviations from the target event rate in one batch of data can lead to proportionate adjustments to trigger thresholds in the subsequent batch.

The Proportional-Integral-Derivative (PID) controller is a natural extension of this approach found across a broad range of applications in daily life where parameters must be regulated in evolving environments \cite{Bradu:2018cryogenics, Steinhagen:2007orbit}.
The PID algorithm computes corrections based on the deviation  $e(t)$ from a target value:
\begin{equation}
    \Delta u(t) = K_p \,e(t) + K_d \frac{de(t)}{dt} + K_I \int_0^t e(t') dt', \nonumber
\end{equation}
where $K_p$, $K_d$, and $K_I$ are tunable constants governing proportional, derivative, and integral corrections. Typically, the proportional term sets the dominant behavior, while the derivative and integral terms contribute to predictive adjustments and error accumulation corrections, respectively. In our toy model, the regulated parameter is the trigger threshold for a single algorithm (either \HT or AD), which seeks to maintain a fixed background rate, taken to be $r_t=100$~kHz as in Section~\ref{sec:perf_fixed_menu}.

Through comparisons of controllers with various gain value configurations, we find no significant impact from the presence of integral term feedback ($K_I [\sum_{t=t_0}^{t=t_N} e(t)]$).
Consequently, we implement the simpler PD loop to regulate menu items dynamically, noting that even the improvement from including a derivative ($K_d [e(t_{N}) - e(t_{N-1})]$) is relatively modest. Optimal values of the controller gain constants $K_p$ and $K_d$ are obtained from grid searches, which are performed separately for each trigger algorithm.

Figures~\ref{fig:bkg_rate_pid}, \ref{fig:sig_rate_pid}, and~\ref{fig:G_sig_rate_pid} summarize 
the impact of the PD controller on background stability and signal efficiency for both the 
\HT and  AD  triggers.
Figure~\ref{fig:bkg_rate_pid} shows the background trigger rates under PD control compared to 
fixed menus. In both the \HT (Fig.~\ref{fig:bkg_rate_pid_a}) and  AD  (Fig.~\ref{fig:bkg_rate_pid_b}) 
cases, the controller compensates for the time dependence of the collision environment and maintains 
the rate within the desired tolerance band, while the fixed-threshold menus exhibit a clear drift.
The batch-to-batch variations in rate observed in the controller scenarios are due to the statistical uncertainty associated with the finite batch size.

The effect on the signal is illustrated in Fig.~\ref{fig:sig_rate_pid}, which presents the relative 
change in instantaneous signal efficiency as a function of the fraction of run time. For both the 
\HT and  AD  triggers, the PD-controlled menus maintain higher and more stable efficiencies than 
their fixed counterparts, mitigating the degradation that would otherwise occur as pileup decreases.
Figure~\ref{fig:G_sig_rate_pid} reports the relative change in cumulative signal efficiency 
over the full run. Here, the benefit of the dynamic strategy becomes particularly pronounced: 
PD-controlled thresholds enhance the total number of recorded signal events by up to about 70\% 
with respect to fixed menus, while still 
maintaining the background rate constraint.

\begin{figure}[htbp]
    \centering
    \begin{subfigure}{0.45\textwidth}
        \includegraphics[width=\linewidth]{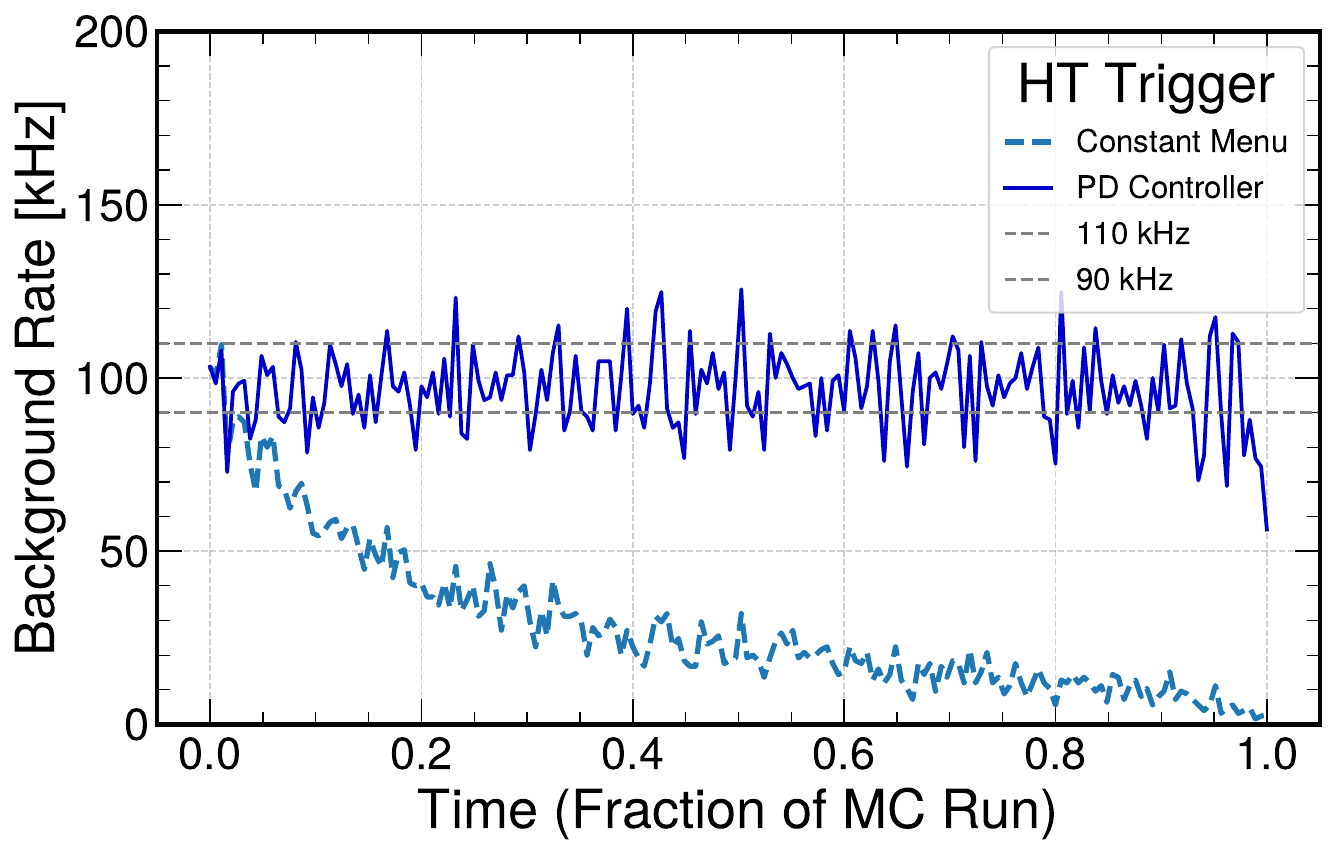}
        \caption{\HT trigger.}
        \label{fig:bkg_rate_pid_a}
    \end{subfigure}
    \qquad
    \begin{subfigure}{0.45\textwidth}
        \includegraphics[width=\linewidth]{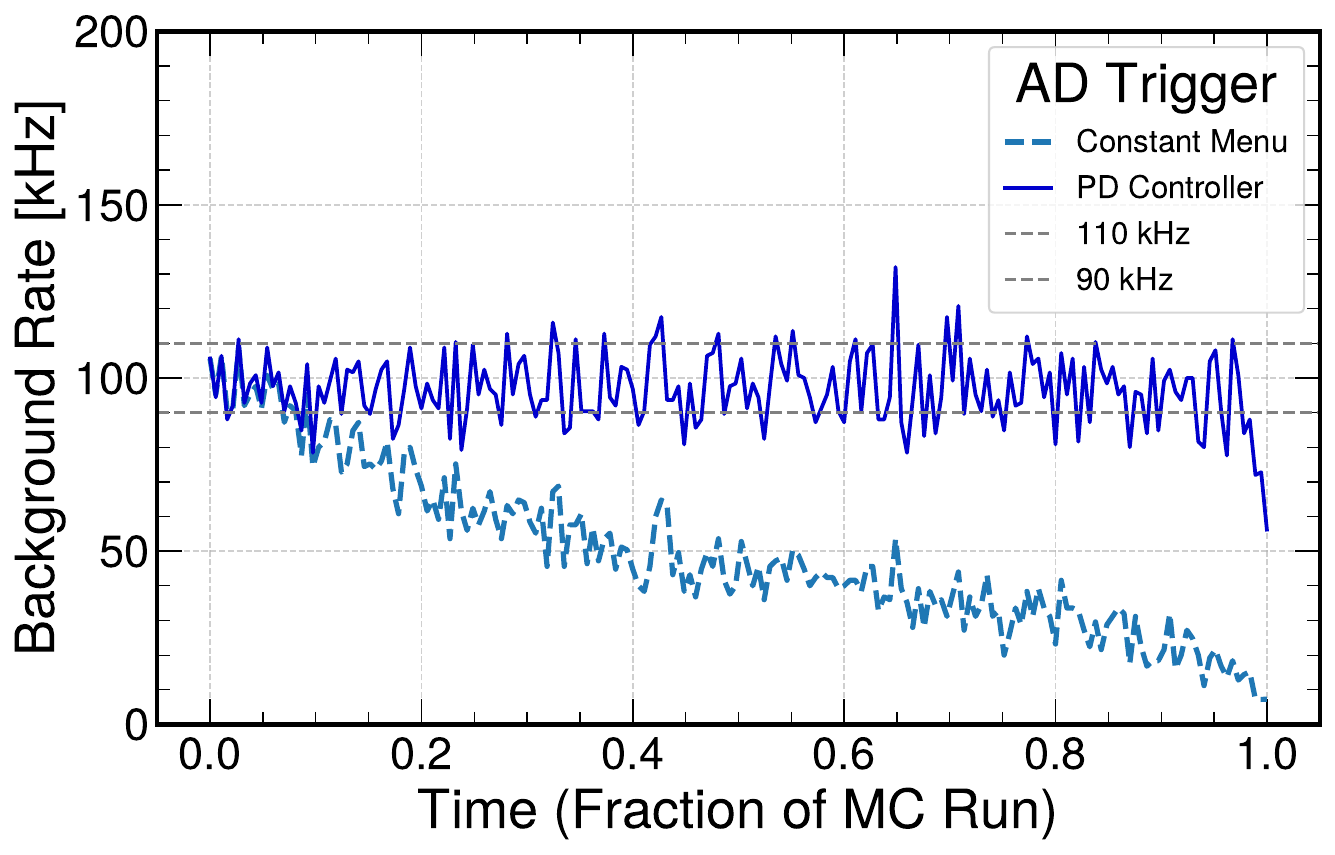}
        \caption{ AD   trigger.}
        \label{fig:bkg_rate_pid_b}
    \end{subfigure}

    \caption{
        Background trigger rates under PD control. 
        (a) \HT trigger,
        (b)  AD   trigger. 
        The controller stabilizes the background rate within the target tolerance band and mitigates the drift observed with fixed menus.
    }
    \label{fig:bkg_rate_pid}
\end{figure}

\begin{figure}[htbp]
    \centering
    \begin{subfigure}{0.45\textwidth}
        \includegraphics[width=\linewidth]{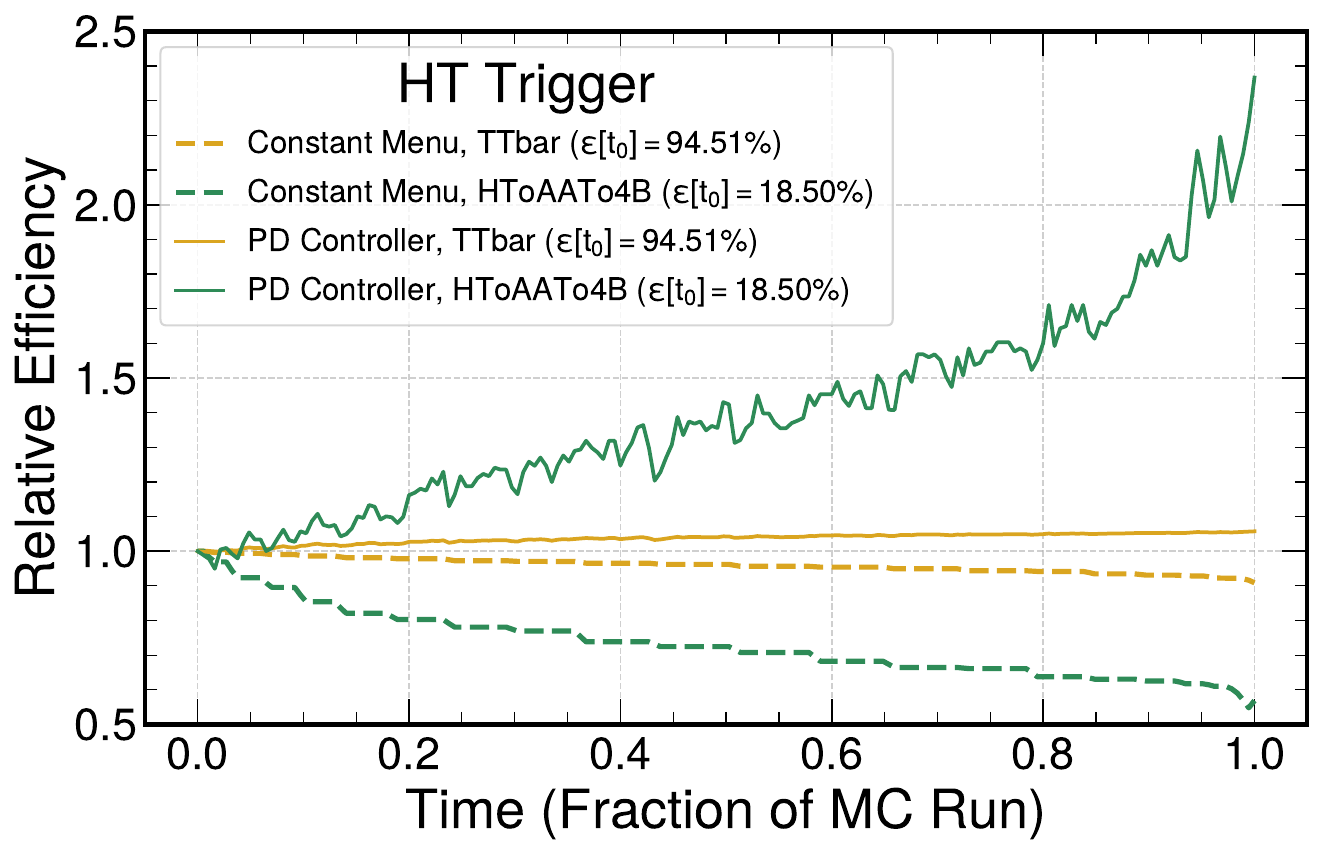}
        \caption{\HT trigger.}
        \label{fig:sig_rate_pid_a}
    \end{subfigure}
    \qquad
    \begin{subfigure}{0.45\textwidth}
        \includegraphics[width=\linewidth]{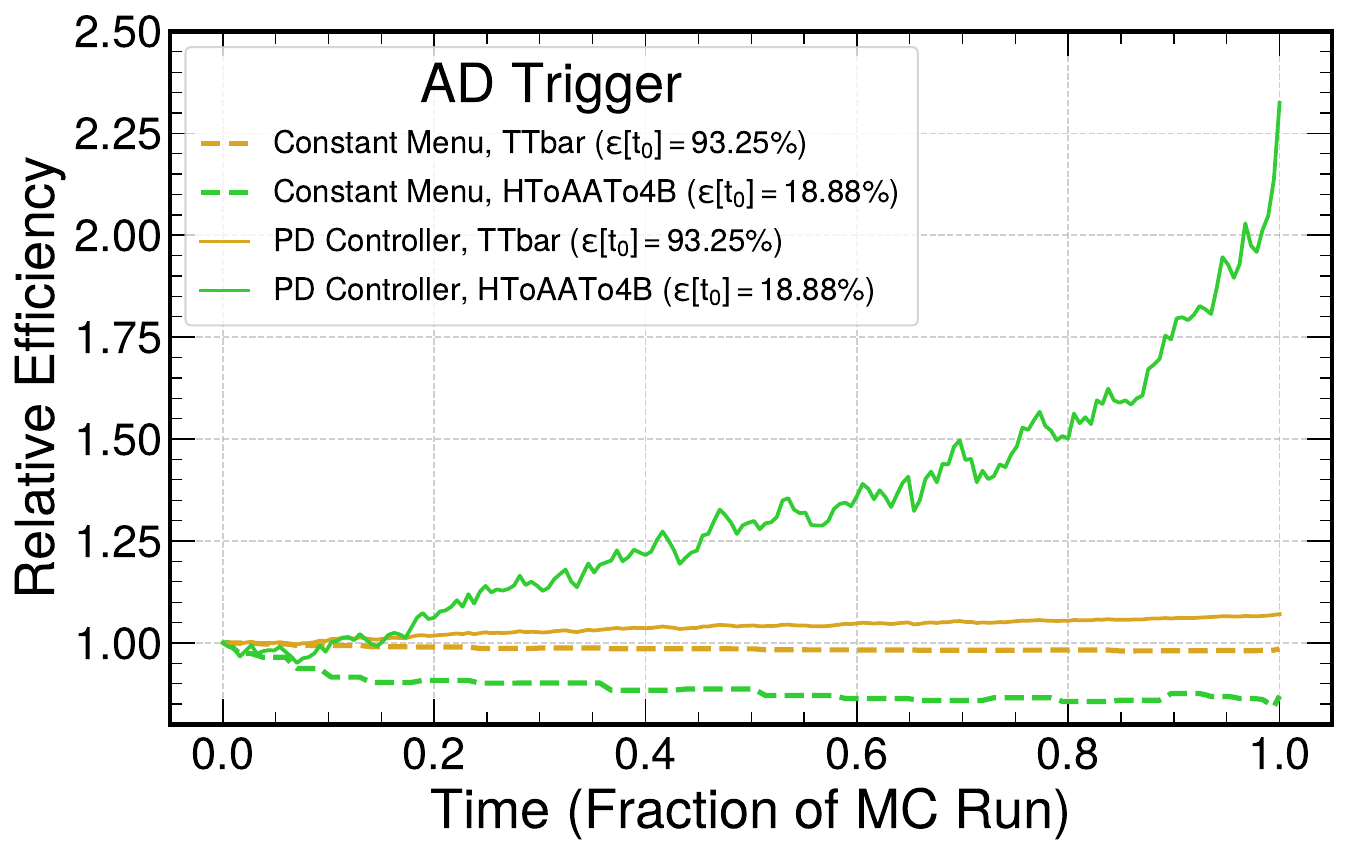}
        \caption{ AD   trigger.}
        \label{fig:sig_rate_pid_b}
    \end{subfigure}

    \caption{
        Relative change in instantaneous signal efficiency over time. 
        (a) \HT trigger,
        (b)  AD   trigger. 
        PD-controlled menus maintain higher and more stable signal efficiencies compared to fixed menus.
    }
    \label{fig:sig_rate_pid}
\end{figure}

\begin{figure}[htbp]
    \centering
    \begin{subfigure}{0.45\textwidth}
        \includegraphics[width=\linewidth]{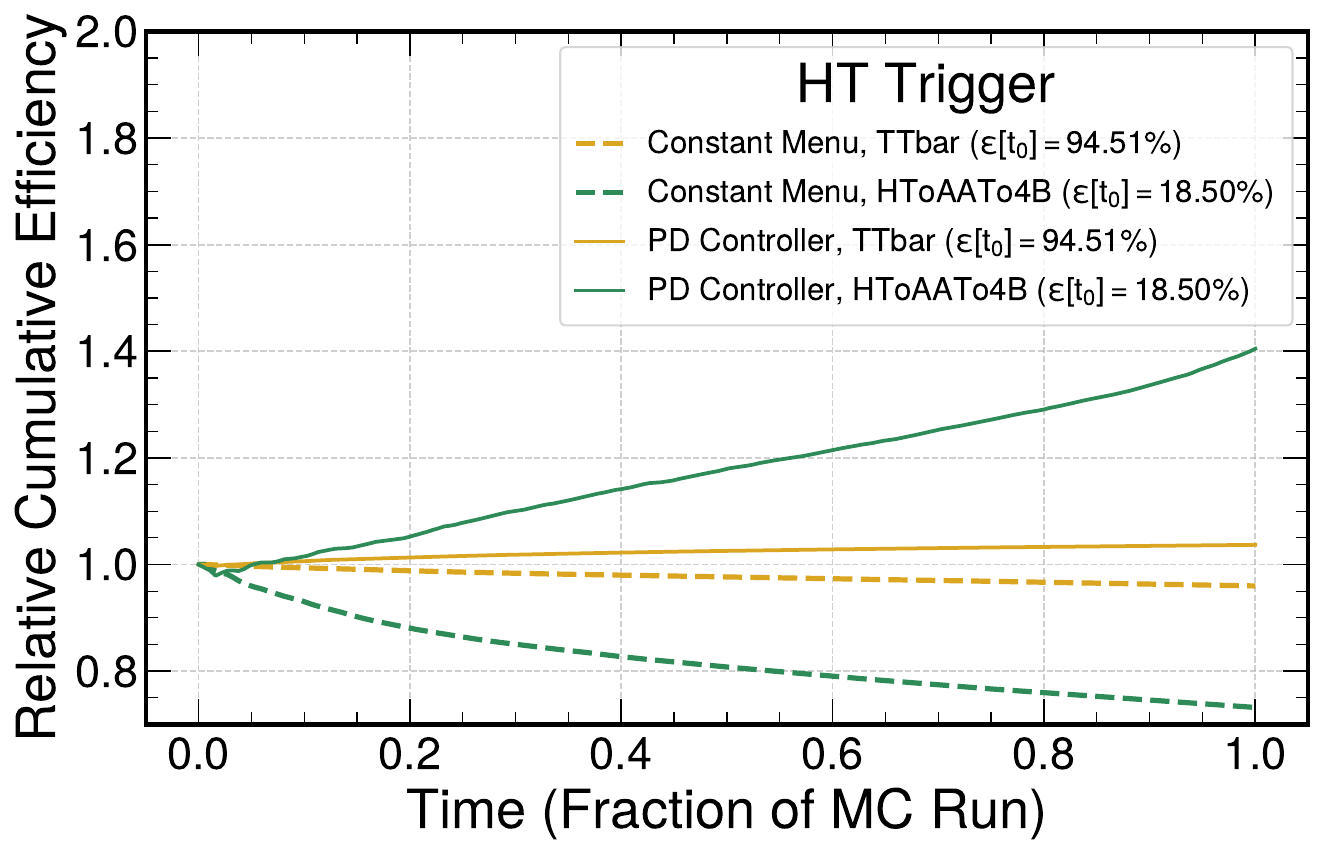}
        \caption{\HT trigger.}
        \label{fig:G_sig_rate_pid_a}
    \end{subfigure}
    \qquad
    \begin{subfigure}{0.45\textwidth}
        \includegraphics[width=\linewidth]{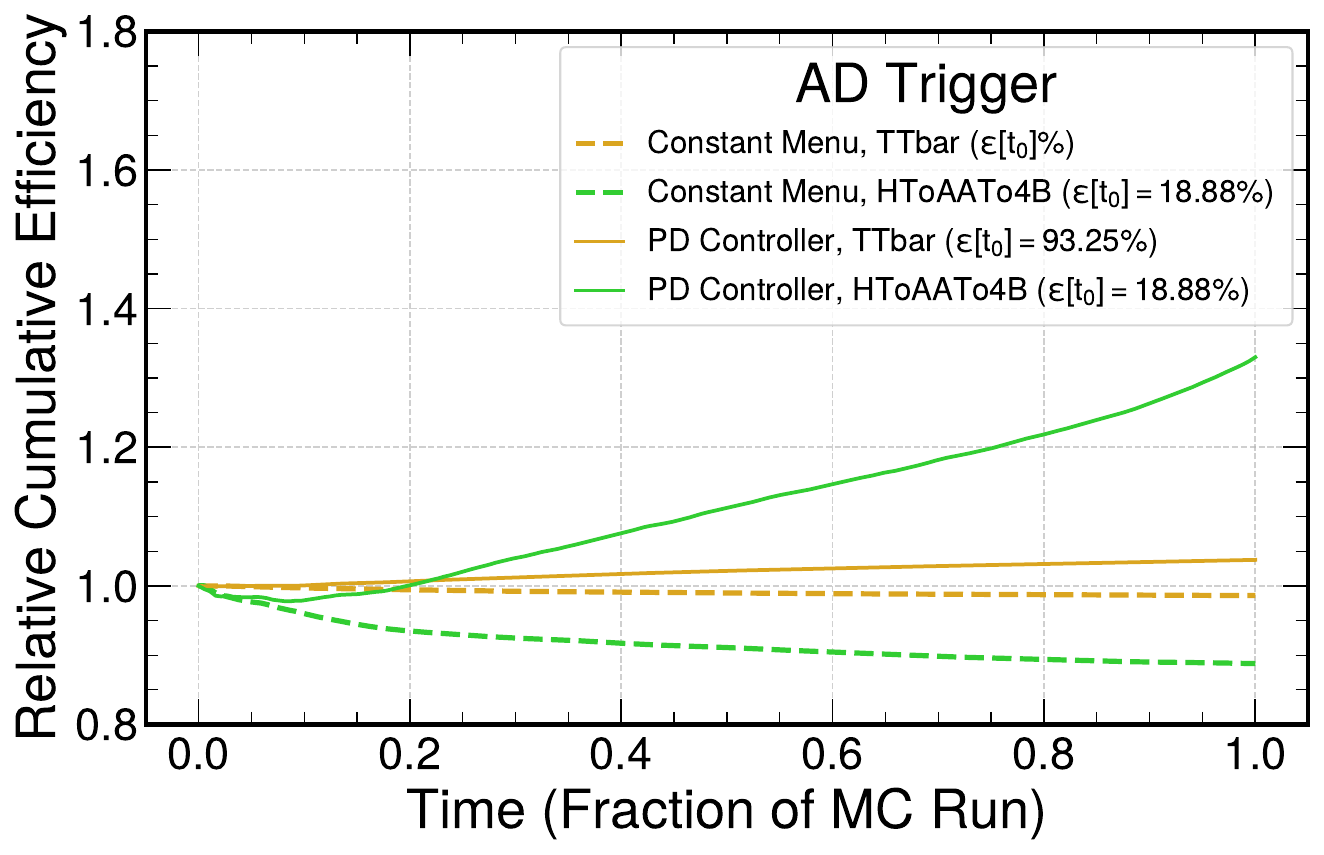}
        \caption{ AD   trigger.}
        \label{fig:G_sig_rate_pid_b}
    \end{subfigure}

    \caption{
        Relative change in cumulative signal efficiency over time. 
        (a) \HT trigger, 
        (b)  AD   trigger. 
        PD-controlled strategies substantially (up to 70\%) improve the accumulation of signal events compared to fixed threshold menus, especially for pileup-dependent triggers.
    }
    \label{fig:G_sig_rate_pid}
\end{figure}


\section{Multi Trigger Control Framework}\label{section6}

While the PID loop approach demonstrated effectiveness in maintaining steady trigger rates for intrinsically unstable input features, this simple framework is inherently limited to control of a single target parameter using a single tunable parameter. 
It furthermore relies on tuning several gain constants of the model, which can themselves vary over time in principle, and may also belie a deeper dependence on multiple effects contributing on different timescales. 
In a real experimental setup, multiple trigger paths operate concurrently, deciding whether to accept a common stream of events interpreted under different hypotheses, requiring a more sophisticated control strategy. 
Viewing each trigger algorithm as a simple function of a single parameter, the action space for a controller manipulating an $n$-path trigger system is an $n$-dimensional space of thresholds.
The primary operational objective of constraining the total data taking rate as near as possible to some predetermined target cuts is an $n-1$ hypersurface upon which the controller's prescription should lie.

Though one may consider constructing independent PID loops for each trigger, this approach is cumbersome and challenging to define in a straightforward way.
It is not trivial to define distinct target rates for each trigger independently, or more generally assign value to their positive and negative correlations. 
Instead, inspired by PID's principle of minimizing the error term at each time step through corrective actions, we extend this concept to defining a global cost function $C$, as a function of a set of variables and hyperparameters, that can be dynamically assessed and minimized throughout a run. 
Though a neural network is not required to perform this optimization, such a cost function would be a natural candidate to direct the training of an agent-based controller.

The cost function can be defined based on a set of constraint functions, \( f_i(\vec x; \theta)  \), each representing a system requirement (e.g., background rate below a threshold). A strict control strategy would enforce that all constraints are exactly satisfied, mathematically corresponding to requiring \( f_i = 0 \) for all i, at $\vec{x}=\vec{x_0}$, where each function is defined to be zero when the corresponding condition is fulfilled.
However, exact satisfaction may not always be feasible in practice. Instead, a soft-constrained approach is adopted, where deviations from the ideal values are tolerated and prioritized through a set of weights. In other words, each requirement is formulated as a scalar function of control variables and associated hyperparameters (constraint functions), and the cost decreases as the constraints are better satisfied. For this purpose, it seems natural to define the terms such that negative cost is not allowed and each term is dimensionless.

To this end, each constraint function \( f_i\) is assigned a weight \( w_i \), which reflects the relative importance of that constraint. 
Another natural interpretation of the weights is to consider their inverses as setting a tolerance scale $\sigma_i = \frac{1}{w_i}$, with $\sigma_i$ representing the size of an `acceptable' deviation for the i-th constraint.
Any sub-constraint $f_i$ that varies from the ideal target by this magnitude will result in an identical penalty, i.e., contribution to the total cost function:
\begin{equation}
    \text{C} = \sum_{i=1}^{N} w_i f_i = \sum_{i=1}^{N} \frac{f_i}{\sigma_i}. \nonumber
\end{equation}
This formulation enables the controller to balance multiple competing objectives, each contributing according to an acceptable tolerance scale, while avoiding hand-tuning the many gain constants that would be necessary in a generalized PID approach. 

In this scheme, the first cost function that one might consider is $C_0 = |r_b-r_t|/\sigma_b$, in which $r_b$ is the total data taking rate dominated by background processes and $r_t$ is its target. The scale $\sigma_b$ would serve to normalize the contribution of $C_0$ if it were summed with other contributions to the total cost. However, in the presence of multiple trigger paths, $C_0$ is minimized by many degenerate actions of the controller. It is natural to break this degeneracy by extending $C$ to incorporate more sophisticated operational priorities into the full control strategy.

The remainder of this work is dedicated to studying this framework in simulated and real collision data sets, making use of our toy model of a hadronic trigger menu that is based on an \HT and an  AD   path. In this context, the controller design must consider trade-offs between competing goals, such as maximizing signal efficiency from the MC library, allocating resources to exploratory paths, and respecting resource limits. To investigate how different priorities influence controller behavior, we define a set of representative case studies, each built around a distinct objective function:

\begin{itemize}
    \item  Case 1 focuses on maximizing the efficiency to collect target signal processes while also maintaining a fixed total background rate.
   \item  Case 2 introduces dedicated bandwidth allocation for an exploratory path (e.g., anomaly detection), while ensuring that the trigger performance for a known benchmark signal (here \ttbar) is not sacrificed.
    \item Case 3 considering the highly asymmetric computational cost of each trigger path in the optimization, highlighting trade-offs between idealized physics performance and finite computational resources.
\end{itemize}

Each case illustrates how the controller's overall menu configuration and response to shifting conditions are shaped by the chosen optimization strategy.
To decouple the impact of the cost function selection from the design of the control algorithm itself, we first study the behavior of an (impossibly) ideal controller, which `looks into the future' to set trigger thresholds for each batch that will minimize each cost. This step of the study makes sure the cost function definition and format are reasonable to be used in a real control framework. 

\subsection{Case 1: Background-Signal Tradeoff}

In a multi-path trigger configuration, a fundamental goal is to optimize signal collection while maintaining a stable background rate within bandwidth constraints. This trade-off can be formalized through a simple cost function that simultaneously prioritizes both objectives:
\begin{equation}
C_1 = \frac{|r_b - r_t|}{\sigma_b} + \frac{1-\epsilon_s}{\sigma_s}. \nonumber
\end{equation}
Here, $r_b$ is the total rate of background events, $r_t$ is the target rate, and $\epsilon_s$ denotes the overall signal efficiency, calculated on the total set of signal processes ($t\bar t$ and \haaFourB). The scales $\sigma_b$ and $\sigma_s$ represent allowed deviations for each constraint term that determine the relative importance of background stability and signal efficiency.
As such, these weights inherently encode experimental priorities, for which we must make representative choices.

Figure~\ref{fig:ideal_controller1} summarizes trigger performance under the Case 1 cost, for a choice of $\sigma_b = 4\,\text{kHz}$ and $\sigma_s = 0.05$. Panel \ref{fig:case1_a} shows the total background rate and its decomposition into the \HT and  AD   trigger paths, illustrating how the ideal controller holds the combined background near the target. Panel \ref{fig:case1_b} reports the instantaneous and cumulative signal efficiencies, indicating how the policy prioritizes signal acquisition over time. Panel \ref{fig:case1_c} focuses on the BSM benchmark \haaFourB, illustrating its total efficiency and the split between \HT and  AD. Panel \ref{fig:case1_d} similarly breaks down the \ttbar\ signal efficiency, showing contributions from both paths and the overall high efficiency for this well studied process.

\begin{figure}[htbp]
    \centering
    \begin{subfigure}[t]{0.45\textwidth}
        \centering
        \includegraphics[width=\linewidth,trim={0 0 2mm 0},clip]{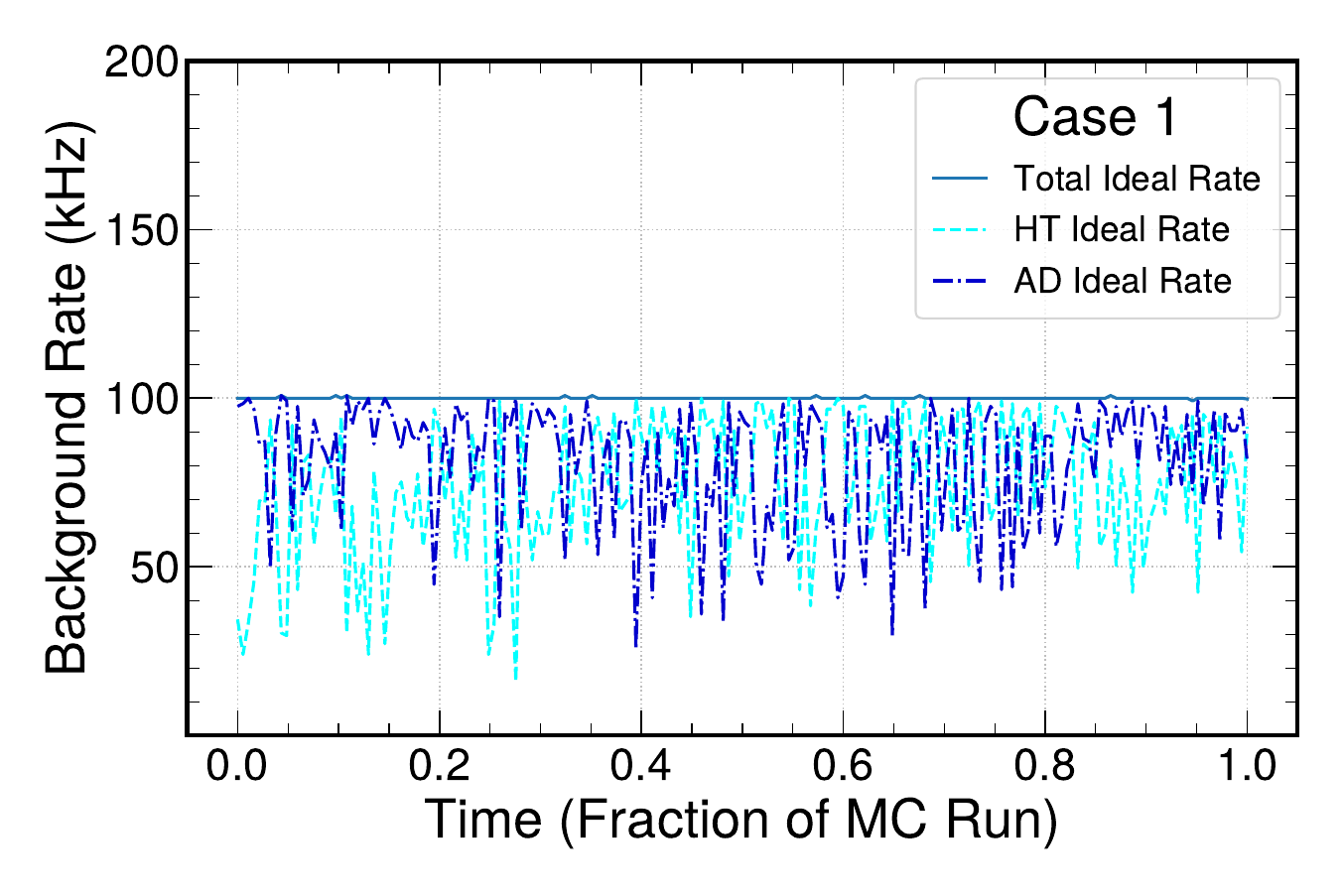}
        \caption{Background rates: total (solid blue) and contributions from \HT (cyan dashed) and  AD   (blue dash-dotted), showing stable control near the target.}
        \label{fig:case1_a}
    \end{subfigure}\qquad
    \begin{subfigure}[t]{0.45\textwidth}
        \centering
        \includegraphics[width=\linewidth]{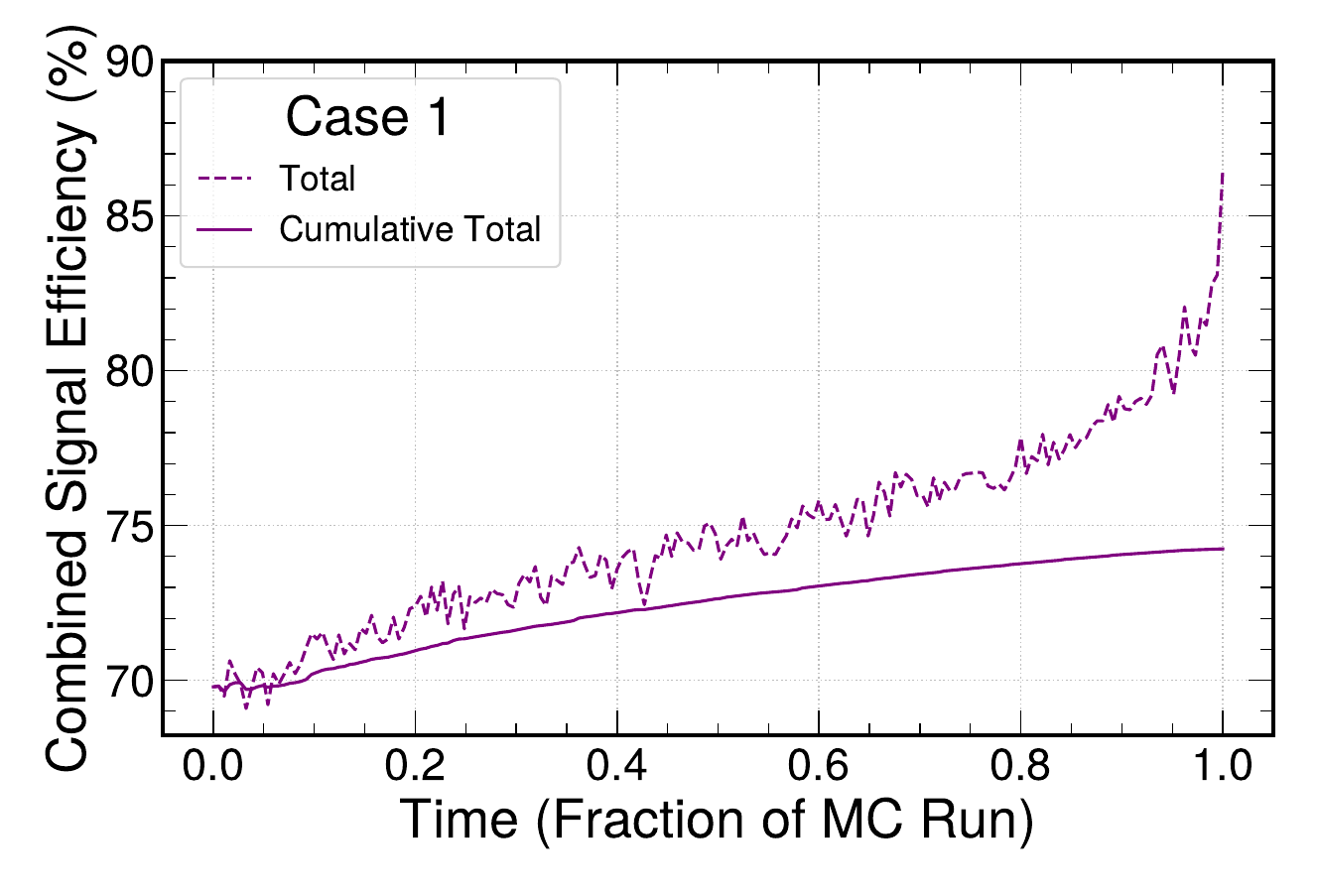}
        \caption{Signal efficiency: instantaneous (purple dashed) and cumulative (solid), illustrating the trade-off between prompt response and overall yield.}
        \label{fig:case1_b}
    \end{subfigure}
    
    \begin{subfigure}[t]{0.45\textwidth}
        \centering
        \includegraphics[width=\linewidth]{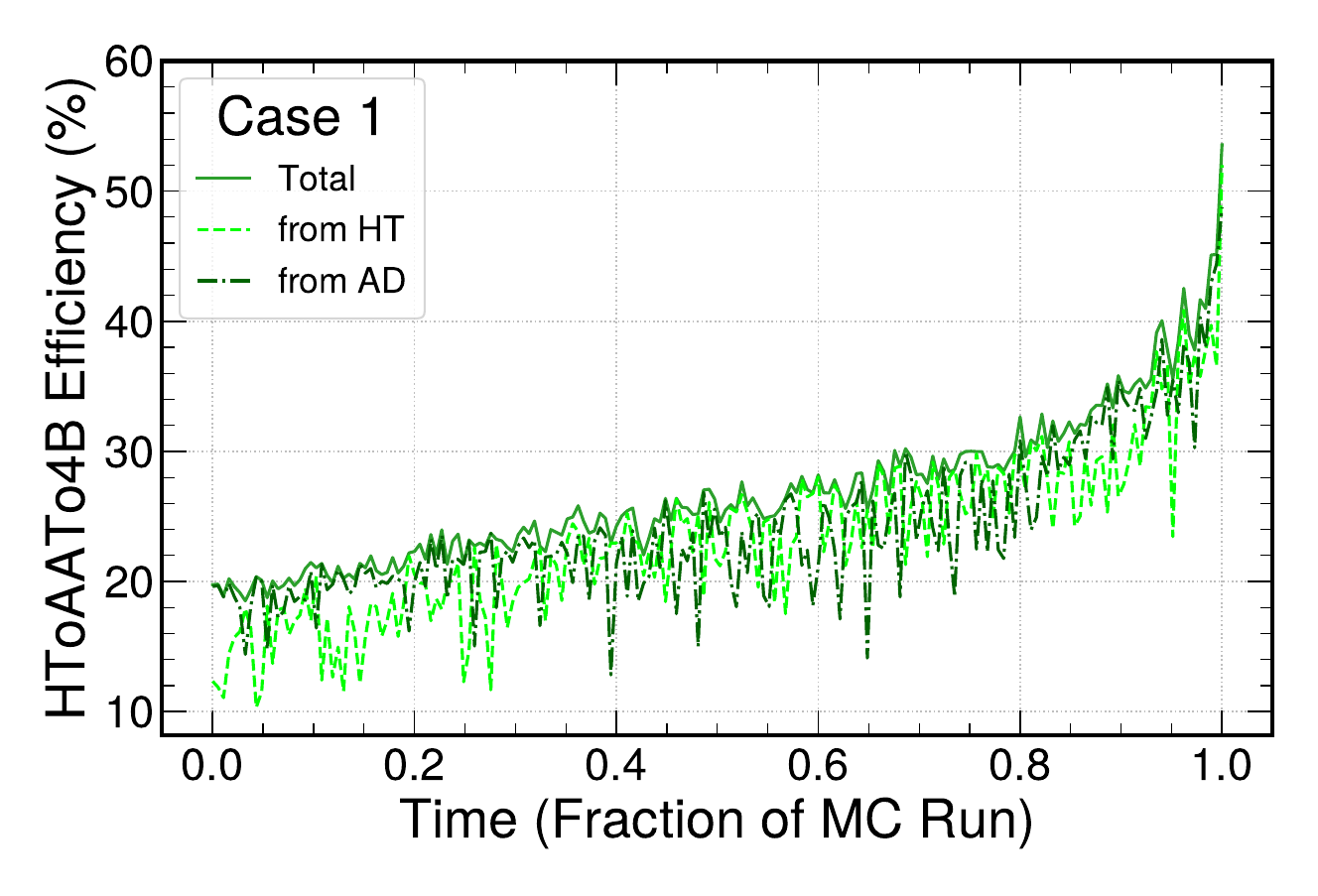}
        \caption{BSM signal \haaFourB\ efficiency breakdown by path, increasing over time.}
        \label{fig:case1_c}
    \end{subfigure}\qquad
    \begin{subfigure}[t]{0.45\textwidth}
        \centering
        \includegraphics[width=\linewidth]{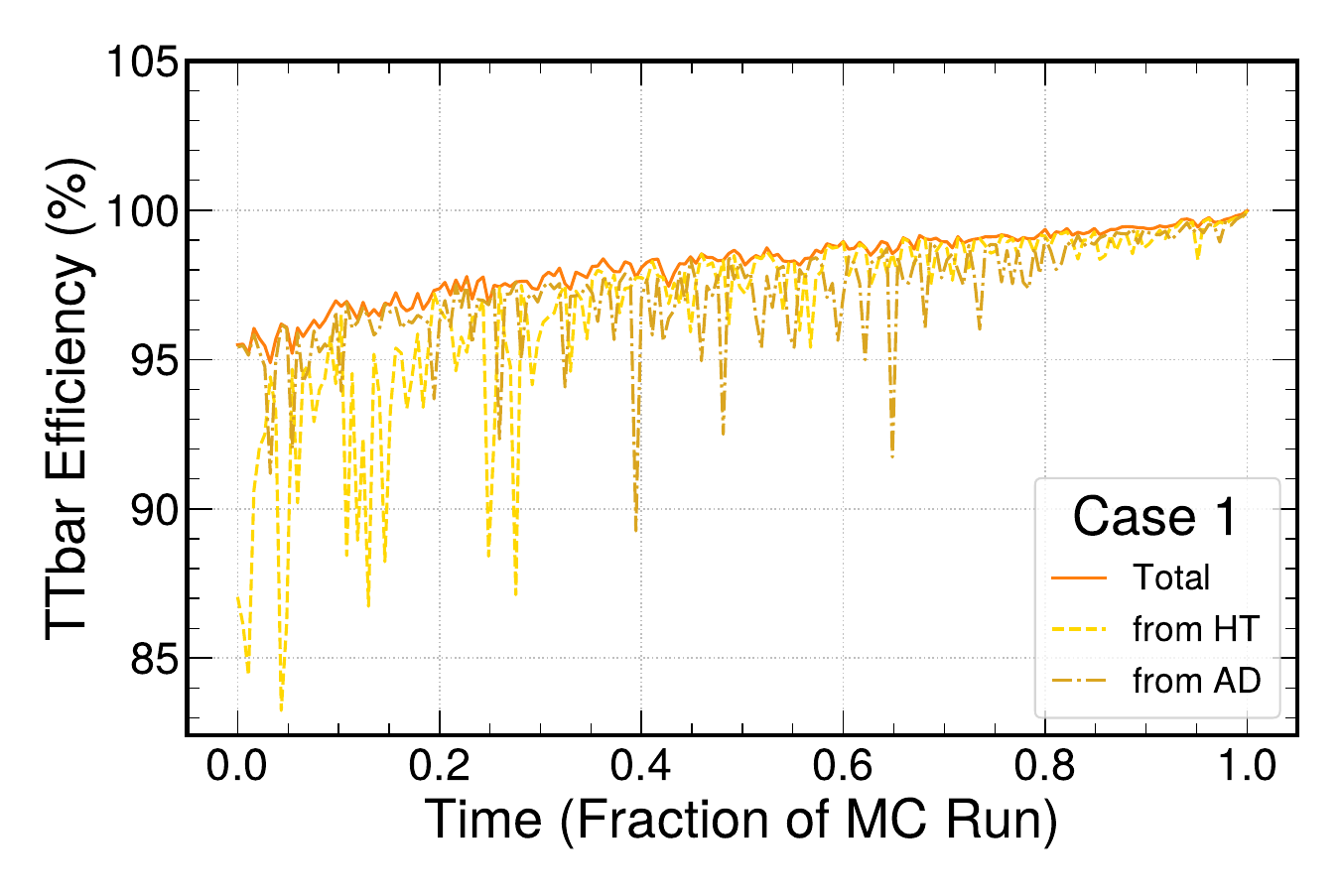}
        \caption{\ttbar\ signal rate, decomposed by path, confirming high efficiency across both triggers.}
        \label{fig:case1_d}
    \end{subfigure}

    \caption{Performance of the Ideal multi-path trigger system under Case~1 across four panels: 
    (a) Background rates, 
    (b) Signal efficiency, 
    (c) \haaFourB\ efficiency, 
    (d) \ttbar\ signal rate.}
    \label{fig:ideal_controller1}
\end{figure}

\subsection{Case 2: Specific Path Bandwidth Allocation}

For a new AI-enhanced trigger whose selection behavior is still uncertain, the path should operate in parallel with proven triggers, perhaps with a fixed bandwidth allocation earmarked for exploration. At the same time, to ensure continued coverage of high value SM processes, one might also like to enforce some minimum baseline efficiency on these signals.

For instance, we may require the trigger to maintain a high, but not necessarily maximal, efficiency for an important signal such as \ttbar, allowing the remaining bandwidth to be flexibly prioritized toward the  AD   path.
To implement such a strategy, we define the following cost function:
\begin{equation}
C_2 = \frac{|r_b - r_t|}{\sigma_b}  + \frac{|\epsilon_{t\bar{t}} - 0.9|}{\sigma_{t\bar{t}}} + \frac{|r_{\text{AD}}^{\text{(ex)}} - p \cdot r_t|}{\sigma_{AD}} \nonumber
\end{equation}
where: \( \epsilon_{t\bar{t}} \) is the total efficiency for \ttbar\ events, \( r_{\text{AD}}^{\text{(ex)}} \) is the exclusive rate for the  AD   path and \( p \) represents the allocated fraction of bandwidth.
As an example configuration, we consider  $\sigma_b = 4\, \text{kHz}$, $\sigma_{t\bar t}=0.05$, $\sigma_\text{AD}= 16\, \text{kHz}$, and \( p = 0.5 \). While this formulation focuses on exclusive AD rate to capture events that would not otherwise be collected, the inclusive AD rate may also be prioritized to better study correlations between trigger paths, as one example.

\begin{figure}[htbp]
    \centering
    \begin{subfigure}[t]{0.45\textwidth}
        \centering
        \includegraphics[width=\linewidth, trim={0 0 2mm 0},clip]{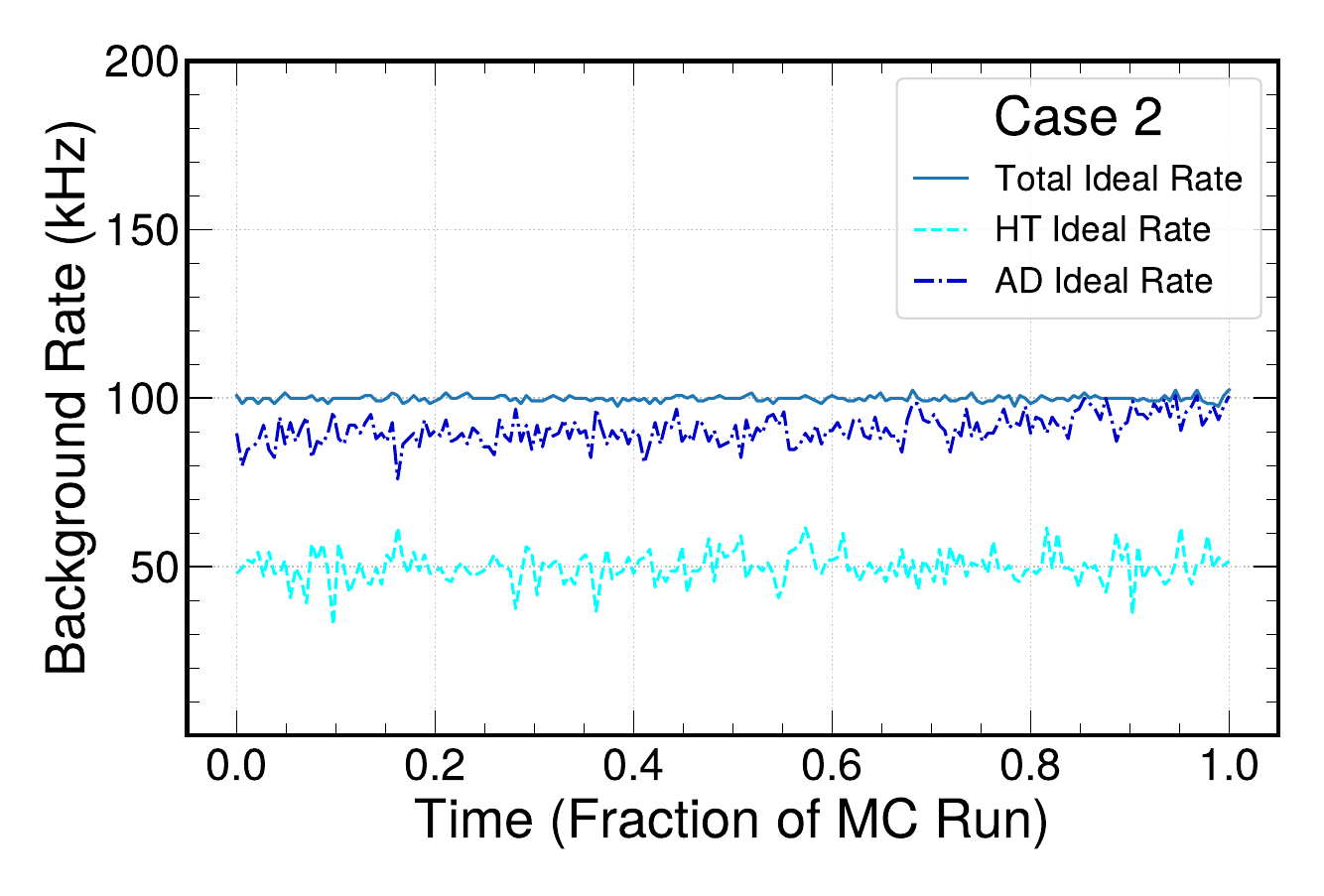}
        \caption{Background rates: total controlled near 100\,kHz, with \HT held close to 50\,kHz to ensure $\sim$50\,kHz exclusive rate for the  AD   trigger ($p=0.5$).}
        \label{fig:case2_a}
    \end{subfigure}\qquad
    \begin{subfigure}[t]{0.45\textwidth}
        \centering
        \includegraphics[width=\linewidth,trim={0 2mm 0 0},clip]{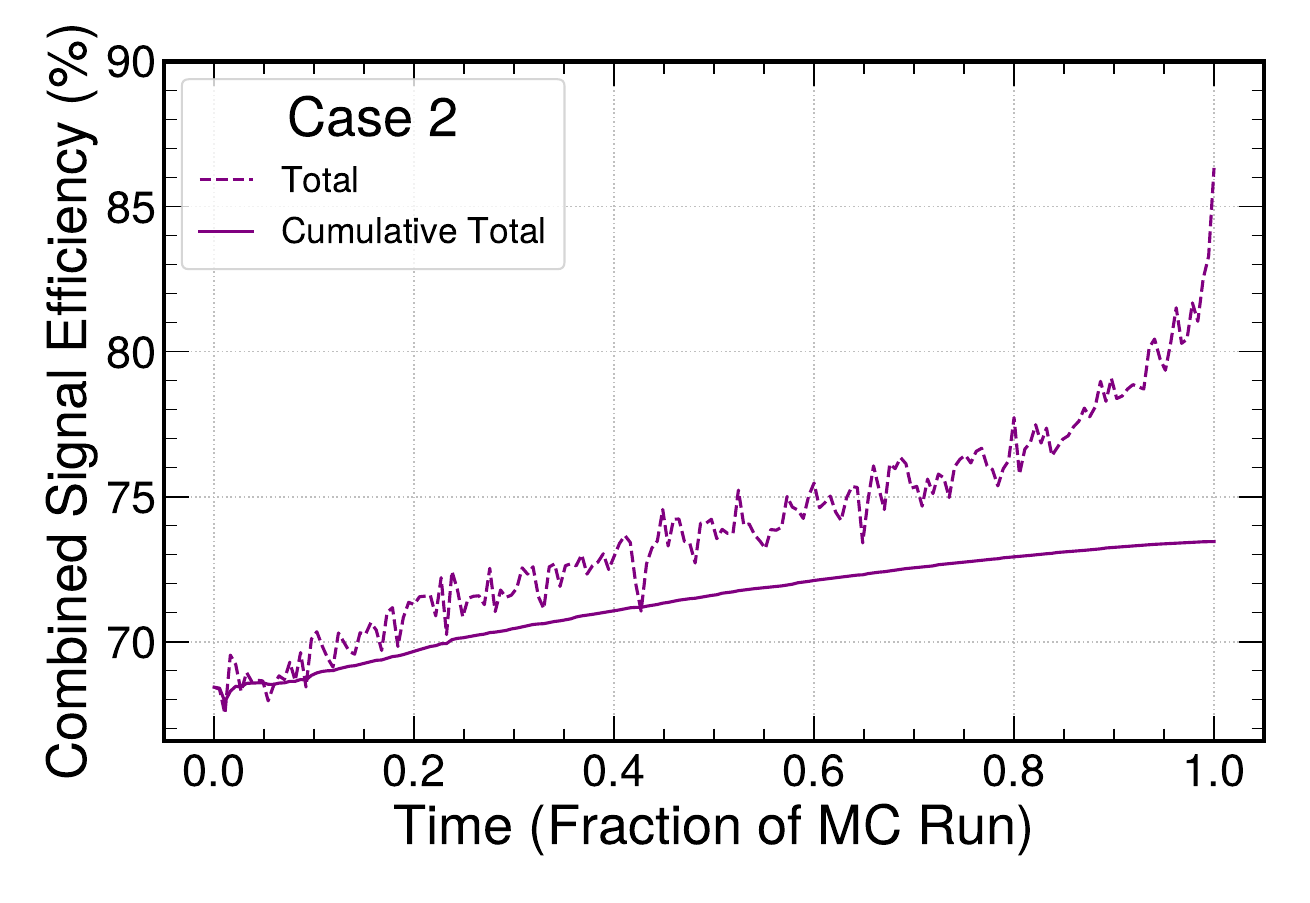}
        \caption{Signal efficiency: changing the cost format has not significantly altered the trend shown in Case 1, as it correlates mostly with total bandwidth accessibility.}
        \label{fig:case2_b}
    \end{subfigure}
    
    \begin{subfigure}[t]{0.45\textwidth}
        \centering
        \includegraphics[width=\linewidth]{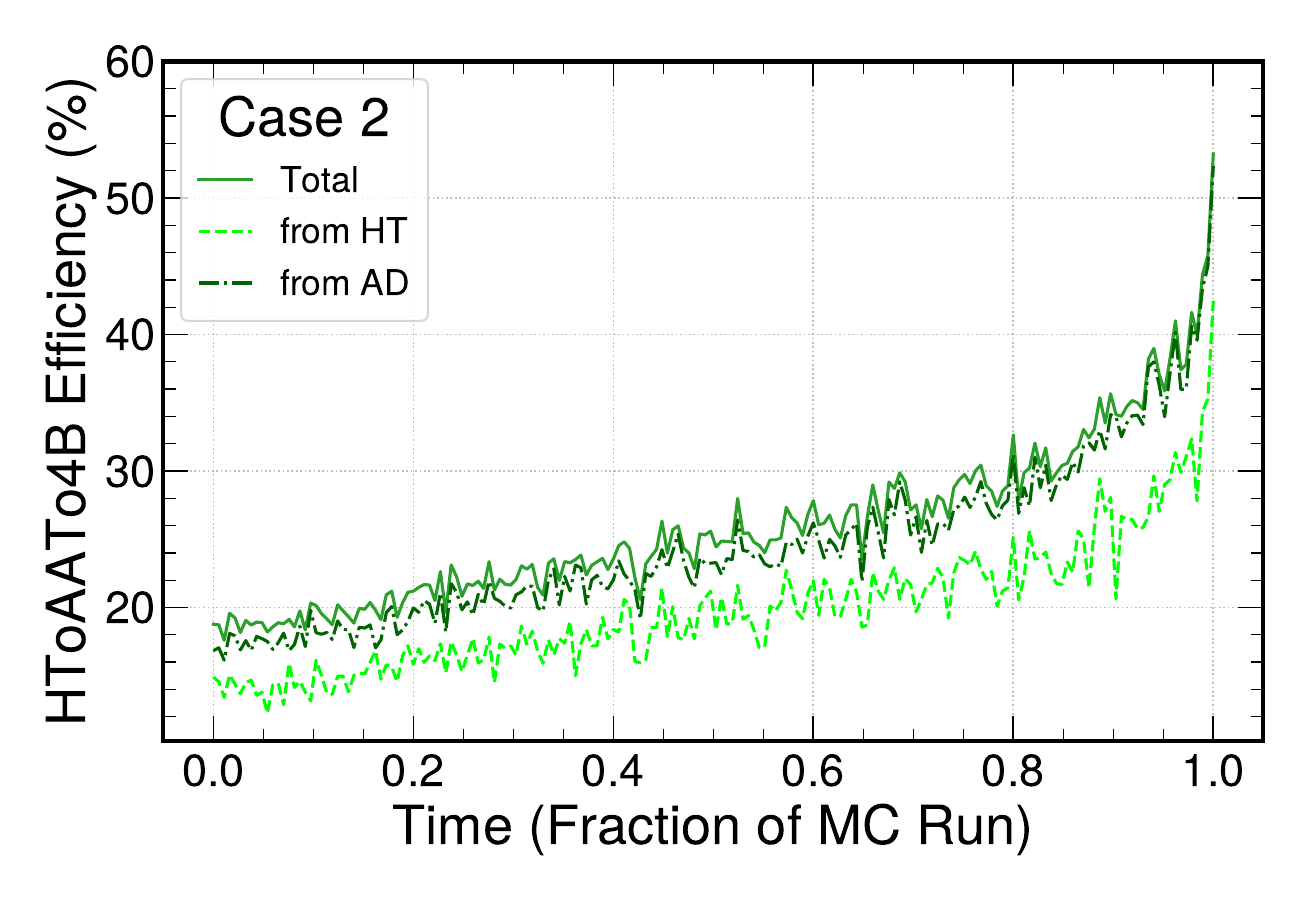}
        \caption{BSM signal \haaFourB\ efficiency breakdown by path, showing increased  AD   contribution due to bandwidth allocation.}
        \label{fig:case2_c}
    \end{subfigure}\qquad
    \begin{subfigure}[t]{0.45\textwidth}
        \centering
        \includegraphics[width=\linewidth]{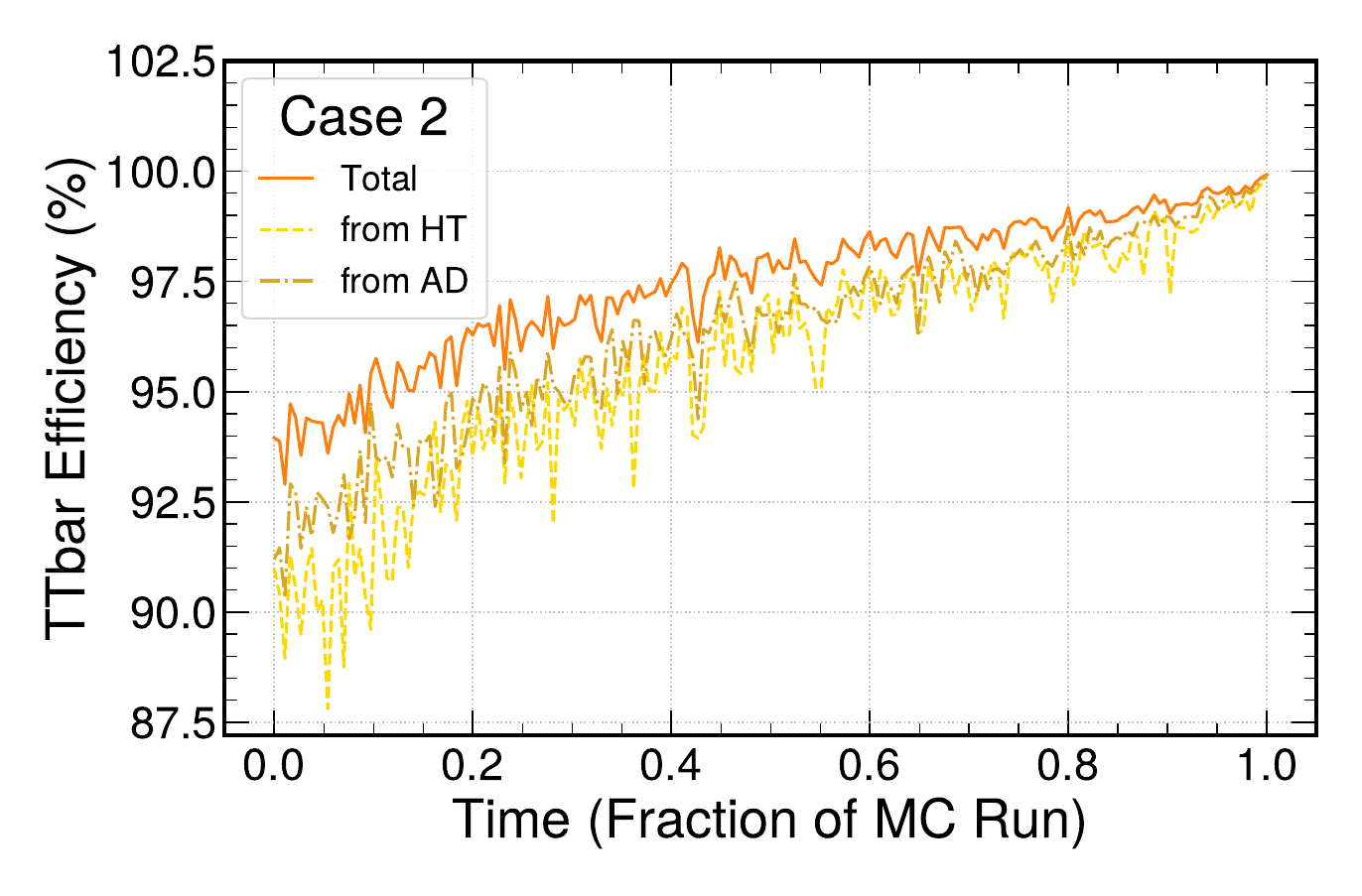}
        \caption{\ttbar\ signal rate by path, dominated by  AD   but nearly failing the 90\% efficiency criterion imposed by the cost.}
        \label{fig:case2_d}
    \end{subfigure}

    \caption{Ideal Optimization outcome for Case~2 across four panels: 
    (a) Background rates, 
    (b) Signal efficiency, 
    (c) \haaFourB\ efficiency, 
    (d) \ttbar\ signal efficiency.}
    \label{fig:ideal_controller2}
\end{figure}

In Fig.~\ref{fig:ideal_controller2}, we observe the heightened  AD  trigger activity in collecting both signal and background events compared to Case~1, as a result of the 50~kHz ($p=0.5$) exclusive allocation to this path. The cost function is also designed to maintain the \ttbar\ efficiency at 90\%. However, this term is correlated with the total bandwidth constraint in the first part of the cost. Specifically, we find in Fig.~\ref{fig:sig_rate} that the \ttbar\ efficiency exceeds 90\% for both \HT and  AD in a fixed menu, and controlling it requires very tight cuts. However, if both triggers were forced to adopt such cuts, the total rate would deviate from the target, and this is penalized in the cost function through a larger sensitivity to bandwidth variations. 
With this more complex set of priorities, it is not possible for the `ideal' controller (separately minimizing the cost for each batch) to prescribe actions that perfectly satisfy each item at all times. This is a good example of demonstrating that the hierarchical priority setup in the cost function is effective. 
On the other hand, this richer problem space opens the opportunities for improvements by more sophisticated controllers that may globally optimize cost over longer timescales.

\subsection{Case 3: Computational Cost Awareness}

In parallel with our physics-driven goals for data collection, it is important to recognize that sending an event through a given trigger path incurs a considerable computational cost. 
To attempt to quantify how this could impact data-taking decisions, we recall that the L1 trigger is only the first step in the two-stage trigger architectures used by ATLAS and CMS.
Events selected by the L1 trigger are sent to the High Level Trigger (HLT), which performs a more comprehensive event reconstruction in a commodity PC farm, to further reduce the population of selected events for permanent storage.
The processing steps that are required to be carried out in this software-based trigger vary, depending on which L1 trigger item has already accepted the event.
As an example, the calorimeter subdetector system may be ignored by the HLT for events that only pass muon-related triggers in the L1 step.

As such, one accounting of the `computational cost' of an L1 trigger is the level of resources required to execute a given algorithm, such as the fraction of a Field-Programmable Gate Array (FPGA) required to compute an anomaly score.  
However, perhaps even more critical is the cumulative cost of all downstream HLT algorithms that must be executed when the event is accepted by that path.
If a certain trigger path allows an event to be reconstructed and accepted for relatively low computational cost, the freed resources can instead be utilized to evaluate more sophisticated trigger rules.
Therefore, the computational cost one expects to incur for each accepted event can be an important factor for the trigger agent to consider during optimization.

While the active monitoring of computational costs during data taking is a high priority, at this point such constraints are only implicitly reflected in the handmade trigger menus by the experimental collaborations~\cite{ATLAS-Run2-Trig, ATLAS-Run3-Trig,CMS-HLT-Run2}. If the optimization were to minimize only this computational cost, the solution would be trivial: the agent would favor filling the bandwidth with the ``cheaper" path, disregarding physics objectives. 
To study a more realistic scenario, we incorporate the computational cost into a composite cost function that also includes physics goals, such as signal efficiency. 
This introduces a trade-off between maximizing physics performance and maintaining computational feasibility, allowing for a more meaningful outcome.

For the context of this open data benchmark, we utilize a toy model of these computational costs that does not aim to emulate the underlying technical details relevant to any given experiment. 
A more realistic and dynamic integration of these factors could be developed and incorporated into the framework in future work. 
As such, we consider two subcategories of these costs that represent different loads on the HLT CPU farm to further process events selected by the L1 trigger:
\\
\indent \textit{Event Level Computational Cost}: reflects the `complexity' of the event and is quantified here by its jet multiplicity. While a high number of jets could indicate interesting physics, it could equally be a pileup artifact.
Therefore, event cost can be defined as the average number of jets in accepted events per batch. For example, if 2 events pass in a batch, with 2 and 3 jets respectively, the event-level cost associated with that batch is equal to 2.5.

\textit{Trigger-Path Level Computational Cost}: accounts for differences in the class of algorithms required to process an event in the HLT depending on the L1 seed, irrespective the event level complexity. Each L1 path may invoke a distinct set of HLT reconstruction algorithms, resulting in varying per-event processing requirements. 

\HT trigger targeting highly energetic jets may only require reconstruction of hadronic and electromagnetic calorimeter information, bypassing track reconstruction completely. However, AD trigger inherently seek to test a wide range of event hypotheses, using all subdetector systems to precisely measure low-\pt\ jets, perform $b$-tagging, and evaluate more powerful AD models.

Based on recent HLT timing estimates from CMS~\cite{CMS-DP} where jet-related algorithms accounted for roughly 150 of the average 600\,ms processing time, our toy model assigns a cost to AD-selected events that is four times larger than those selected by the \HT path alone,i.e., an accepted event costs 4 units if passed by AD, otherwise 1, and the trigger-path cost is computed as the average path-associated cost per batch. 
    
Note that events passing both paths are not double counted in the cost calculation, since the HLT executes each algorithm only once, and the computational cost is calculated solely based on background events.
The final form of the cost that we introduce is:
\begin{equation}
C_3 = \frac{|r_b - r_t|}{\sigma_b}  + \frac{1 - \epsilon}{\sigma_s} + 
\frac{ \max(C_{\text{evt}} - C_{\text{evt}}^{\text{ref}},\, 0)}{\sigma_{\text{evt}}} +
\frac{\max(C_{\text{algo}} - C_{\text{algo}}^{\text{ref}},\, 0)}{\sigma_{\text{algo}}} \nonumber
\end{equation}
where \( C_{\text{evt}} \) and \( C_{\text{algo}} \) denote the average event level and trigger-path level computational costs per batch, respectively.
This formulation only penalizes the use of particularly expensive trigger paths by including costs to those surpassing some prescribed reference values \( C_{\text{ev}}^{\text{ref}} \) and \( C_{\text{algo}}^{\text{ref}}\). 
By focusing on these costly events, this benchmark should guide the optimization toward more efficient configurations. Here, the reference values were determined by analyzing the distribution of computational costs observed in Case 1, a scenario where optimization is guided solely by physics performance and the computational cost is not explicitly constrained. This approach introduces the reference costs as hyperparameters in the cost function, but in principle, they can also be updated by the agent using historical trends in computational cost.

Figure~\ref{fig:cost_histogram} displays histograms of the average event level and trigger-path level costs per batch in that context. 
These represent typical computational loads when both paths are configured without any compute-related feedback. The scaling parameters associated with each of these contributions to the cost function,  \( \sigma_{\text{ev}} \) and \( \sigma_{\text{trig}} \), are selected to be comparable to the width of the cost distributions in Fig.~\ref{fig:cost_histogram}. 

\begin{figure}[htbp]
    \centering
    \begin{subfigure}{0.435\textwidth}
        \centering
        \includegraphics[width=\linewidth]{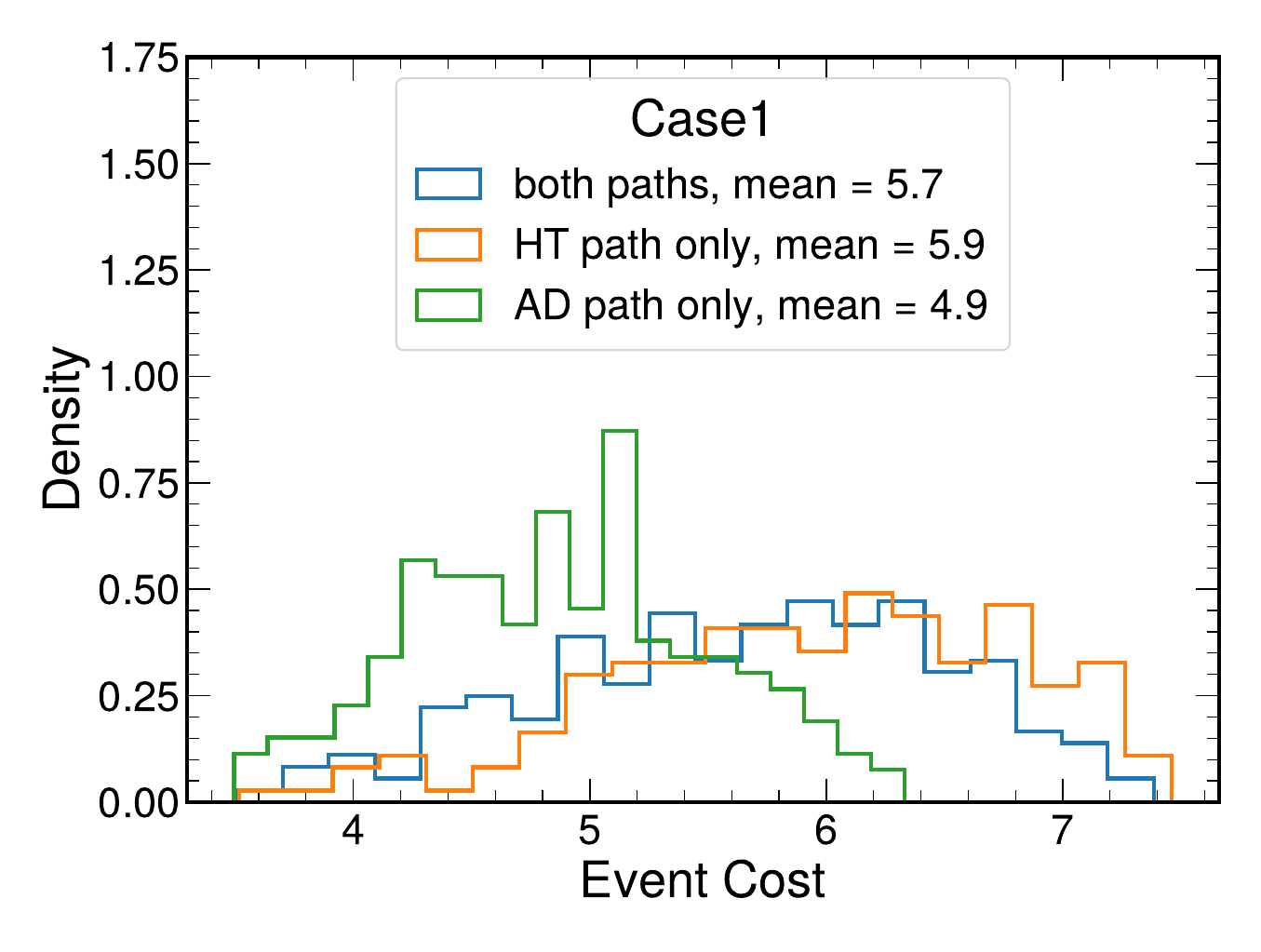}
        \caption{Event-level cost for three configurations: both paths active (blue), only the \HT path (orange), only the  AD   path (green).}
        \label{fig:comp_cost_a_MC}
    \end{subfigure}\qquad
    \begin{subfigure}{0.45\textwidth}
        \centering
        \includegraphics[width=\linewidth]{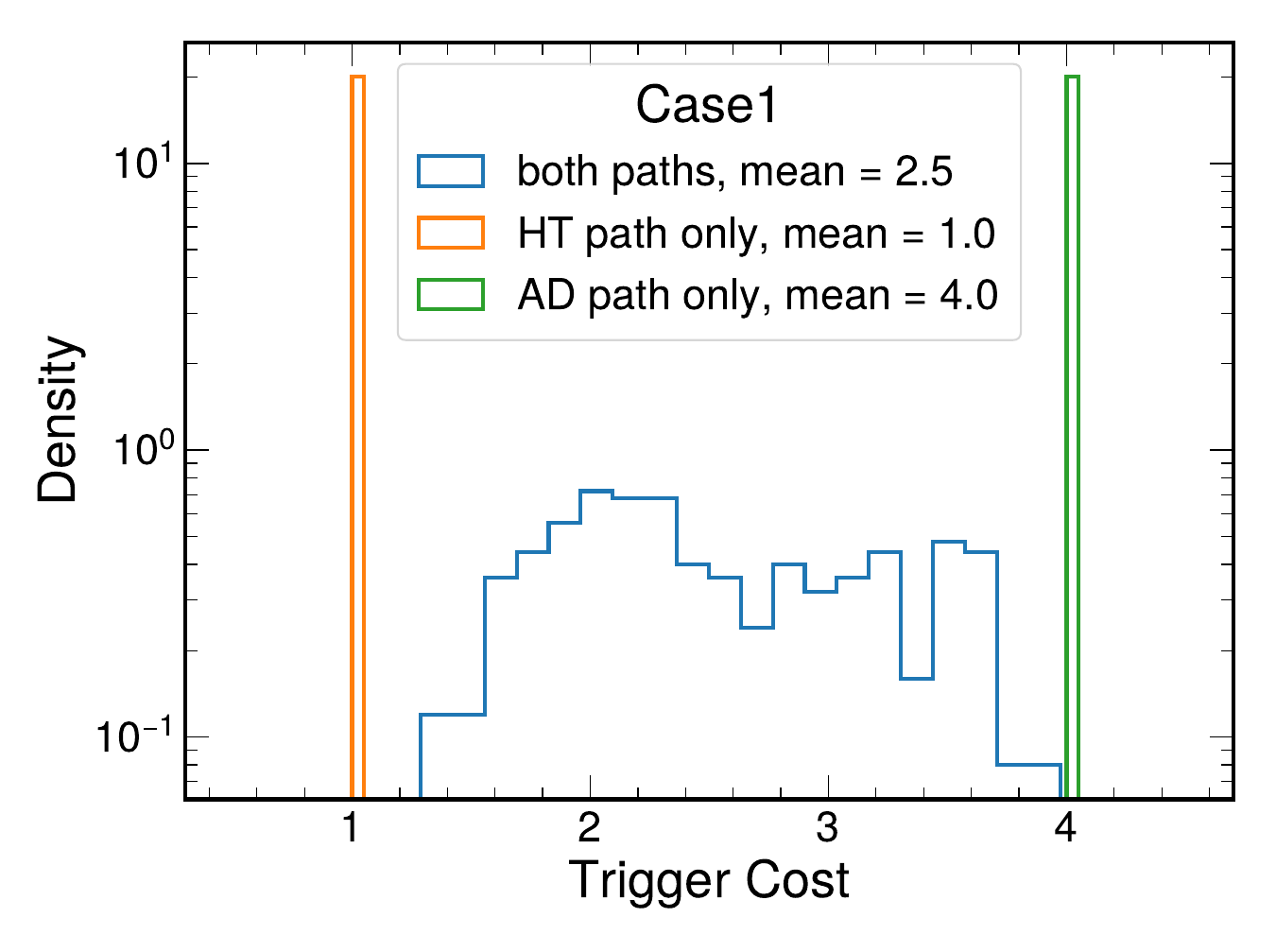}
        \caption{Trigger-path level cost distributions for the same three configurations, showing the relative computational load associated with each path.}
        \label{fig:comp_cost_b_MC}
    \end{subfigure}

    \caption{
    Distributions of computational costs of accepted background events for Case~1 (physics-driven configuration using both trigger paths).  
    Panels (a) and (b) define the reference cost values used in Case 3 optimization.
    }
    \label{fig:cost_histogram}
\end{figure}

The resulting behavior under this strategy is illustrated in Fig.~\ref{fig:ideal_controller3} with below set of parameters: 
\[
\sigma_b = 4\, \text{kHz}, \quad
\sigma_s = 0.05,  \quad
\sigma_\text{evt} = 0.5,  \quad
\sigma_\text{algo} = 0.5,   \quad
C_{\text{evt}}^{\text{ref}} = 5.7, \quad
C_{\text{algo}}^{\text{ref}} = 2.5.  
\]

\begin{figure}[htbp]
    \centering
    \begin{subfigure}[t]{0.45\textwidth}
        \centering
        \includegraphics[width=\linewidth,trim={0 0 2mm 0},clip]{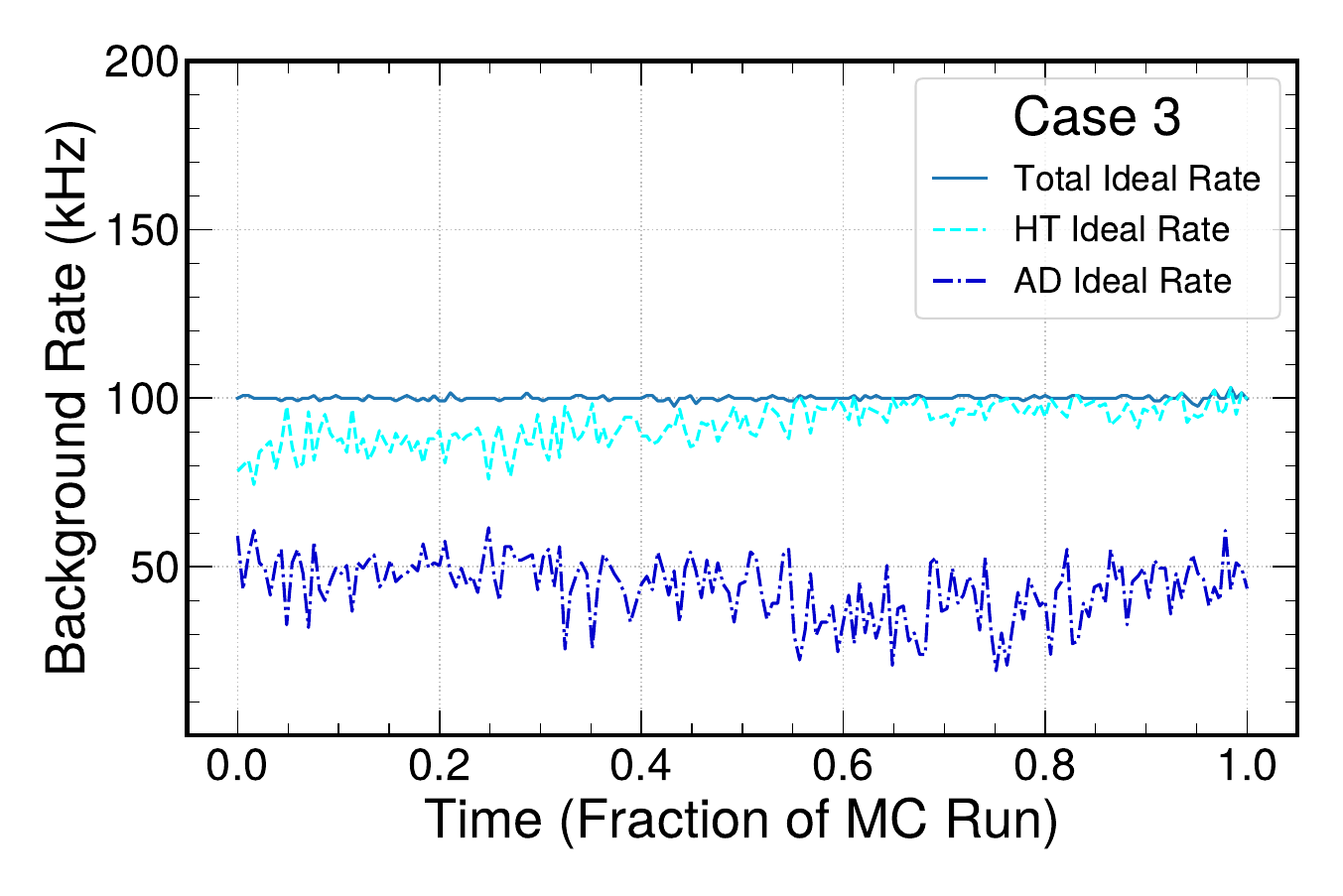}
        \caption{Background rates, demonstrating stable control near the target while showing a different rate decomposition compared to earlier cases.}
        \label{fig:case3_a}
    \end{subfigure}
    \hspace{0.02\textwidth}
    \begin{subfigure}[t]{0.45\textwidth}
        \centering
        \includegraphics[width=\linewidth,trim={0 2mm 0 0},clip]{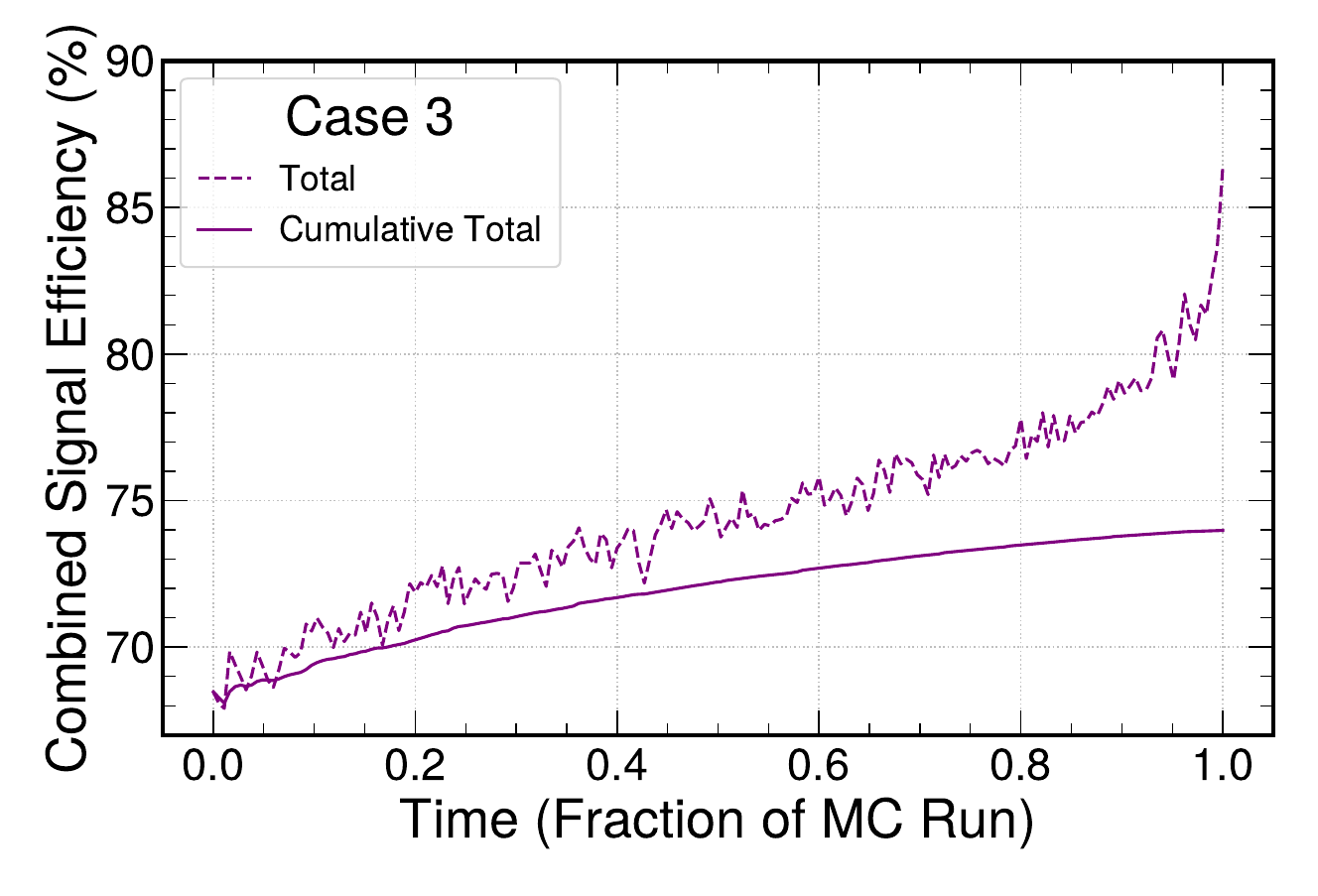}
        \caption{Total signal efficiencies, illustrating sustained gain over time similar to previous cases.}
        \label{fig:case3_b}
    \end{subfigure}
    
    \vspace{2mm}
    
    \begin{subfigure}[t]{0.45\textwidth}
        \centering
        \includegraphics[width=\linewidth]{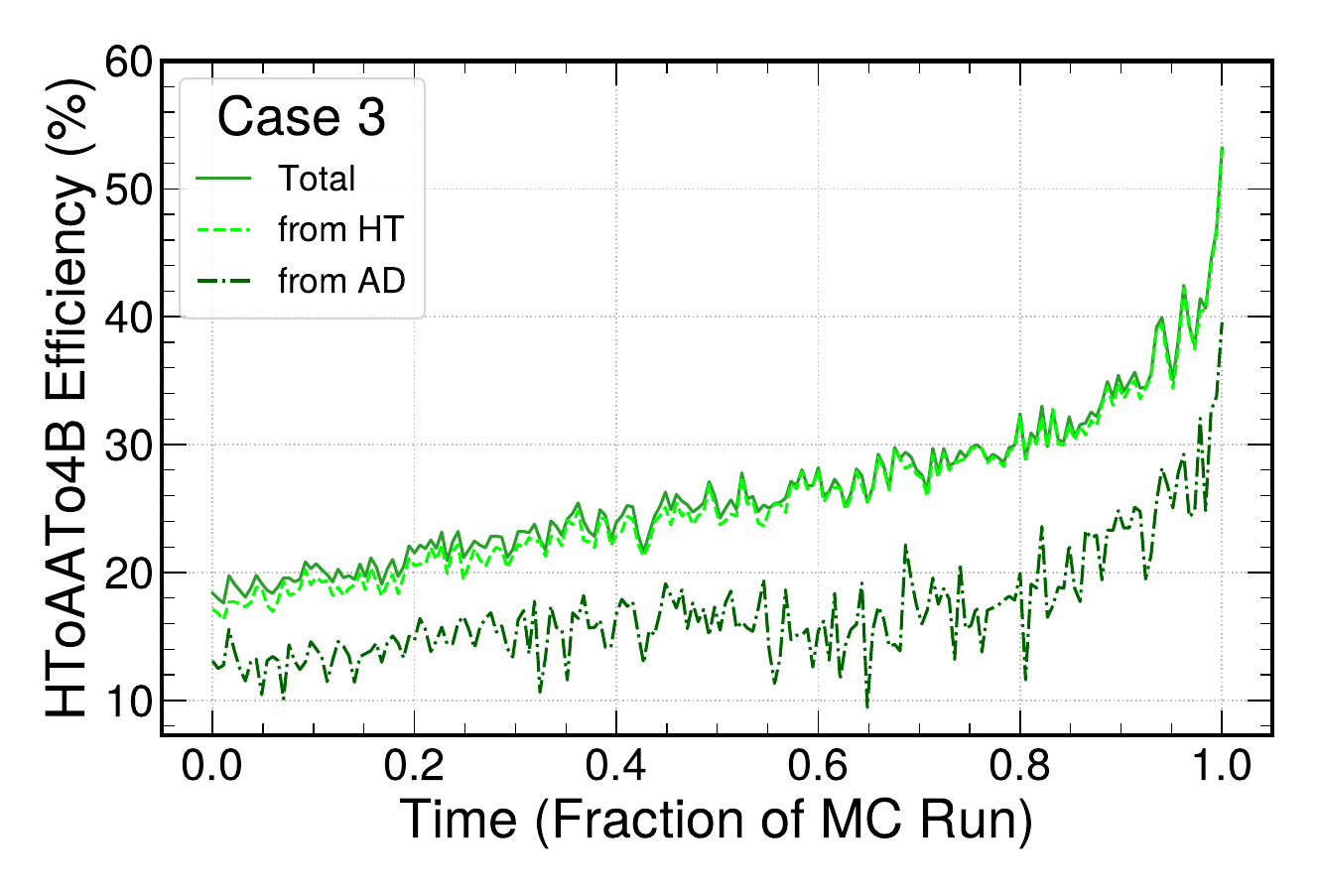}
        \caption{BSM \haaFourB\ efficiency by trigger path, highlighting the controller’s allocation strategy.}
        \label{fig:case3_c}
    \end{subfigure}
    \hspace{0.02\textwidth}
    \begin{subfigure}[t]{0.45\textwidth}
        \centering
        \includegraphics[width=\linewidth]{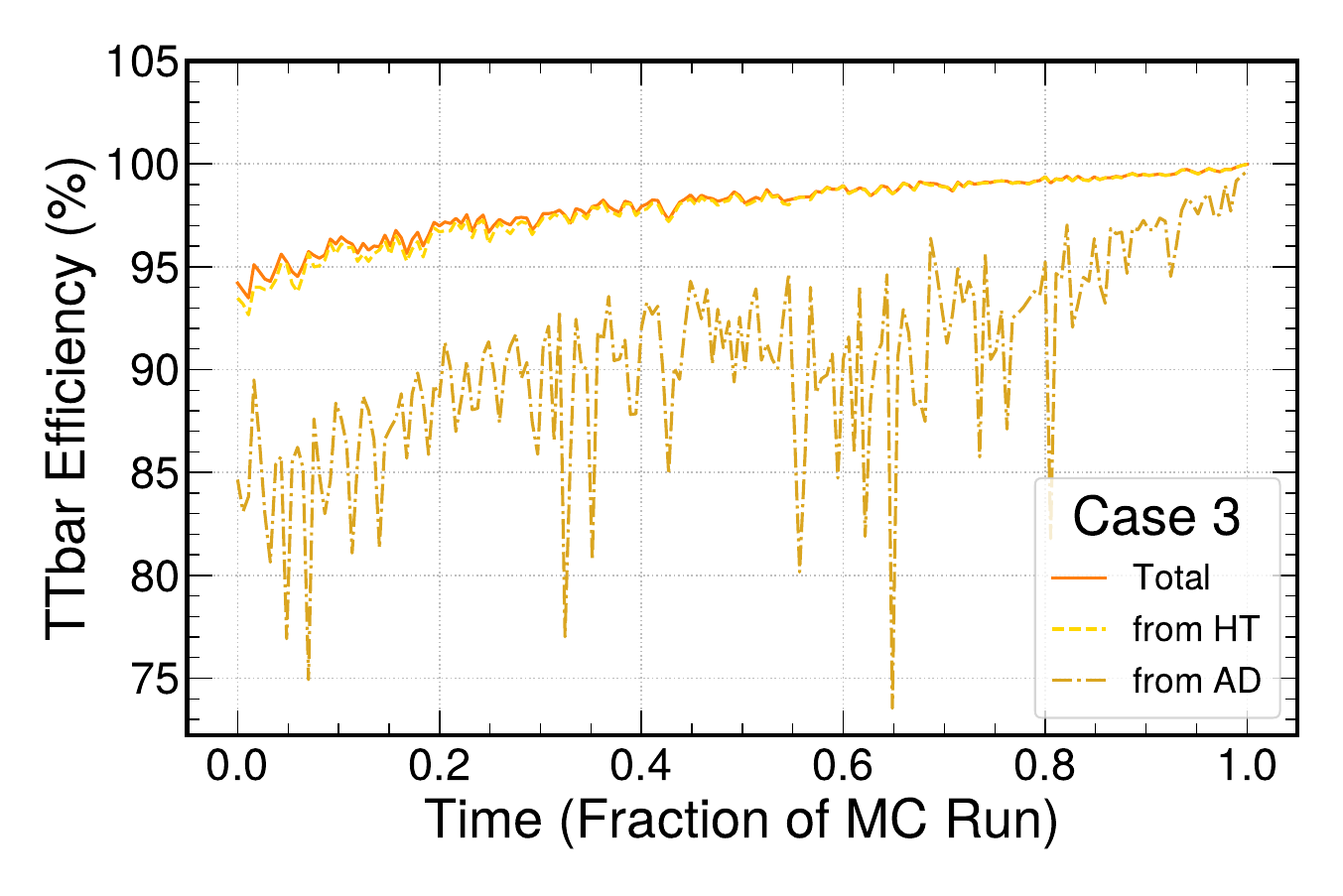}
        \caption{\ttbar\ efficiencies by path, showing the \HT trigger becoming dominant in event collection.}
        \label{fig:case3_d}
    \end{subfigure}

    \caption{Ideal Trigger performance under Case~3 (cost-aware controller):
    (a) Background rates,
    (b) Total signal efficiencies,
    (c) \haaFourB\ efficiency by path,
    (d) \ttbar\ efficiency by path. Together these panels highlight how the controller balances physics goals while managing computational cost, leading to the \HT path becoming the dominant contributor.}
    \label{fig:ideal_controller3}
\end{figure}

In addition to stable total background rates and gain in signal efficiencies with time, Fig.~\ref{fig:ideal_controller3} is also consistent with the choice of \HT trigger as the ``cheaper" path since the collection of events in background and signal samples is dominated by this trigger while no significant loss is observed in total signal efficiency. Furthermore, Fig ~\ref{fig:comp_cost_a_MC} shows that the implemented AD   model tends to collect events with lower jet multiplicities (event level cost), which could in principle compete with the trigger level cost term. However, the difference between the three distributions is not significant enough to drive the optimization in favor of the AD path.

\section{A Simple Local Controller}
\label{sec:local_controller}

In Section~\ref{section6}, we have set the table for three benchmark tasks by codifying competing experimental priorities into explicit functions that can be optimized.
We illustrated the evolution of these cost functions over time in the benchmark MC dataset by invoking an \textit{ideal} controller that acted based on complete knowledge of the upcoming batch of data. 
This agent delineates an important upper bound on performance, achieving the best outcome in each event batch under perfect foresight.

While informative, such an approach is impossible to implement in practice. We therefore introduce a simple and realistic control strategy that relies only on past and current information. In principle, a global grid search across the action space performed on the current batch should approximate the optimal parameter set for the upcoming batch in the adiabatic limit. 
Though this approach may be practical for a simple two-component trigger menu, scaling of the grid volume for a scenario with many paths limits its applicability.
On the other hand, observations of the ideal controller (Fig.~\ref{fig:ideal_evo}) indicate that the optimal parameters evolve smoothly over time and generally remain localized within a constrained region of the parameter space. As a result, a local search strategy becomes both feasible and effective.

\begin{figure}[htbp]
    \centering
    \includegraphics[width=0.31\textwidth,trim={0 2mm 1mm 0},clip]{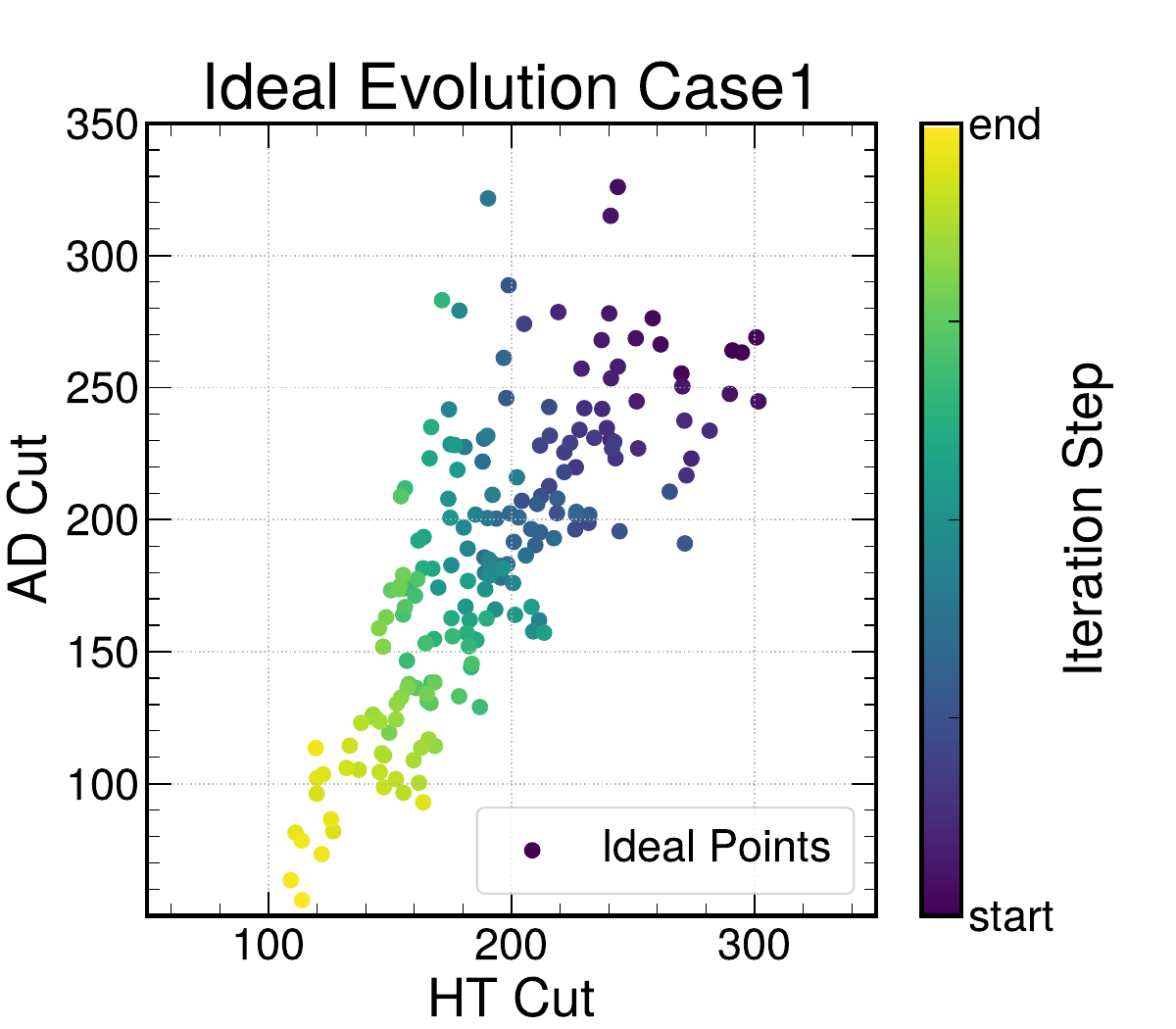}
    \includegraphics[width=0.31\textwidth,trim={0 2mm 1mm 0},clip]{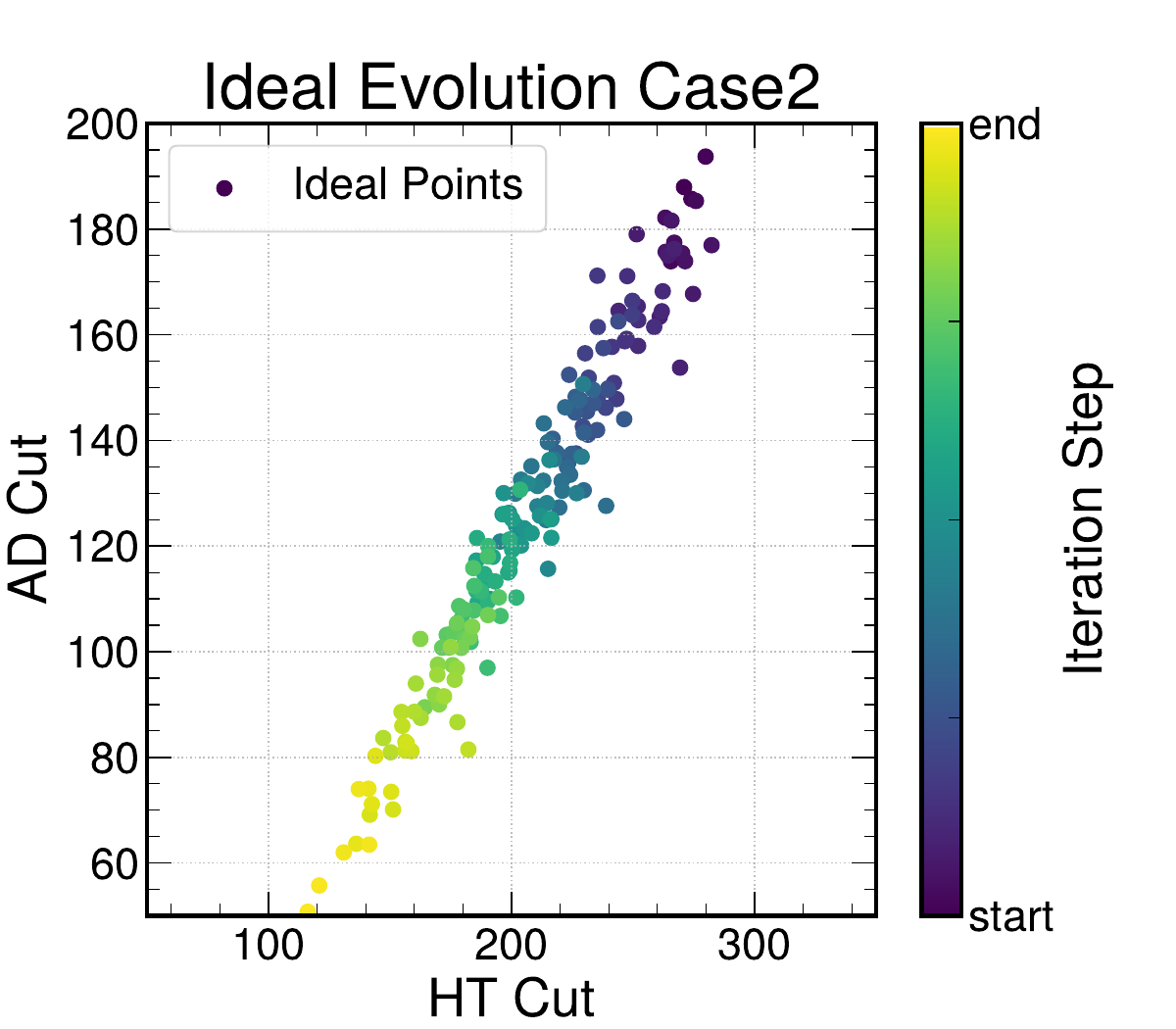}
    \includegraphics[width=0.31\textwidth,trim={0 2mm 1mm 0},clip]{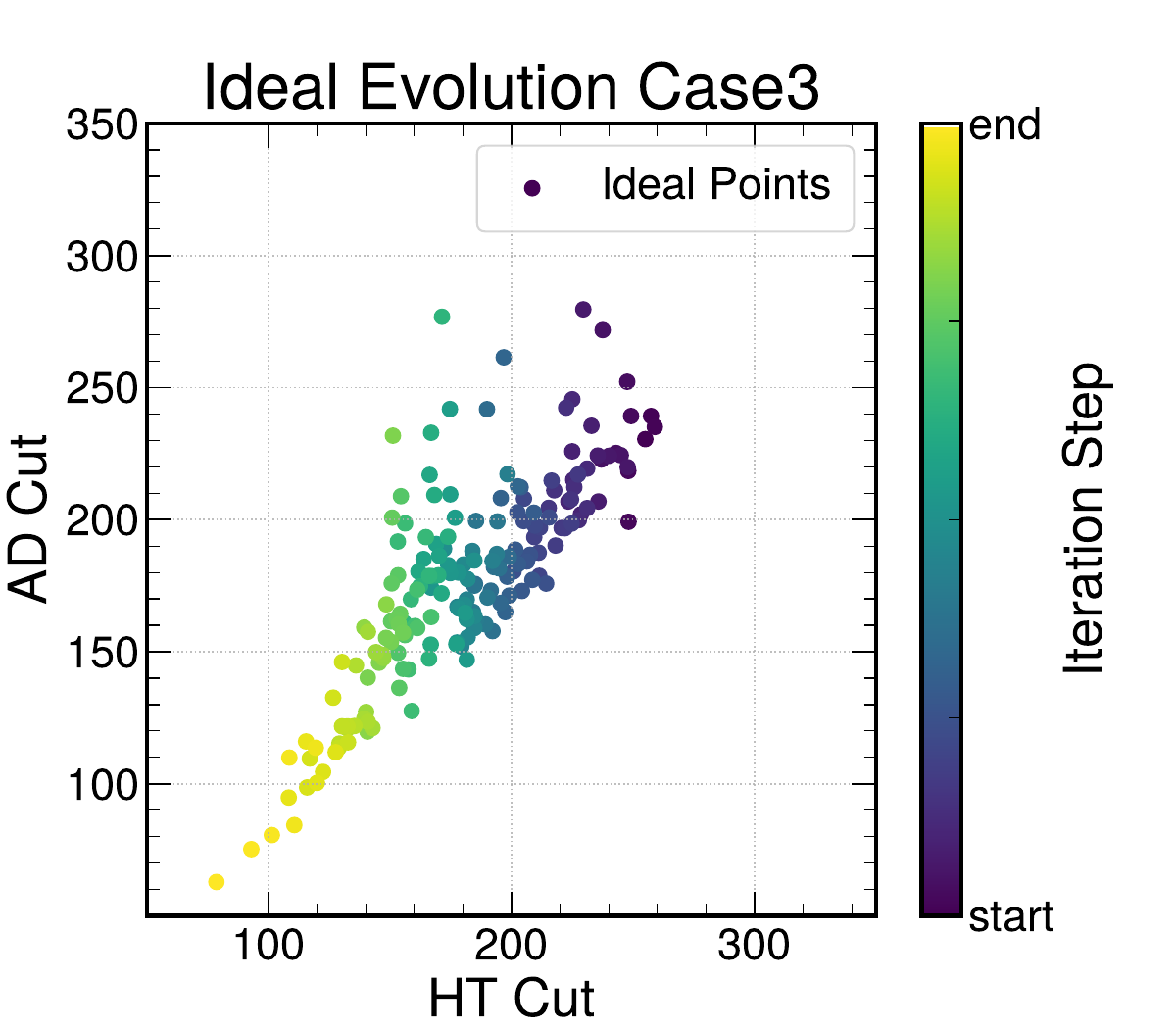}
  \caption{Real time evolution of optimal point of the ideal controllers (minimum of the cost) in the parameter space.}
    \label{fig:ideal_evo}
\end{figure}
Figure~\ref{fig:ideal_evo} generally illustrates how the global minimum of the cost function evolves over time. While we observe a trend toward lower cut values, there is random wandering in the parameter space as well. Based on these observation, a candidate simple controller can be defined to operate through a local grid search over a $10\times 10$ window (versus $100\times 100$ in the ideal case), using a search range of $\pm 20$ units and a memory depth of five batches. Here, memory depth means that each update is computed as an average over the optimal local points from the previous five batches, which improves robustness and mitigates fluctuations. This enables the controller to follow historical trends while preventing abrupt parameter shifts, as shown in Fig.~\ref{fig:real_evo}. 

Figures~\ref{fig:real1},~\ref{fig:real2} and~\ref{fig:real3} compare the performance of this simple local controller with the ideal benchmark across the three case studies introduced earlier. In each case, the time evolution of rates and efficiencies obtained with the controller is shown together with the corresponding ideal curves, allowing a direct comparison. We observe that, when moving from the ideal case to the realistic controller algorithm, the overall behavior of rates and efficiencies, as well as their decomposition across the different trigger paths, is largely unchanged. Importantly, the cumulative signal efficiency in the realistic controller also remains close to that of the ideal case (panels \ref{fig:real1_b}, \ref{fig:real2_b}, \ref{fig:real3_b}). In summary, this controller yields a smooth evolution through the action space and achieves efficiencies very close to the ideal benchmark, demonstrating that the simplified, realistic algorithm does not compromise physics performance.

\begin{figure}[htbp]
    \centering
    \includegraphics[width=0.31\textwidth,trim={0 2mm 1mm 0},clip]{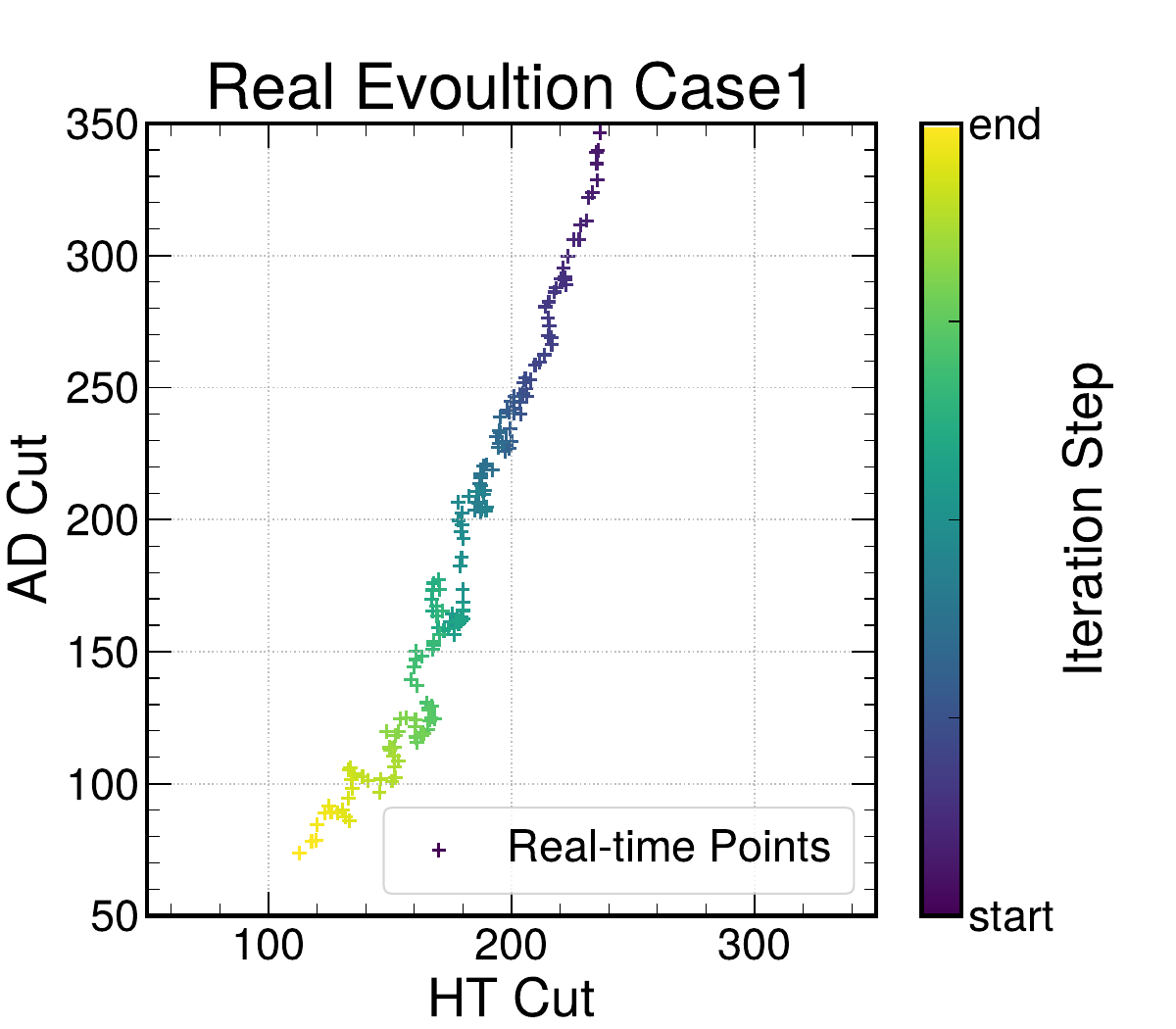}
    \includegraphics[width=0.31\textwidth,trim={0 2mm 1mm 0},clip]{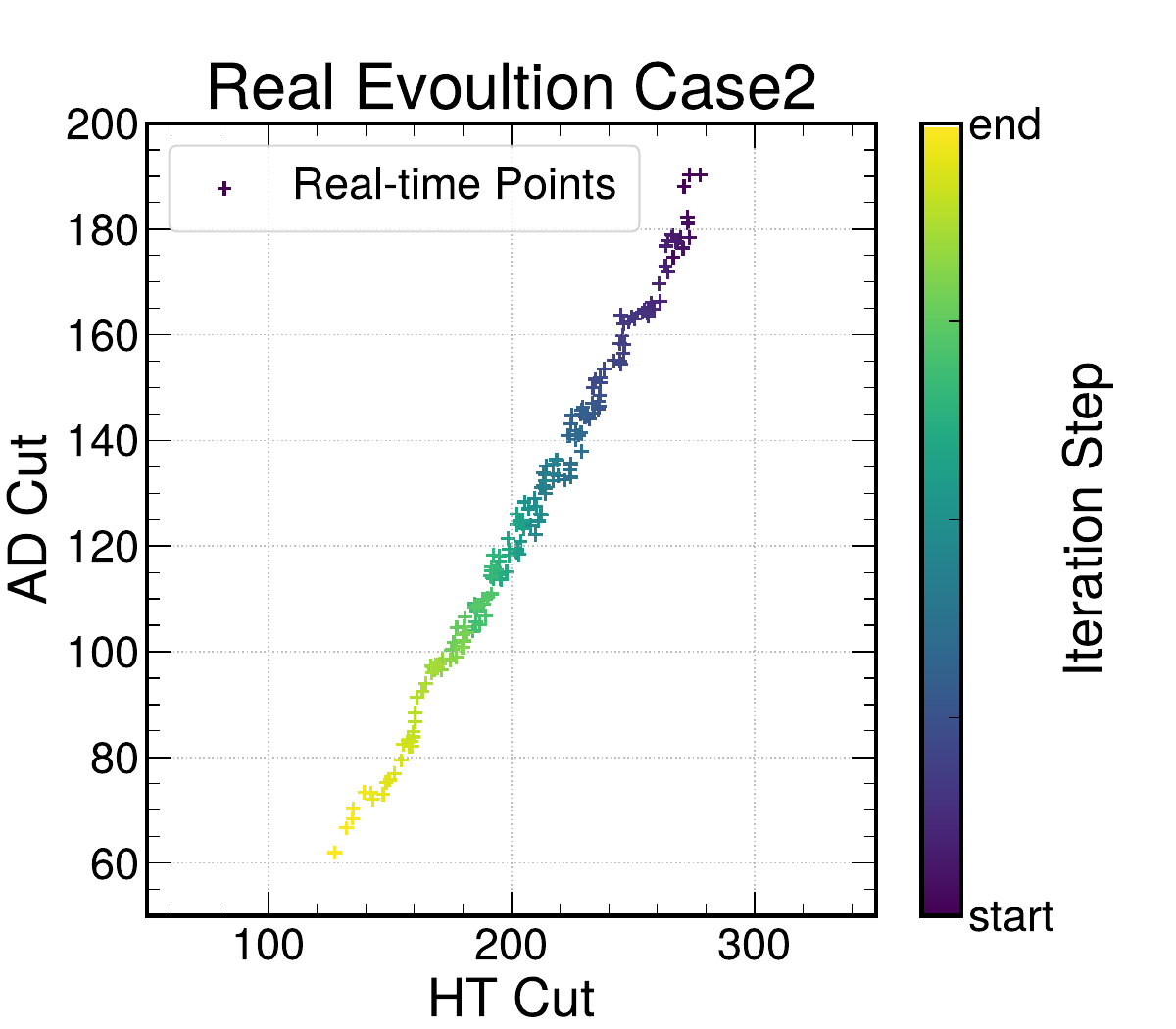}
    \includegraphics[width=0.31\textwidth,trim={0 2mm 1mm 0},clip]{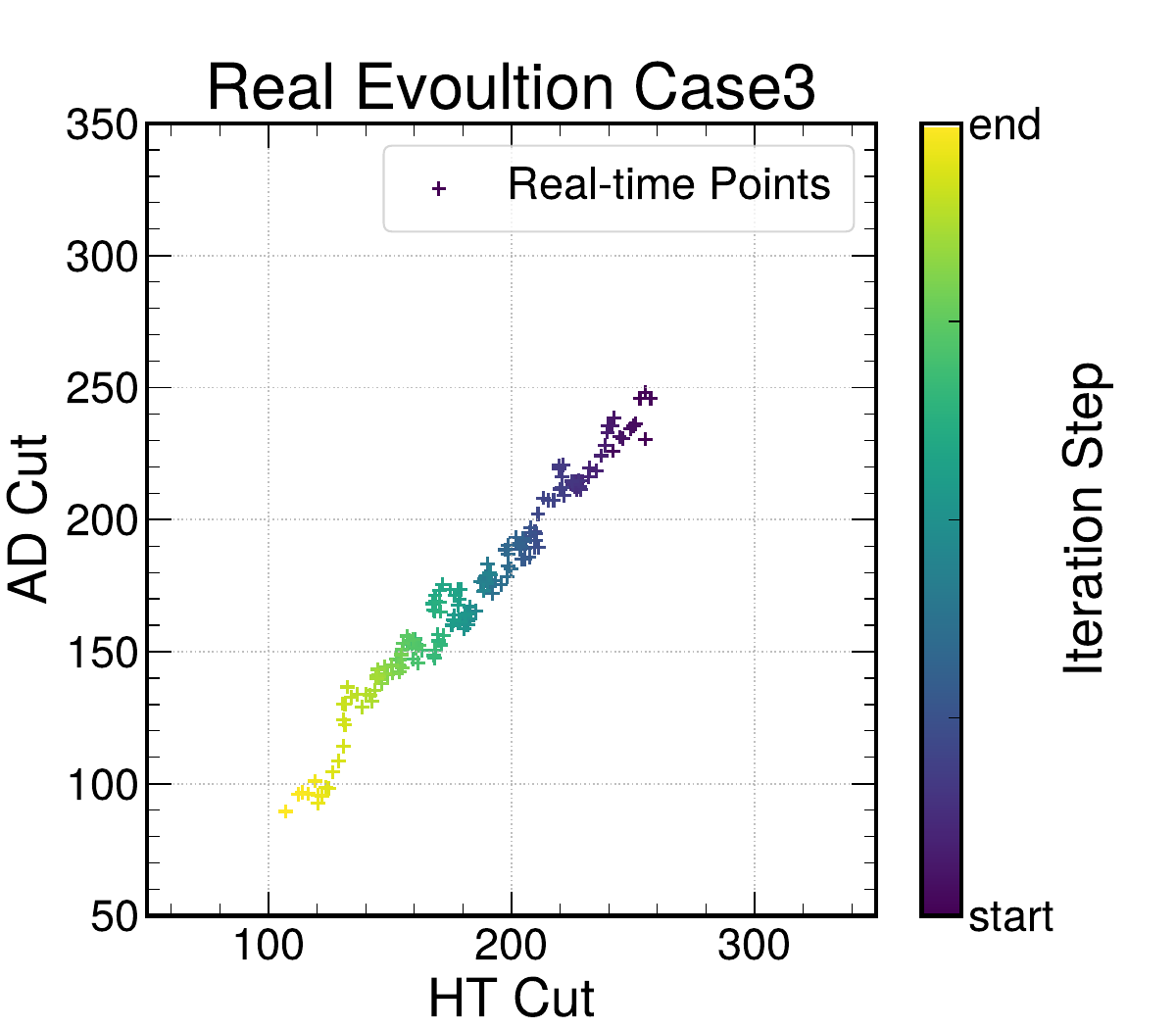}
    \caption{Real time evolution of the cost functions with the simple controller.}
    \label{fig:real_evo}
\end{figure}

\begin{figure}[htbp]
    \centering
    \begin{subfigure}{0.45\textwidth}
        \centering
        \includegraphics[width=\linewidth]{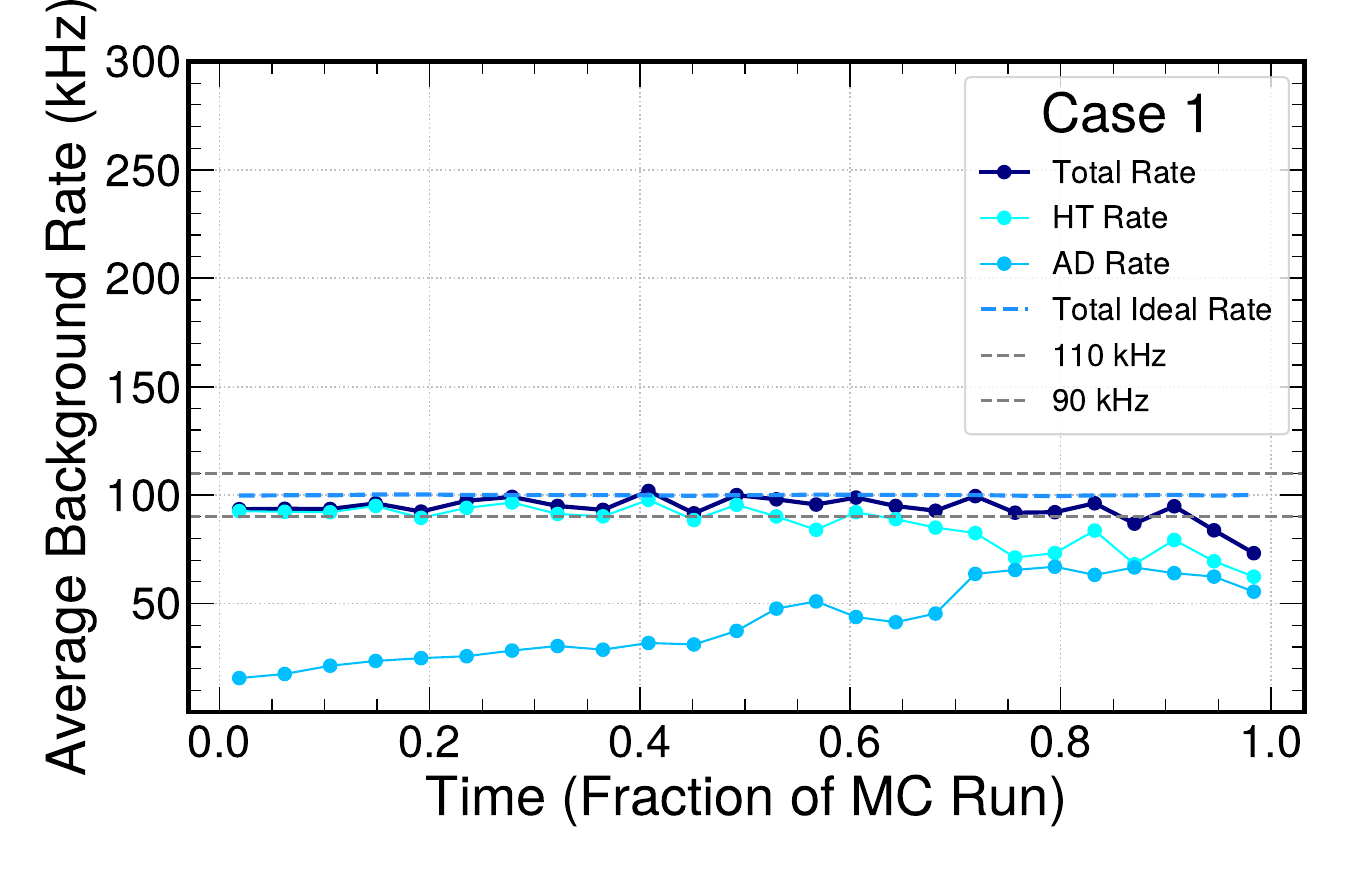}
        \caption{Background Rate.}
        \label{fig:real1_a}
    \end{subfigure}\qquad
    \begin{subfigure}{0.45\textwidth}
        \centering
        \includegraphics[width=\linewidth]{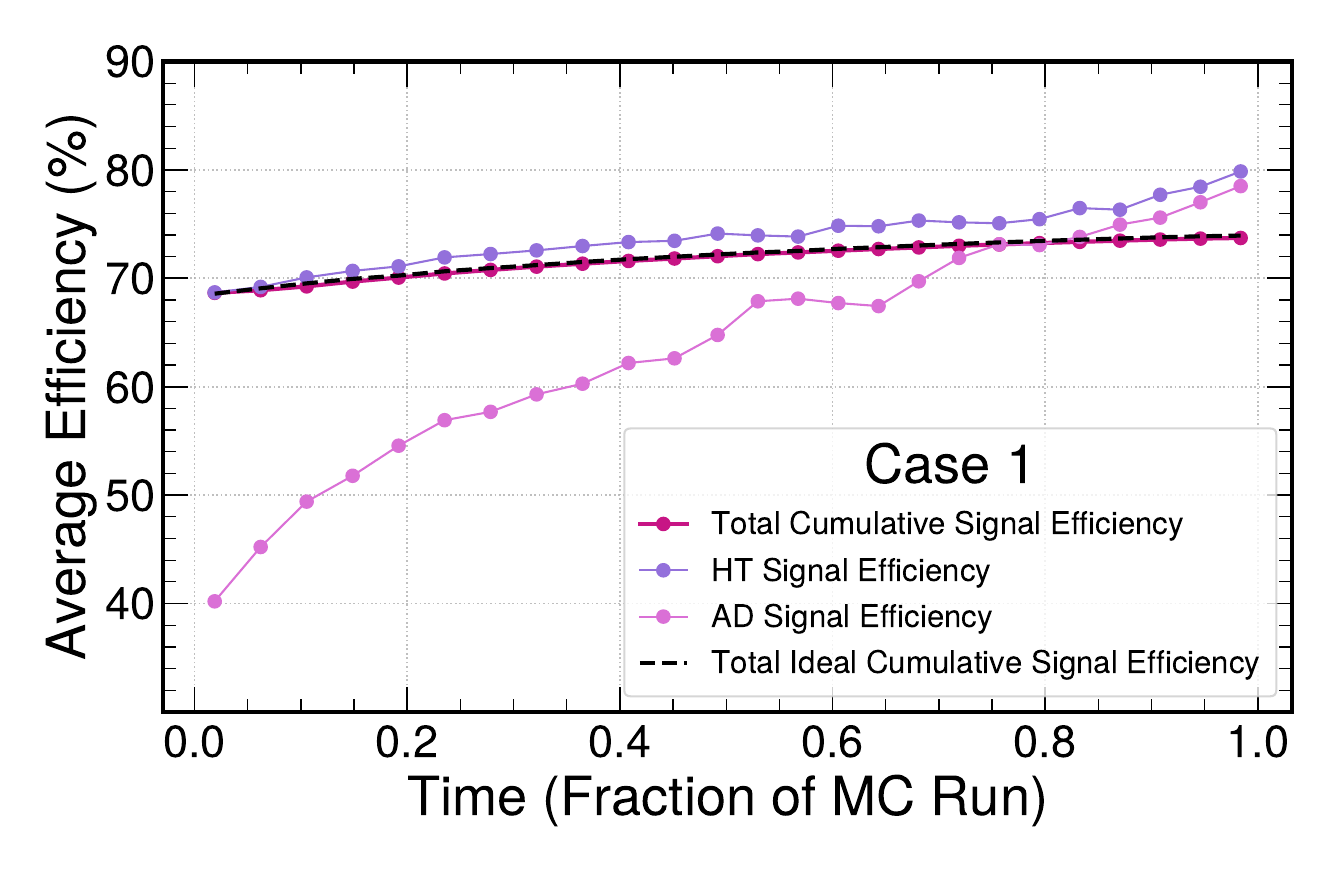}
        \caption{Signal Efficiency.}
        \label{fig:real1_b}
    \end{subfigure}

    \caption{
    Trigger rates for Case~1 under the simple controller. 
    The controller maintains the background rate close to the target while improving signal efficiency over time.
    }
    \label{fig:real1}
\end{figure}

\begin{figure}[htbp]
    \centering
    \begin{subfigure}[t]{0.45\textwidth}
        \centering
        \includegraphics[width=\linewidth]{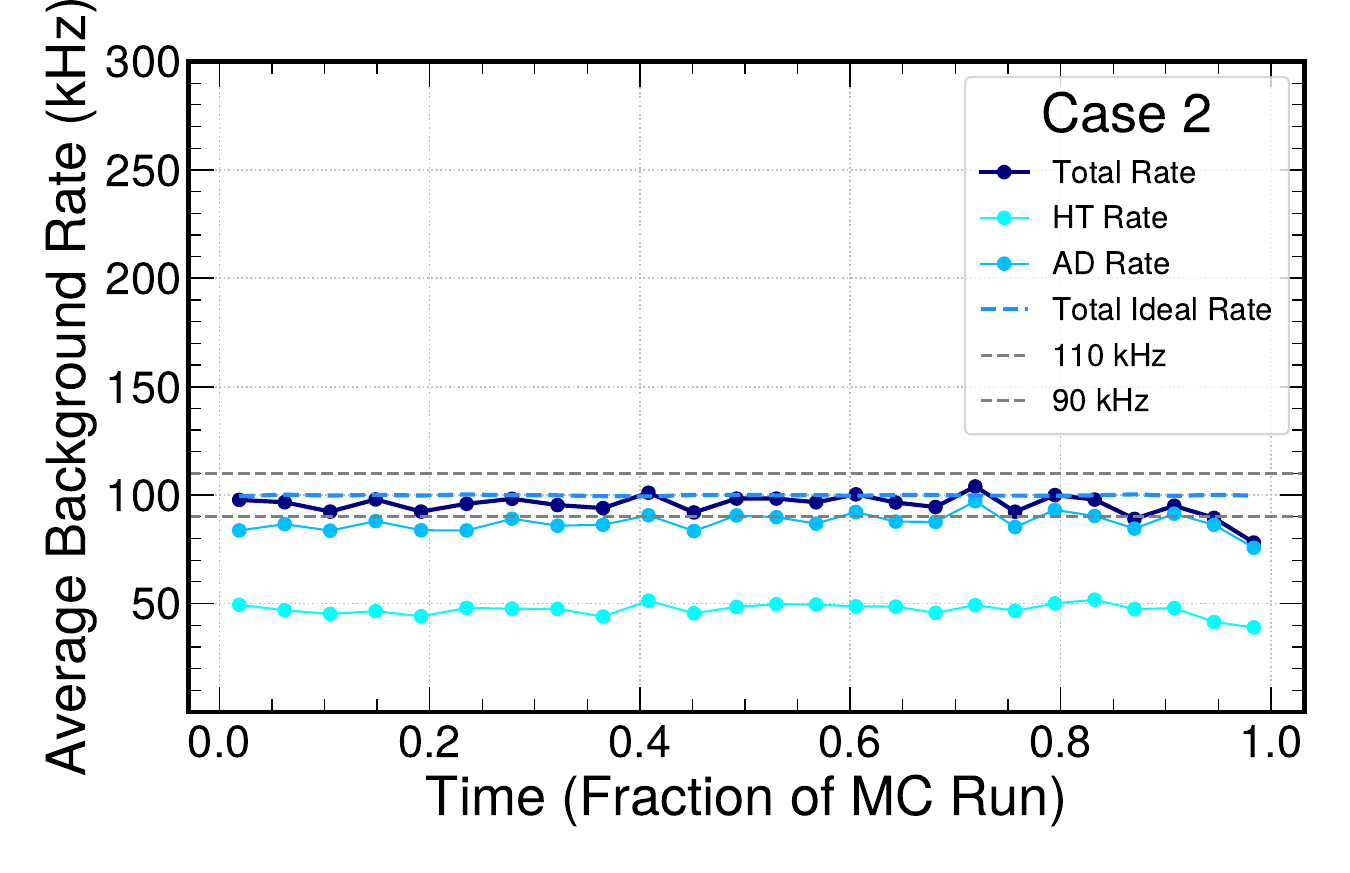}
        \caption{Background Rate.}
        \label{fig:real2_a}
    \end{subfigure}\qquad
    \begin{subfigure}[t]{0.45\textwidth}
        \centering
        \includegraphics[width=\linewidth]{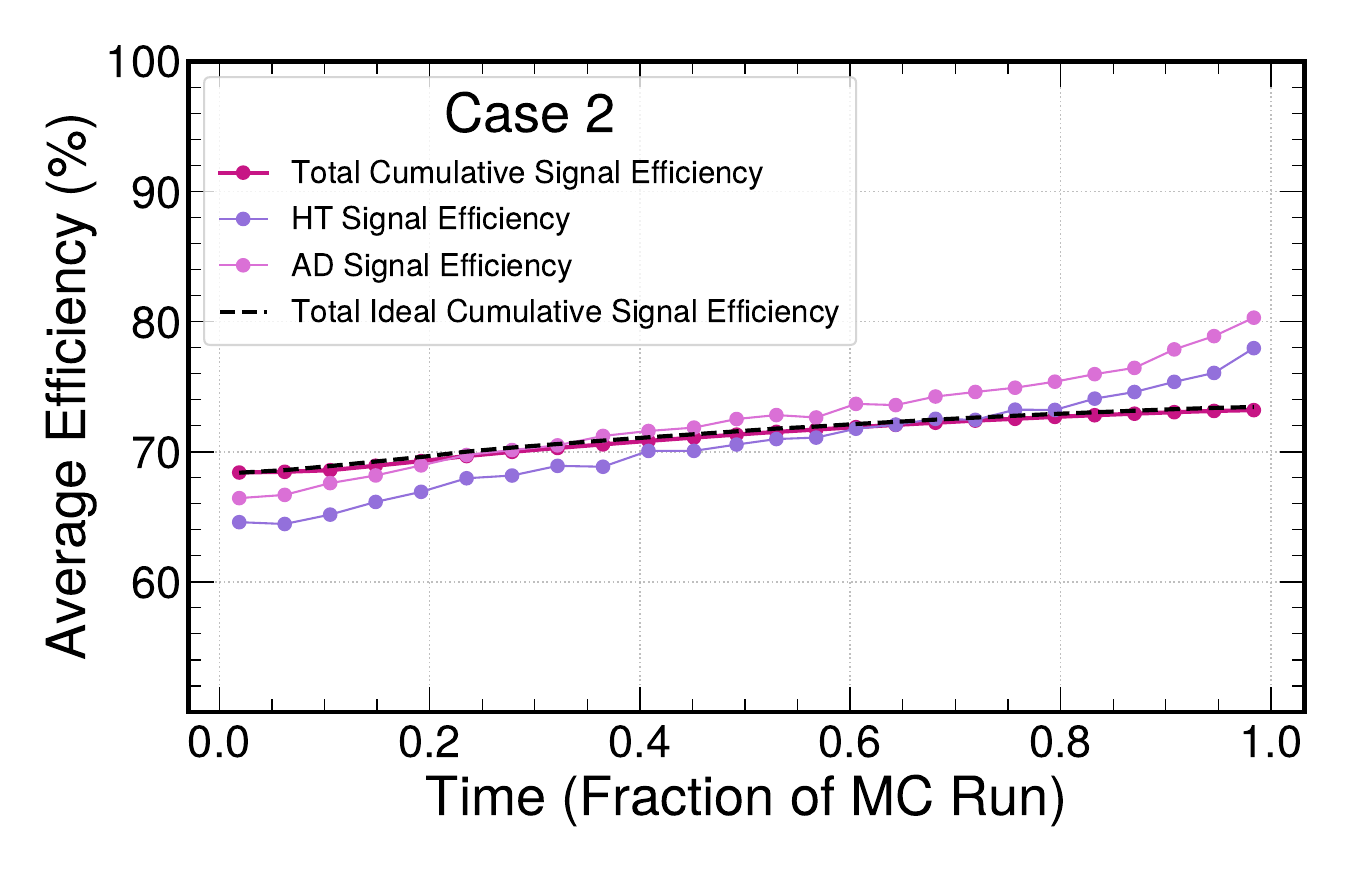}
        \caption{Signal Efficiency.}
        \label{fig:real2_b}
    \end{subfigure}

    \caption{
    Trigger rates for Case~2. 
    The controller successfully achieves specific bandwidth allocation to the anomaly detection path, while keeping the total background rate within limits.
    }
    \label{fig:real2}
\end{figure}

\begin{figure}[htbp]
    \centering
    \begin{subfigure}{0.45\textwidth}
        \centering
        \includegraphics[width=\linewidth]{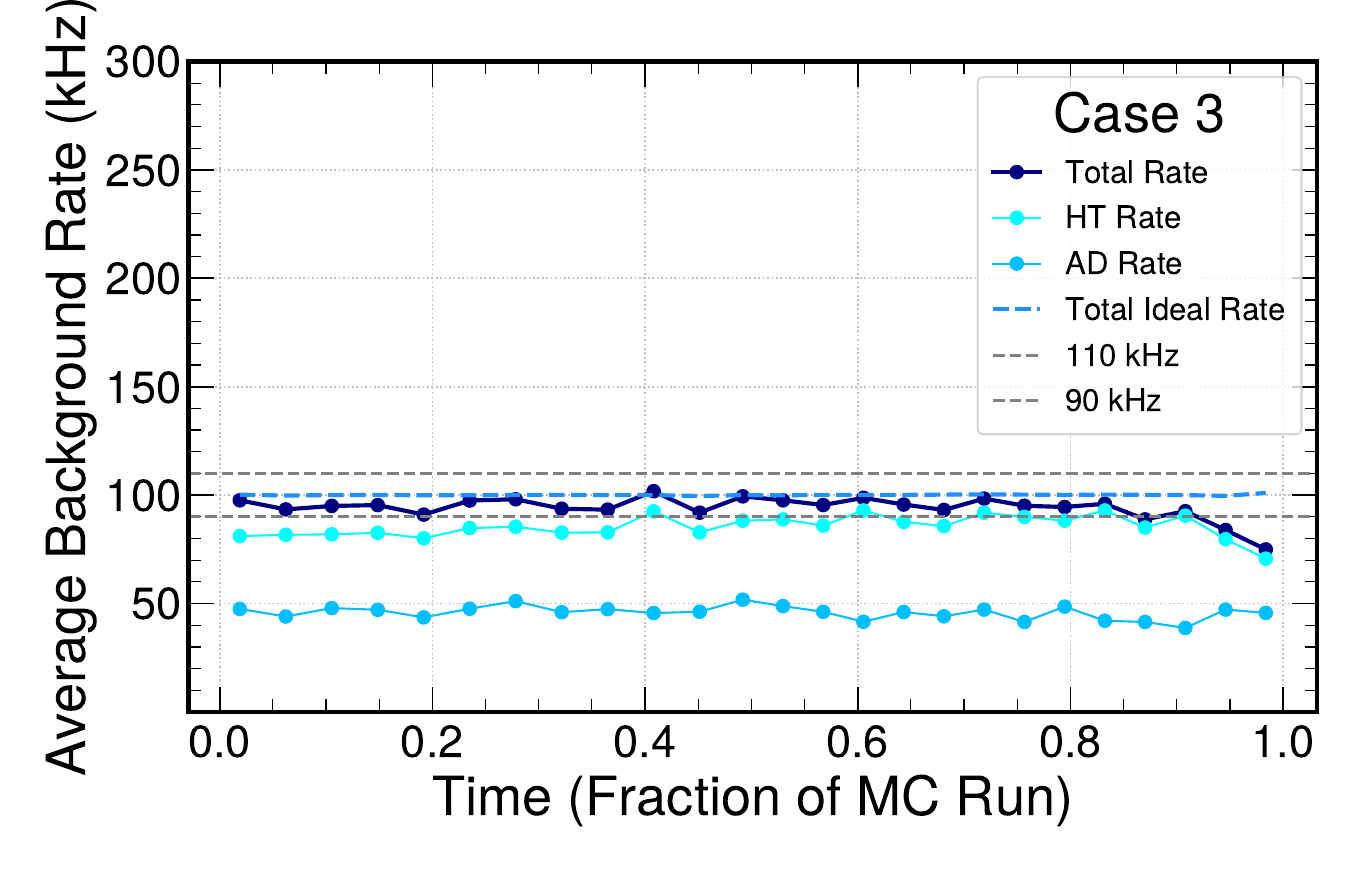}
        \caption{Background Rate.}
        \label{fig:real3_a}
    \end{subfigure}\qquad
    \begin{subfigure}{0.45\textwidth}
        \centering
        \includegraphics[width=\linewidth]{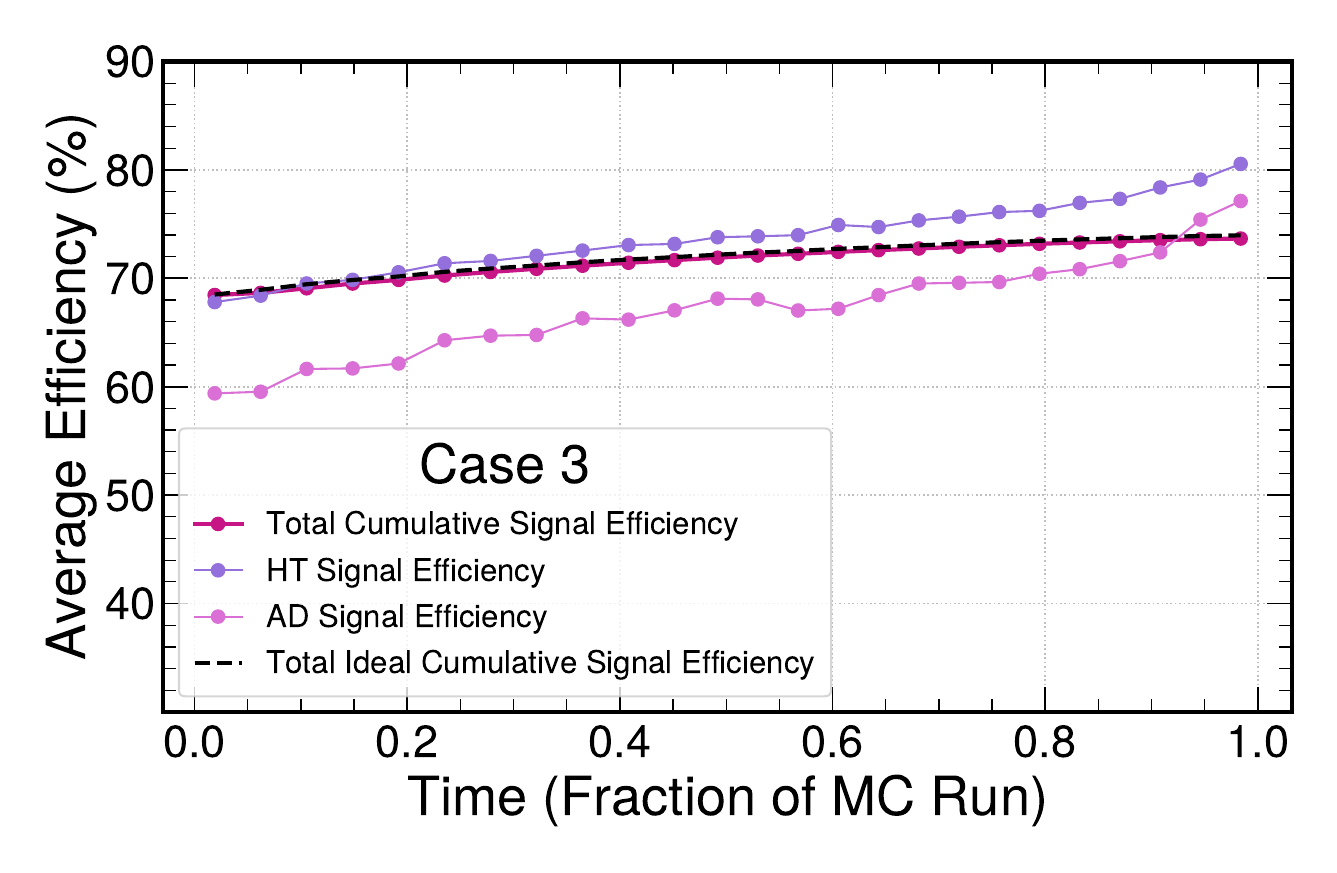}
        \caption{Signal Efficiency.}
        \label{fig:real3_b}
    \end{subfigure}

    \caption{
    Trigger rates for Case~3. 
    The controller incorporates computational cost into the optimization, maintaining physics performance high while keeping the trigger level resource usage under control.
    }
    \label{fig:real3}
\end{figure}

\section{Controller Performance with Data}
\label{sec:controller_perf_data}

Having established the performance of the autonomous trigger framework on simulated datasets, we now turn to real collision data, which provide the decisive test of the controller’s robustness under realistic detector conditions. We consider large samples of randomly triggered proton--proton collision events recorded by the CMS experiment during the 2016 data-taking period. Throughout this section, this sample is referred to as \emph{background data}.

In previous sections, simulated background samples were used to develop and validate the control strategy in a controlled setting. While essential at the design stage, simulation can only approximate real detector behavior and does not fully capture time-dependent effects such as evolving noise patterns, changing hardware configurations, and other subtle features of data taking. By contrast, the background data provide a direct and unbiased snapshot of detector activity, as events are recorded solely on the presence of colliding bunch crossings, without any physics-based selection. This makes them particularly well suited for testing real-time trigger strategies.

The anomaly detection autoencoder is retrained directly on background data, avoiding assumptions inherent to simulated samples, and the controller performance is evaluated under realistic operating conditions. Although these data may contain rare physics signals, their contribution is statistically negligible for both the training of the anomaly detection model and the background rate studies presented here. 

The autoencoder is trained on the second-longest run available in the dataset (Run~283876), while the controller performance is evaluated on the longest run (Run~283408), ensuring a clear separation between training and evaluation samples. Run~283408 spans 16 h 28 min and contains 1.99 million events in the random, prescaled stream used in this study, corresponding to an effective event rate of about 34 Hz. The controller processes events in batches of $2\times 10^{4}$ per update, such that each batch corresponds to an update every $\sim 12~$min. This rate is orders of magnitude lower than the rate seen by an online agent in a real trigger system (40 MHz at the bunch-crossing level), and should therefore be interpreted as a realistic, publicly available proxy for time-evolving conditions, rather than a faithful model of online throughput.

In realistic operation, MC simulation and collision data play complementary roles. Simulated samples, including both background and benchmark signal processes, are used offline to design the anomaly detection model, determine initial thresholds, and tune controller gains. Once deployed on collision data, the controller continuously updates rate estimates and dynamically adjusts trigger thresholds in response to evolving conditions such as changes in instantaneous luminosity. Within this framework, simulation provides predictive expectations, while the data-driven operation constitutes the definitive test of the system under realistic detector and beam conditions.

\subsection{Single-Path Real-Time Control on Data}
We begin by studying the simplest control scenario, in which a single trigger path is regulated in real time to maintain a fixed background rate. As in the simulation studies, we consider the \HT and AD triggers independently, each operating under a proportional--derivative PD feedback loop. For both triggers, the controller target is set to a background acceptance rate of 100~kHz, corresponding to a representative Level-1 bandwidth constraint. The controller updates the trigger threshold after each batch of 20\,000 events, based solely on the observed deviation of the background rate from the target.

As shown in Fig.~\ref{fig:bkg_rate_pid_data}, when applied to real collision data, the PD controller successfully compensates for the gradual decrease in pileup throughout the run. In contrast to a fixed-threshold menu, which exhibits a clear and monotonic rate drift as luminosity decreases, the controlled triggers maintain a stable background acceptance within the desired tolerance band. Residual batch-to-batch fluctuations are dominated by statistical uncertainties associated with the finite batch size rather than systematic trends.

The impact of dynamic control on signal performance is assessed by evaluating the relative signal efficiency as a function of time. For both the \ttbar\ and $h \to 4b$ benchmark processes, the controlled menus preserve a substantially higher instantaneous efficiency than fixed menus, particularly in the later stages of the run when pileup has decreased significantly (Fig.~\ref{fig:sig_pid_data}). This improvement is further reflected in the cumulative signal efficiency, shown in Fig.~\ref{fig:sig_cum_pid_data}, which quantifies the total fraction of signal events accepted over the full run. In all cases studied, real-time control mitigates the progressive loss of signal acceptance that is unavoidable with static thresholds. These results demonstrate that the PD controller, originally tuned on simulation, transfers robustly to real data without retuning, effectively stabilizing rates and preserving signal efficiency under realistic operating conditions.

\begin{figure}[htbp]
    \centering
    \begin{subfigure}{0.45\textwidth}
        \includegraphics[width=\linewidth]{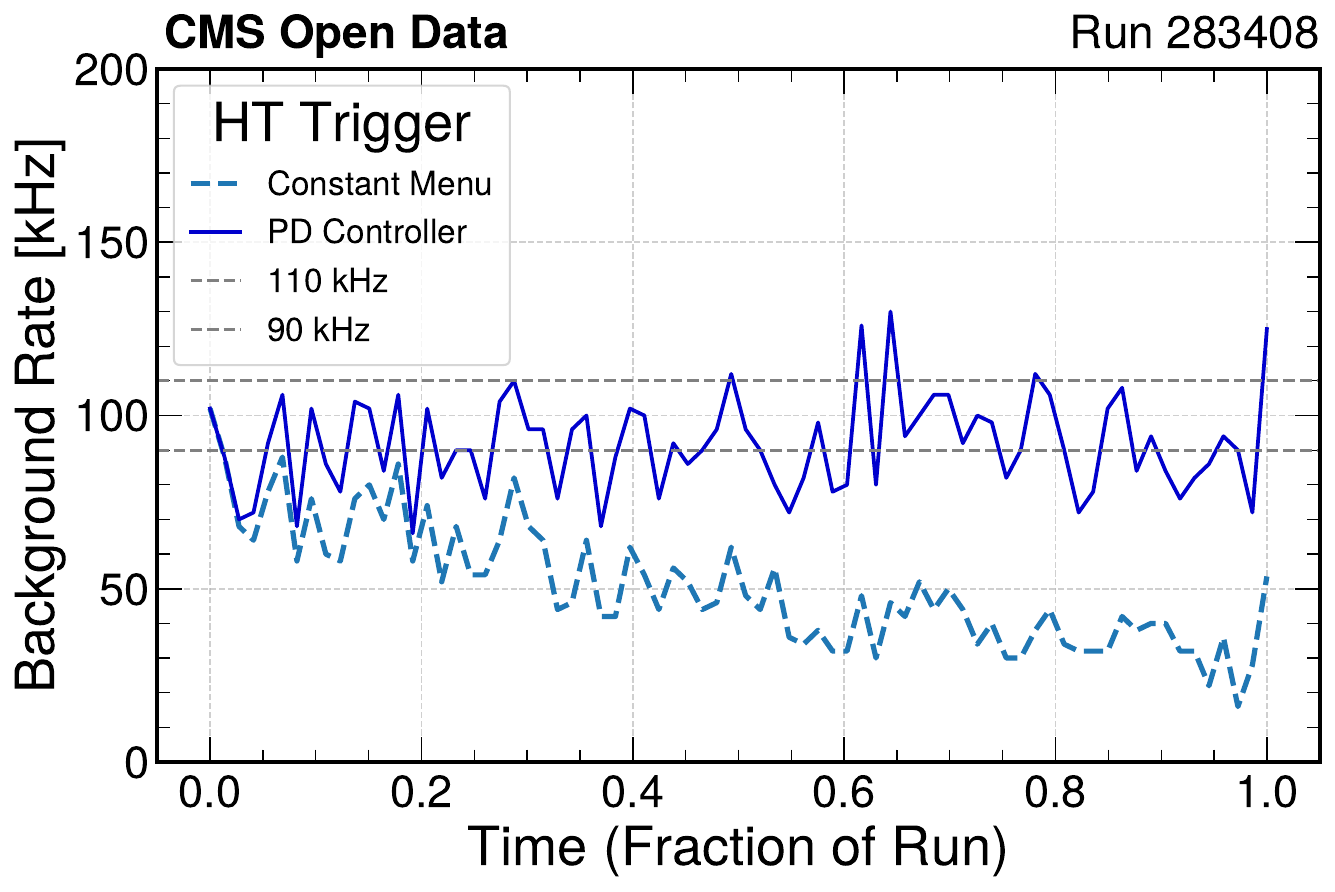}
        \caption{\HT trigger.}
        \label{fig:bkg_rate_pid_data_a}
    \end{subfigure}
    \qquad
    \begin{subfigure}{0.45\textwidth}
        \includegraphics[width=\linewidth]{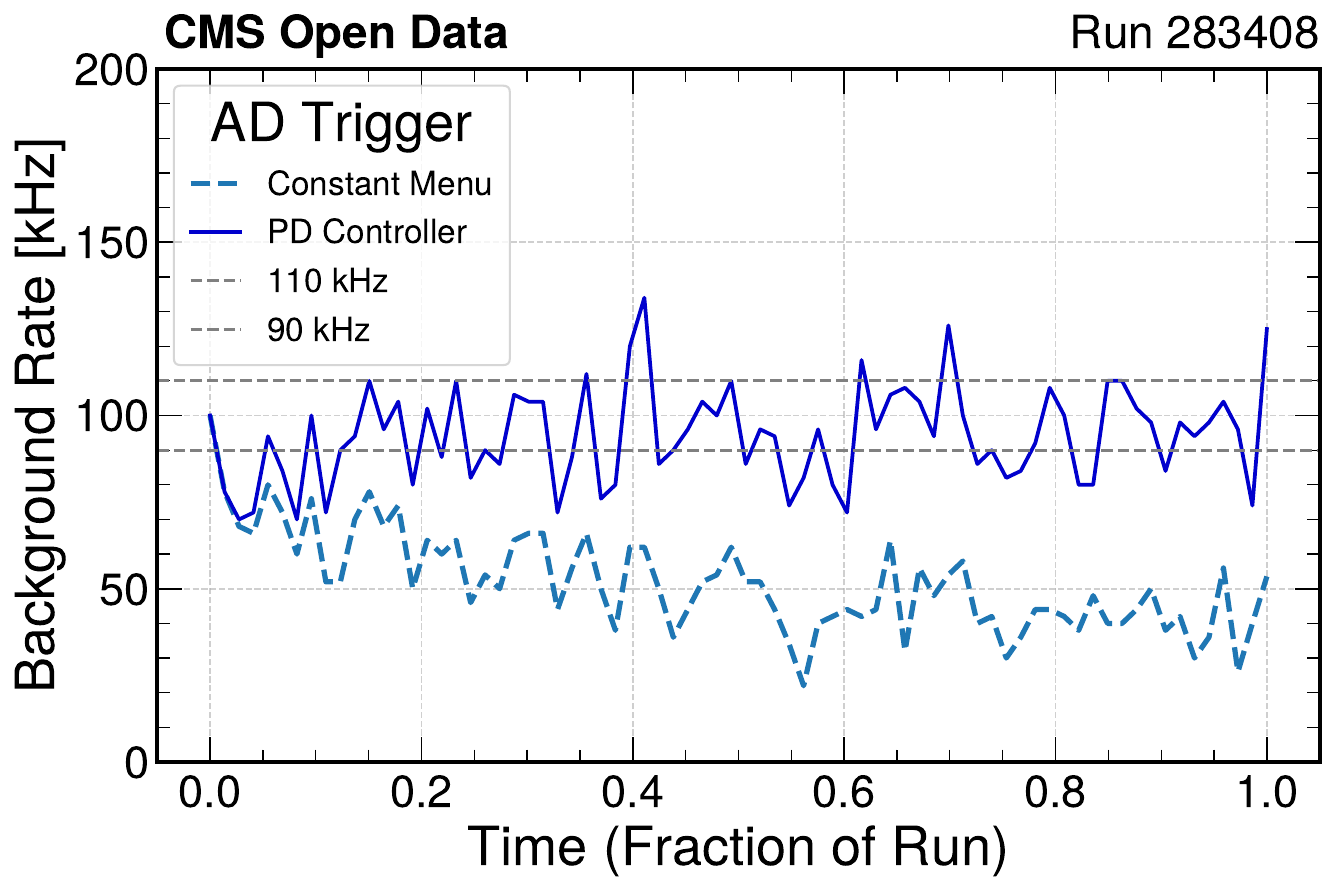}
        \caption{ AD   trigger.}
        \label{fig:bkg_rate_pid_data_b}
    \end{subfigure}

    \caption{
        Background trigger rates in real data under PD control.
        (a) \HT trigger,
        (b)  AD   trigger.
        The PD controller maintains the background rate within the target tolerance band while the constant menu decreases with time.
    }
    \label{fig:bkg_rate_pid_data}
\end{figure}

\begin{figure}[htbp]
    \centering
    \begin{subfigure}{0.45\textwidth}
        \includegraphics[width=\linewidth]{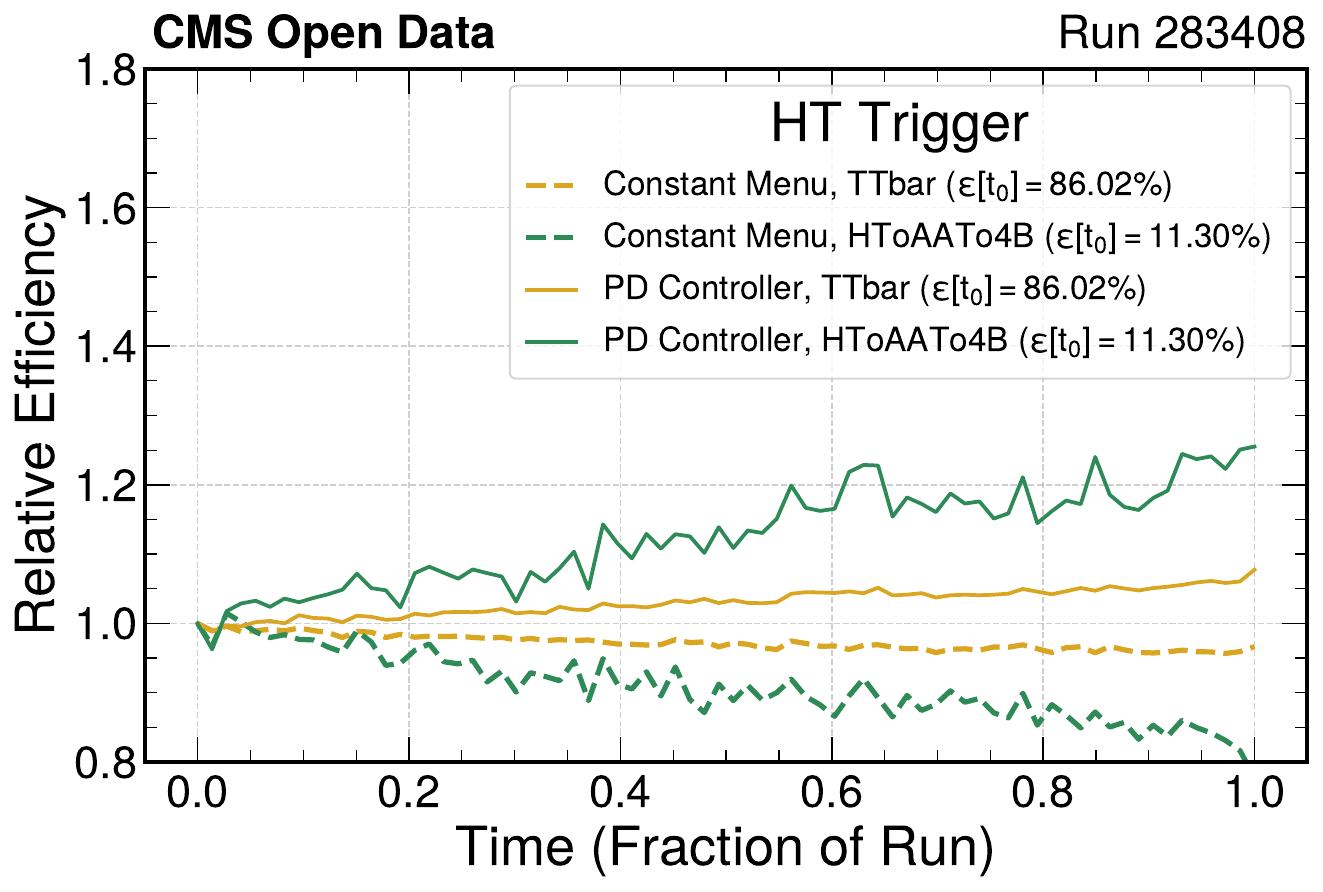}
        \caption{\HT trigger.}
        \label{fig:sig_inst_pid_data_a}
    \end{subfigure}
    \qquad
    \begin{subfigure}{0.45\textwidth}
        \includegraphics[width=\linewidth]{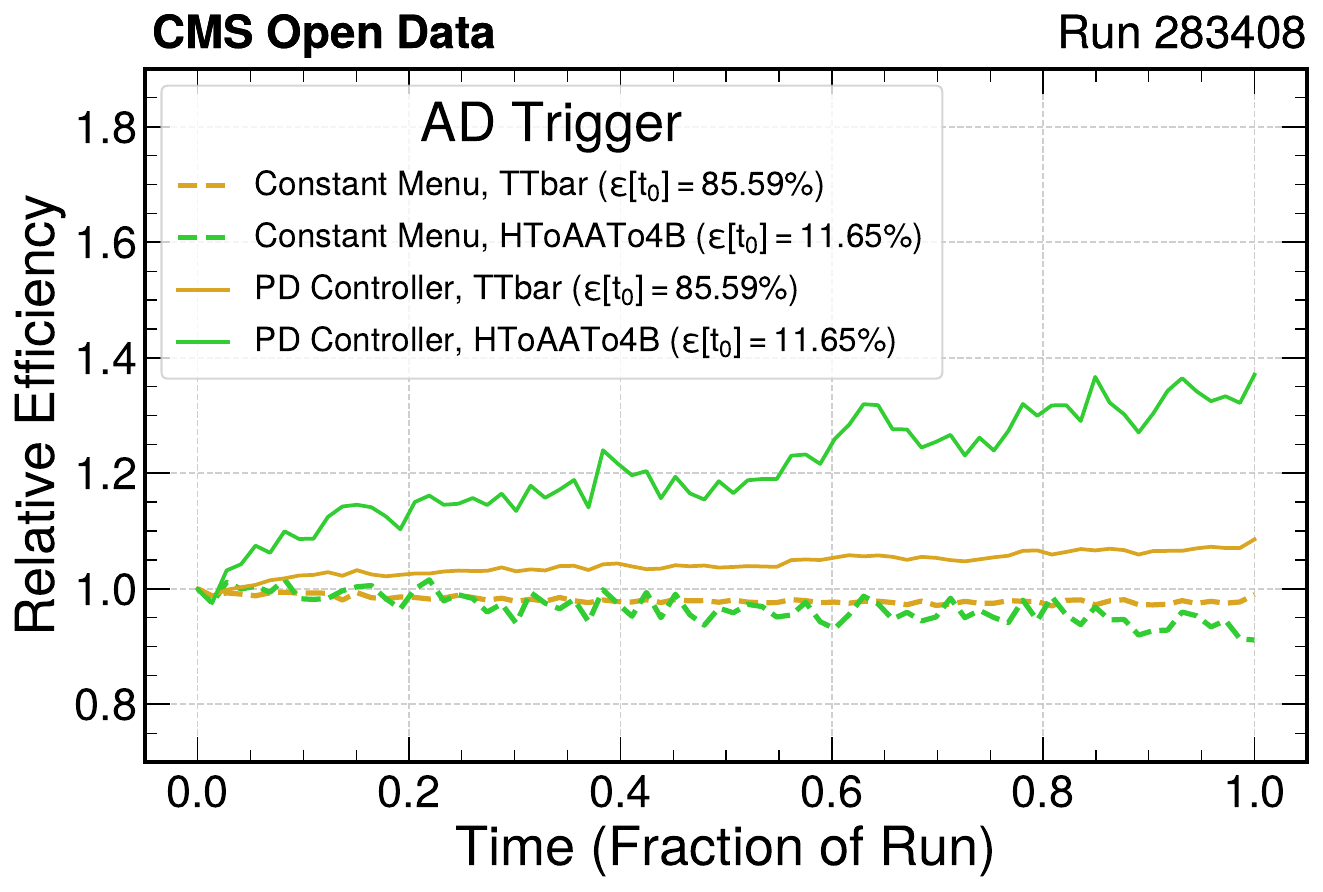}
        \caption{ AD   trigger.}
        \label{fig:sig_inst_pid_data_b}
    \end{subfigure}

    \caption{
        Relative change in instantaneous signal efficiency over time, in data.
        (a) \HT trigger,
        (b)  AD   trigger.
        PD controlled menus achieve higher and more stable signal efficiencies compared to static menus.
    }
    \label{fig:sig_pid_data}
\end{figure}

\begin{figure}[htbp]
    \centering
    \begin{subfigure}{0.45\textwidth}
        \includegraphics[width=\linewidth]{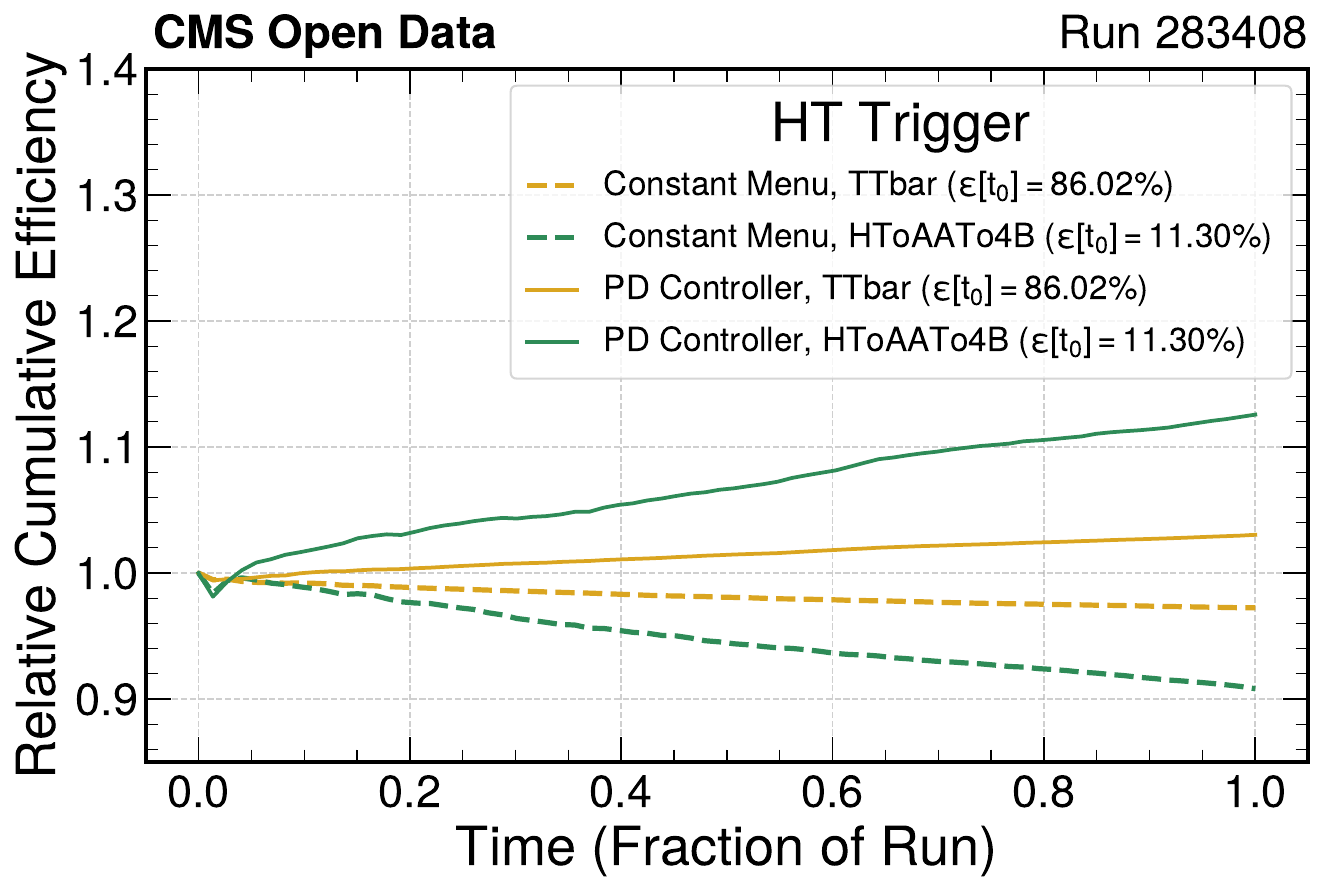}
        \caption{\HT trigger.}
        \label{fig:sig_cum_pid_data_a}
    \end{subfigure}
    \qquad
    \begin{subfigure}{0.45\textwidth}
        \includegraphics[width=\linewidth]{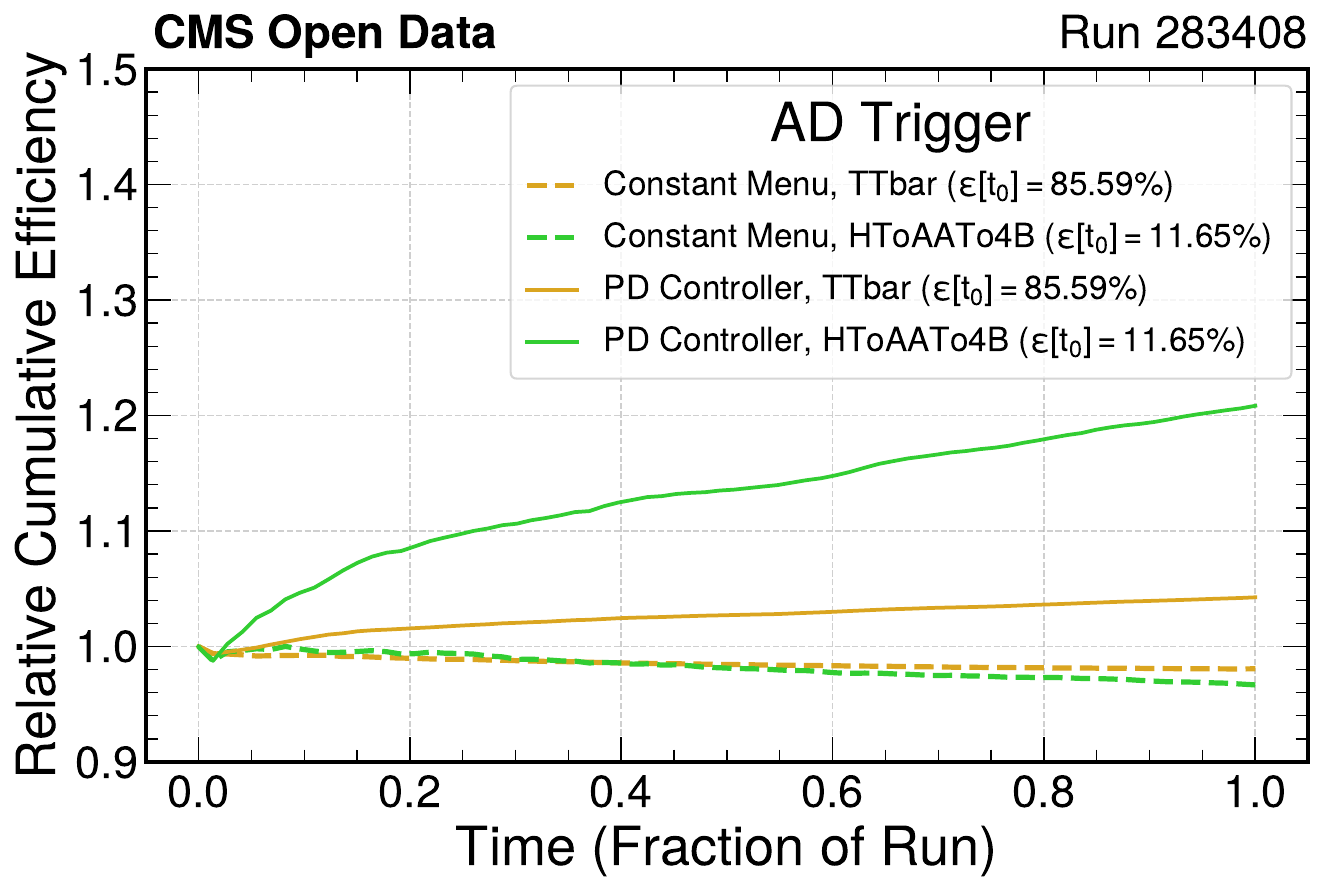}
        \caption{ AD   trigger.}
        \label{fig:sig_cum_pid}
    \end{subfigure}

    \caption{
        Cumulative signal efficiency in real data.
        (a) \HT trigger,
        (b)  AD   trigger.
        The PD controller preserves a higher integrated signal yield (up to 25\% gain in comparison to the fixed menu) while keeping background rates stable throughout the run.
    }
    \label{fig:sig_cum_pid_data}
\end{figure}

\subsection{Multi-Path Real-Time Control on Data}

We now extend the study to the more realistic scenario in which multiple trigger paths operate simultaneously and compete for a shared bandwidth. In this configuration, the controller adjusts the thresholds of both the \HT and AD triggers in concert, guided by a global cost function that encodes multiple objectives. As in Section~\ref{sec:local_controller}, the ideal controller, assuming full knowledge of future data, though not physically realizable, serves as a valuable baseline for assessing how closely practical strategies approach the theoretical optimum.
For this reason, we also use it on data as a benchmark reference, however, we focus on evaluating the behavior of the simple controller in the three representative configurations previously introduced in Section~\ref{section6}, namely Case 1, Case 2, and Case 3. The corresponding results are shown in Figs.~\ref{fig:real1_data}, ~\ref{fig:real2_data} and \ref{fig:real3_data}, with each plot illustrating a different control strategy:
\begin{itemize}
\item \textit{Case 1}  represents the simplest configuration, where the \HT and  AD   paths are treated symmetrically, with no explicit prioritization. The controller responds effectively to background fluctuations, maintaining stable rates over time and achieving improved signal.

\item  \textit{Case 2}  introduces an exclusive fixed bandwidth allocation for the  AD   trigger. In this case, it corresponds to 50\% of the total bandwidth.

\item \textit{Case 3} implements the full cost-aware strategy, simultaneously accounting for computational cost, signal efficiency, and bandwidth constraints. 

Figure~\ref{fig:cost_histogram_data}
illustrates the cost reference distribution derived from the Case 1 configurations. The new reference values and cost weights used in the real data optimization are:
\[
\sigma_b = 4\, \text{kHz},  \quad
\sigma_s = 0.05, \quad
\sigma_\text{evt} = 0.5, \quad
\sigma_\text{algo} = 0.5,  \quad
C_{\text{evt}}^{\text{ref}} = 4.3,  \quad C_{\text{algo}}^{\text{ref}} = 3.5. 
\]

\end{itemize}

The results, shown in the aforementioned plots, provide compelling evidence that the autonomous control framework, initially developed in simulation, remains effective when applied to real data.
Although moving to data introduces distributional shifts and amplifies the intrinsic volatility of the anomaly detection, the controller exhibits robust and interpretable behavior across configurations, dynamically adapting to detector fluctuations while consistently achieving key objectives such as rate stabilization, signal efficiency, and cost optimization.
Even in the presence of realistic noise and operational variability, it maintains reliable performance, demonstrating flexibility in accommodating diverse experimental and system constraints. By enabling real-time feedback-driven optimization, the framework offers a path toward more autonomous, intelligent, and efficient trigger systems, capable of reshaping data acquisition at the LHC and meeting the evolving demands of future high-energy physics experiments.

\begin{figure}[htbp]
    \centering
    \begin{subfigure}[t]{0.48\textwidth}
        \centering
        \includegraphics[width=\linewidth]{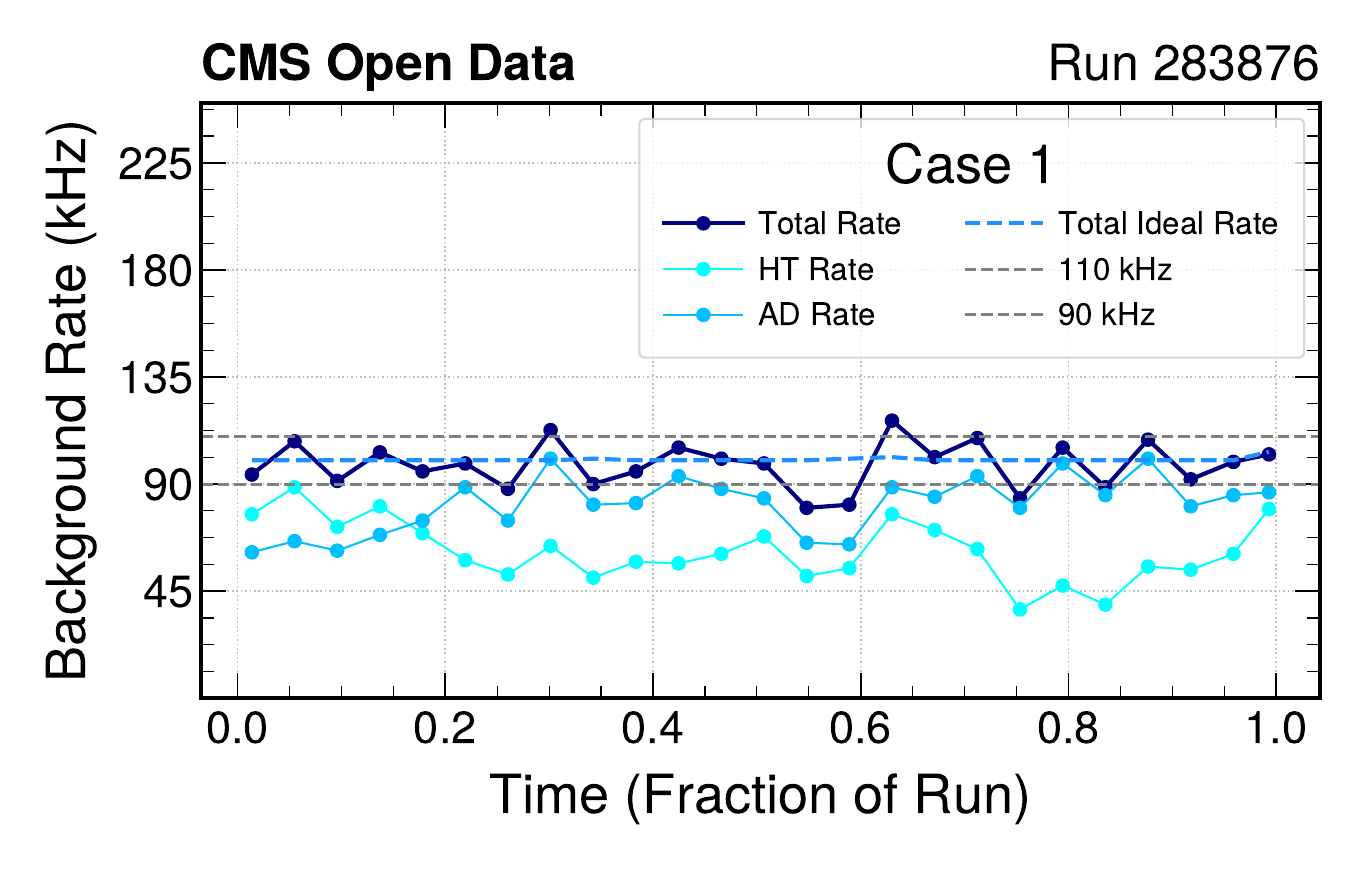}
        \caption{Background Rate.}
        \label{fig:real1_data_a}
    \end{subfigure}\qquad
    \begin{subfigure}[t]{0.47\textwidth}
        \centering
        \includegraphics[width=\linewidth]{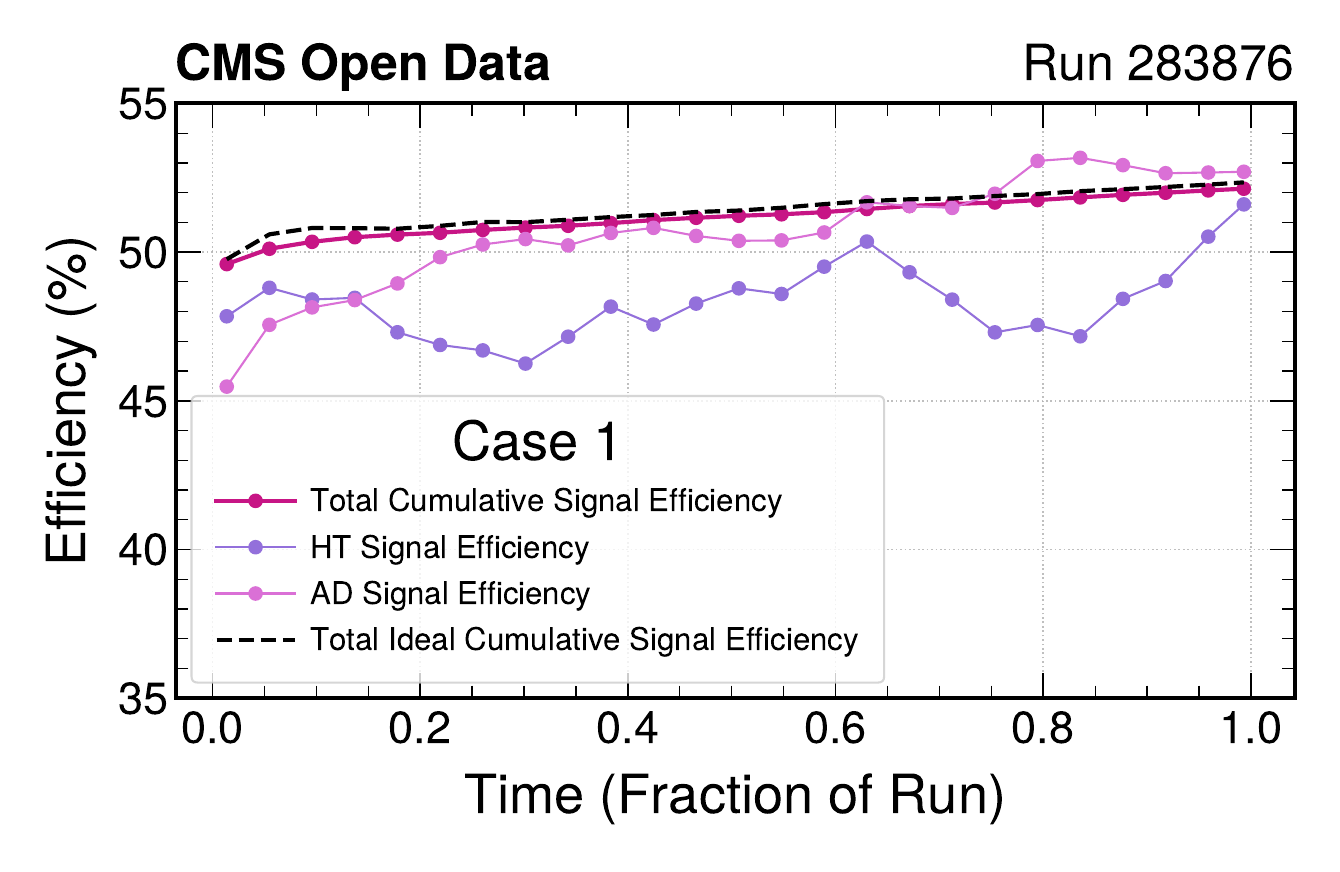}
        \caption{Signal Efficiency.}
        \label{fig:real1_data_b}
    \end{subfigure}

    \caption{
Trigger rate and efficiency for Case 1. The controller maintains the background rate close to the target while improving signal efficiency over time.
    }
    \label{fig:real1_data}
\end{figure}

\begin{figure}[htbp]
    \centering
    \begin{subfigure}[t]{0.48\textwidth}
        \centering
        \includegraphics[width=\linewidth]{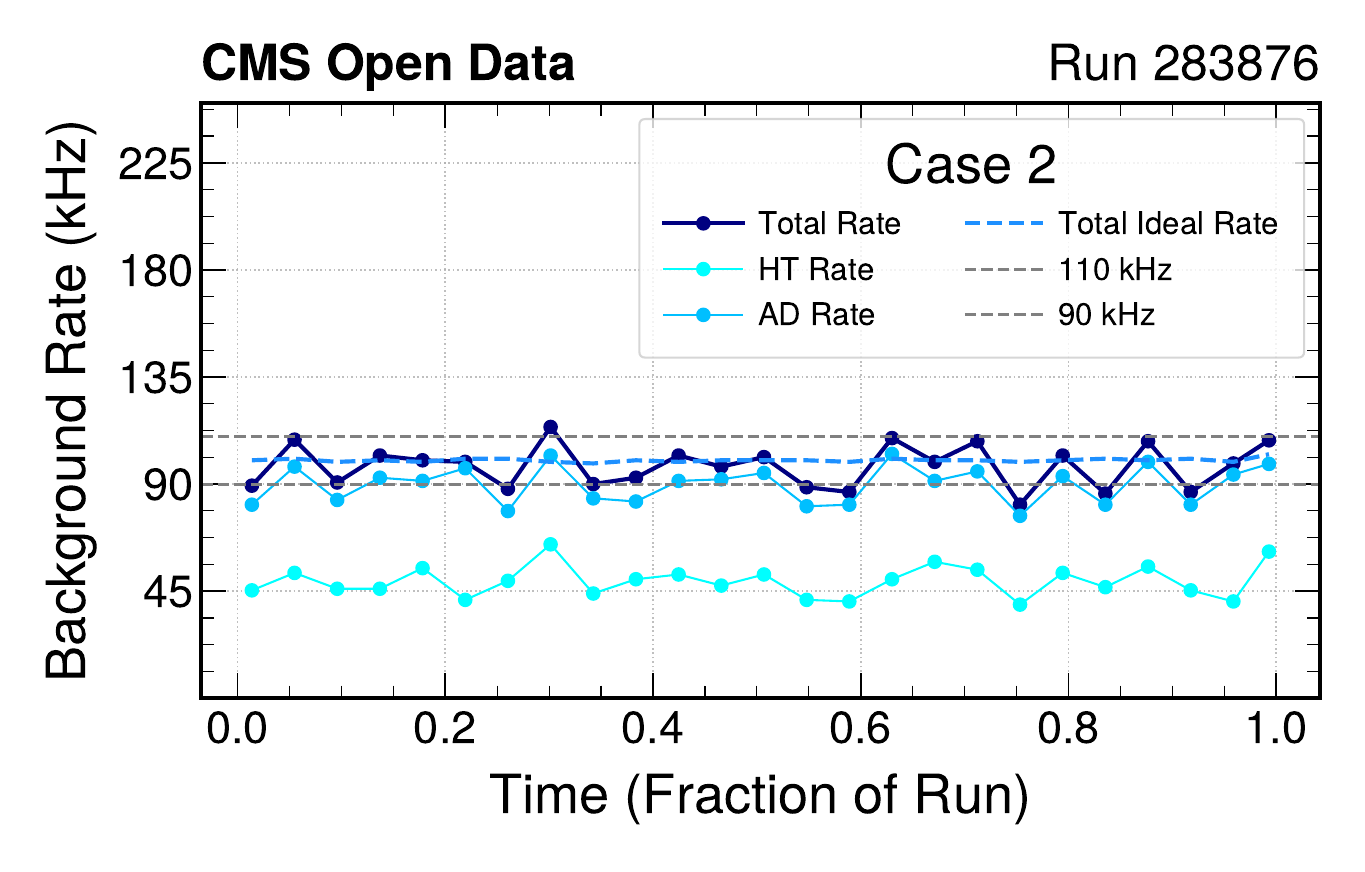}
        \caption{Background Rate.}
        \label{fig:real2_data_a}
    \end{subfigure}\qquad
    \begin{subfigure}[t]{0.47\textwidth}
        \centering
        \includegraphics[width=\linewidth]{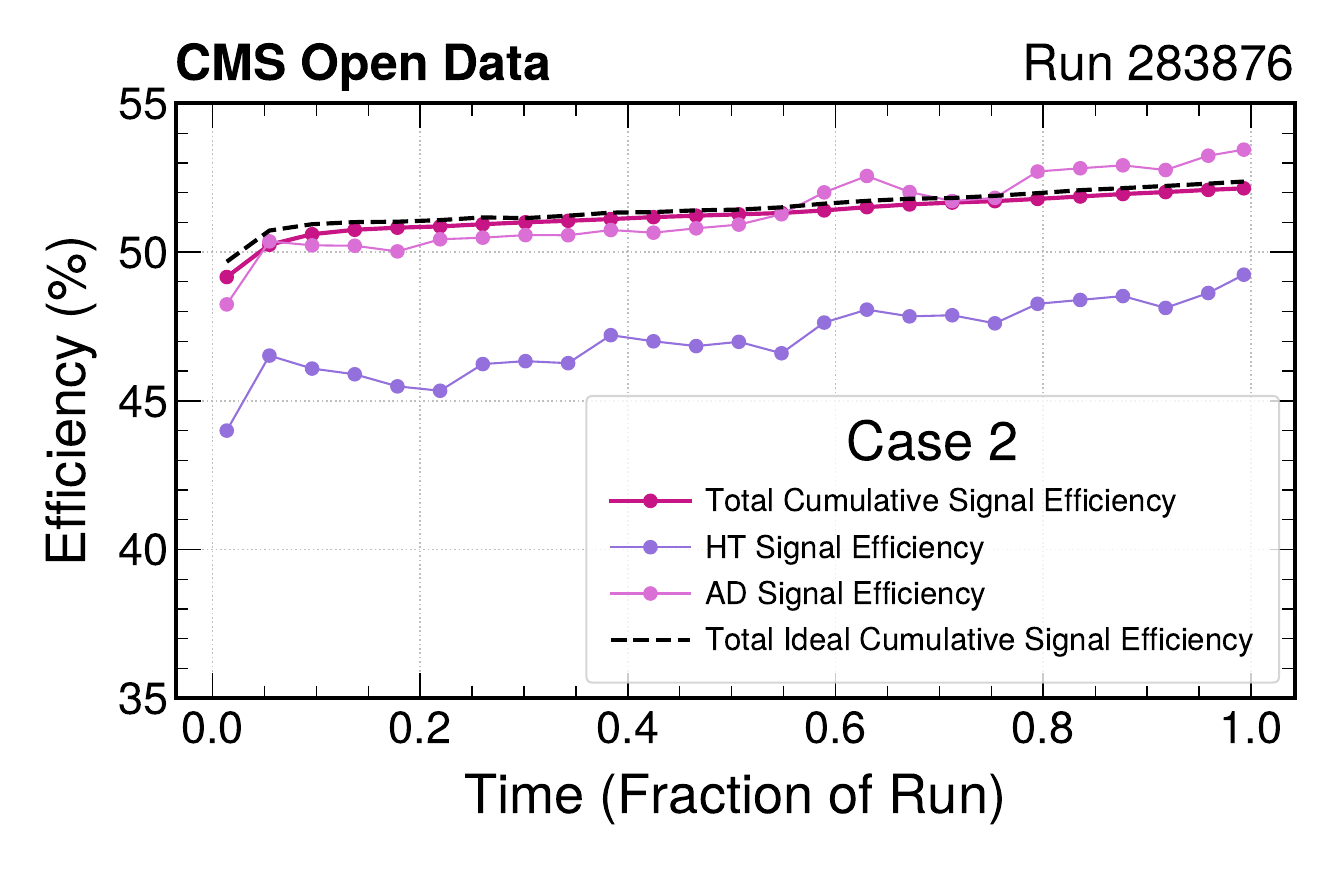}
        \caption{Signal Efficiency.}
        \label{fig:real2_data_b}
    \end{subfigure}
    \caption{
Trigger rates for Case 2 in data. 
    The controller successfully achieves specific bandwidth allocation to the anomaly detection path, while keeping the total background rate within limits.}
    \label{fig:real2_data}
\end{figure}

\begin{figure}[htbp]
    \centering
    \begin{subfigure}[t]{0.48\textwidth}
        \centering
        \includegraphics[width=\linewidth]{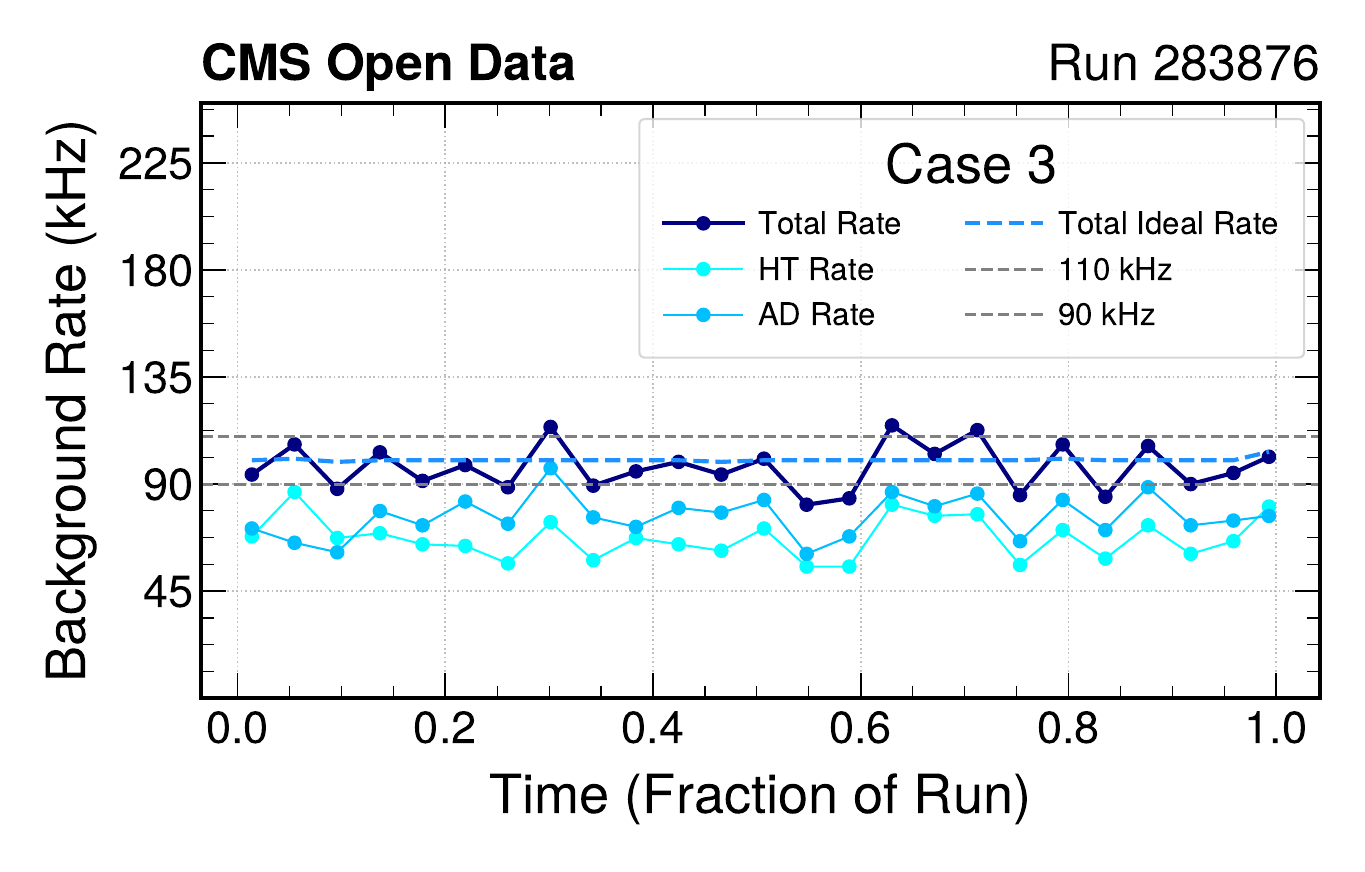}
        \caption{Background Rate.}
        \label{fig:real3_data_a}
    \end{subfigure}\qquad
    \begin{subfigure}[t]{0.47\textwidth}
        \centering
        \includegraphics[width=\linewidth]{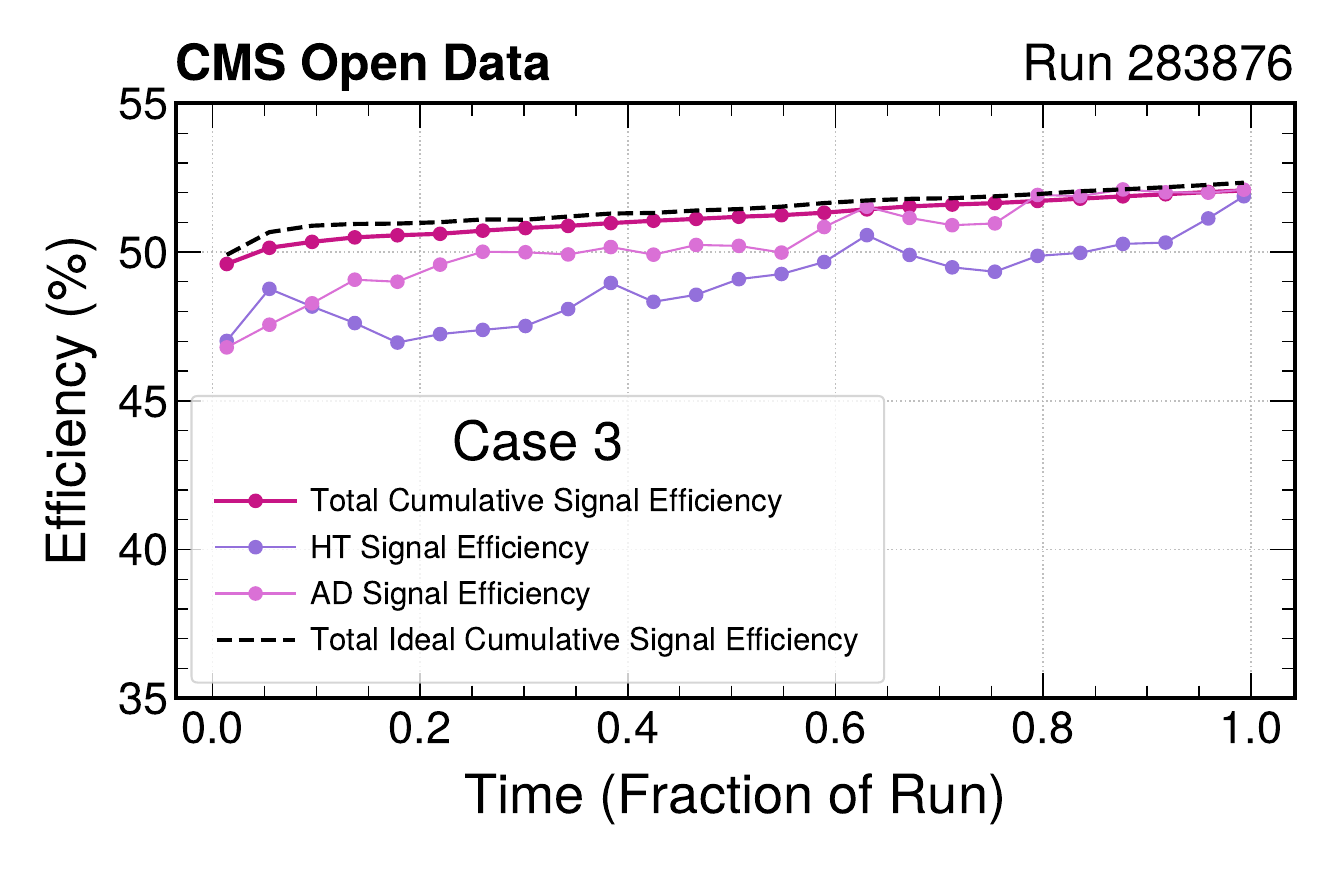}
        \caption{Signal Efficiency.}
        \label{fig:real3_data_b}
    \end{subfigure}

    \caption{
    Trigger rates for Case 3 in data. The controller incorporates computational cost into the optimization, maintaining physics performance while keeping the trigger-level resource usage under control.
    }
    \label{fig:real3_data}
\end{figure}

\begin{figure}[htbp]
    \centering
    \begin{subfigure}[t]{0.4\textwidth}
        \vspace{0pt}
        \centering
        \includegraphics[width=\linewidth]{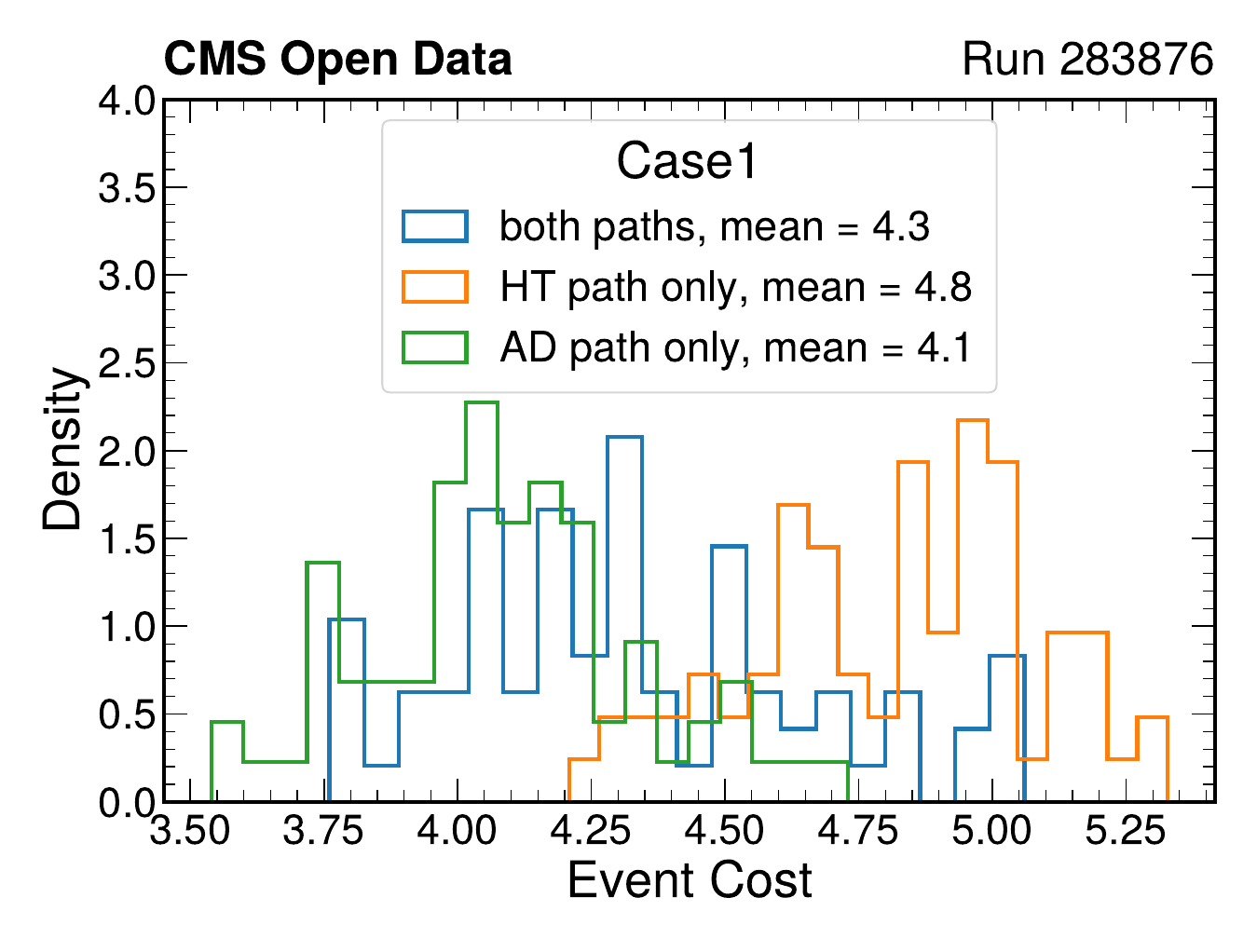}
        \caption{Event-level cost for three configurations: both paths (blue), only the \HT path (orange), only the AD path (green).}
        \label{fig:comp_cost_a}
    \end{subfigure}
     \hspace{0.02\textwidth}
    \begin{subfigure}[t]{0.415\textwidth}
        \vspace{0pt}
        \centering
        \includegraphics[width=\linewidth]{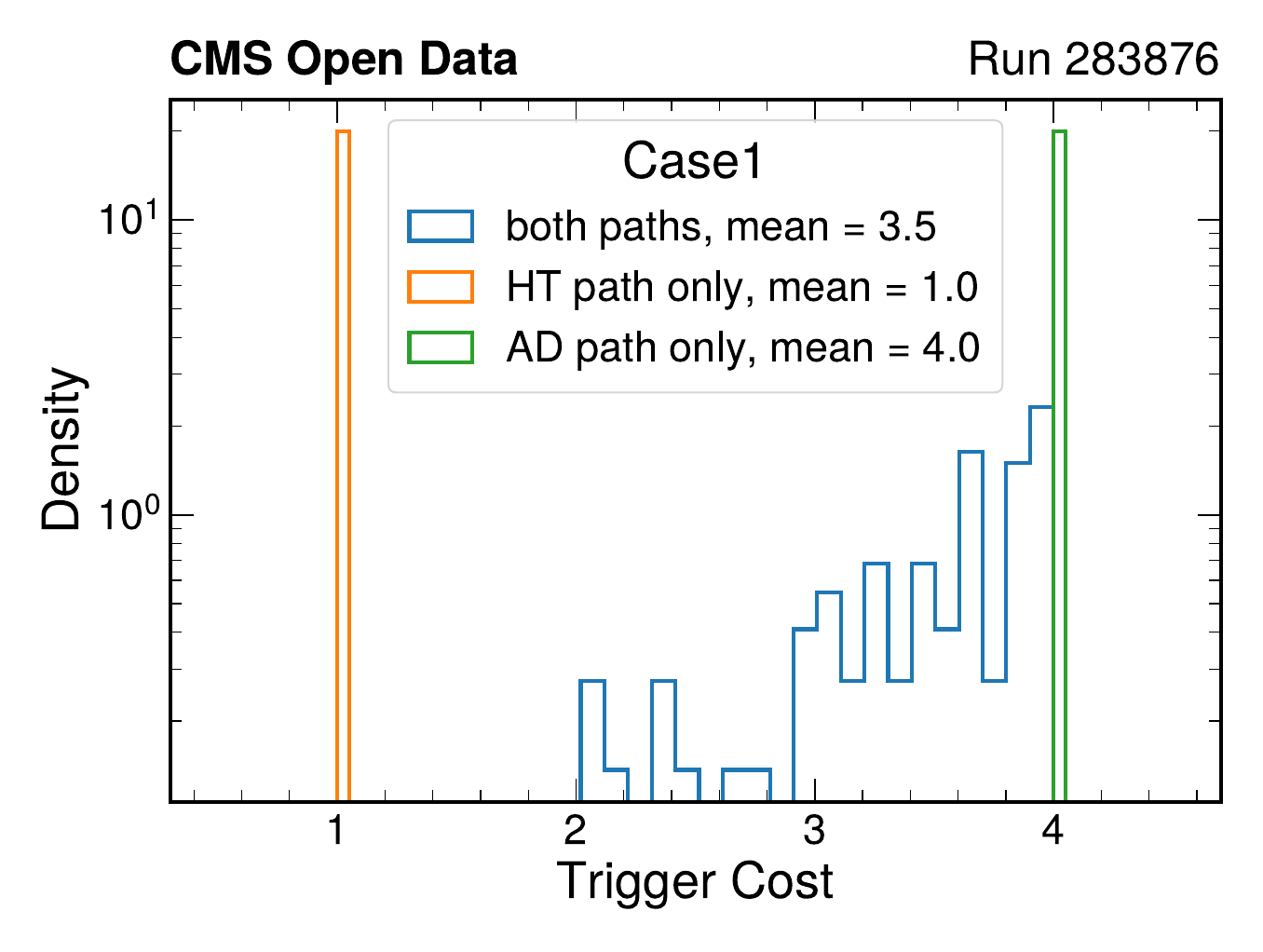}
        \caption{Trigger-path level cost distributions for the same three configurations, showing the relative computational load associated with each path.}
        \label{fig:comp_cost_b}
    \end{subfigure}

    \caption{
    Distributions of computational costs of accepted background events for Case~1.
    Panels (a) and (b) define the reference cost values used in Case 3 optimization.
    }
    \label{fig:cost_histogram_data}
\end{figure}


\section{Summary Plots and Data--Simulation Comparison}
\label{sec:summary_plots}

This section condenses the main outcomes of the benchmark into two compact summaries: one showing the results using the \emph{simulated background} in Fig.~\ref{fig:summary_panel4} and one showing the results using the  \emph{data background} in Fig.~\ref{fig:summary_data}. 
Rather than showing each metric separately as a function of time, these plots represent the trajectory traced by a given control strategy as operating conditions evolve during a run. Each marker corresponds to a time slice, and the color gradient encodes progression through the run (dark $\rightarrow$ light), with the start-of-run point explicitly indicated. The Fixed Menu shown in these figures provides a static reference, while the three adaptive agents, corresponding to Cases 1, 2, and 3, implement feedback control with different optimization objectives.

\subsection{Performance Summary on Simulated Events}
\label{subsec:summary_sim}

Figure~\ref{fig:summary_panel4} summarizes the simulated case study, where the time evolution is emulated through the event ordering to mimic a gradual decrease in instantaneous luminosity. 
Each of the panels illustrates the evolution of paired trigger performance metrics over time as measured in signal and background events, which is strongly dictated by the policy adopted by each control agent.

\begin{figure}[htbp]
  \centering
  \setlength{\tabcolsep}{4pt}
  \renewcommand{\arraystretch}{1.0}
  \begin{tabular}{ccc}
    \begin{subfigure}{0.42\linewidth}
      \includegraphics[width=\linewidth]{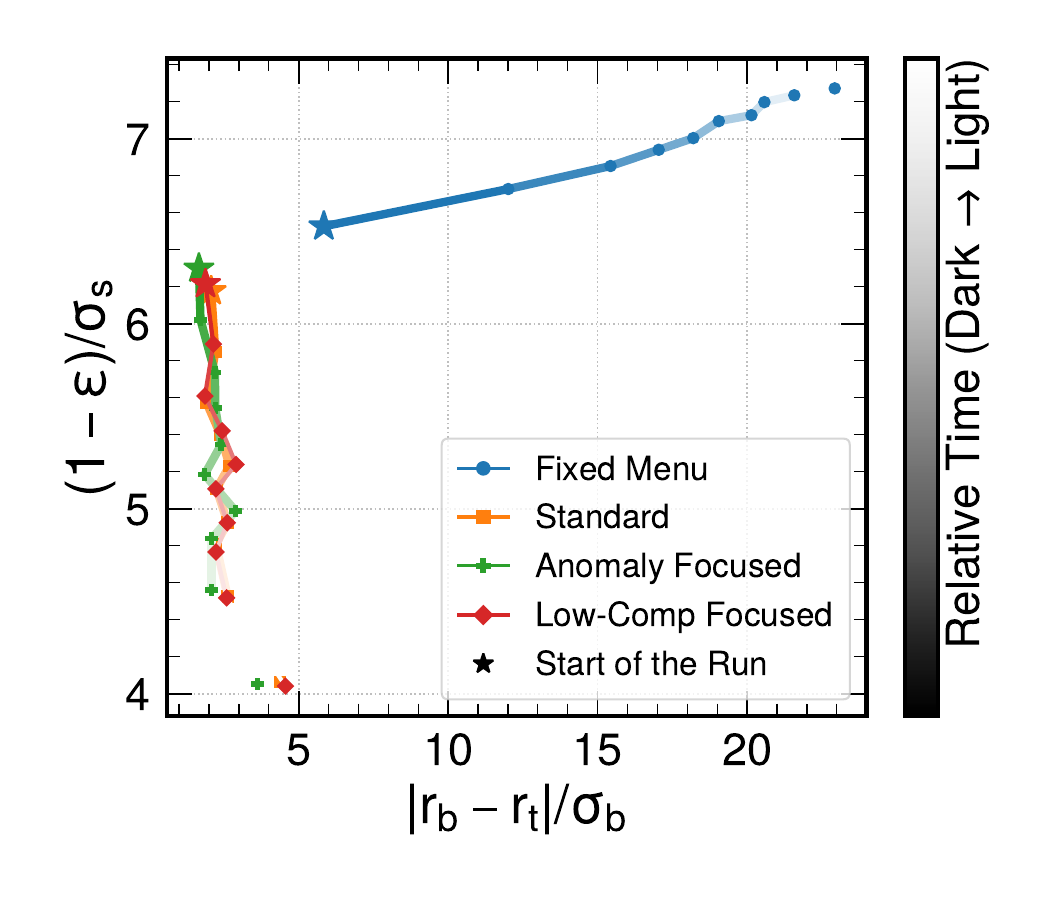}
      \caption{}
      \label{fig:summary_panel4-a}
    \end{subfigure} &
    \begin{subfigure}{0.42\linewidth}
      \includegraphics[width=\linewidth]{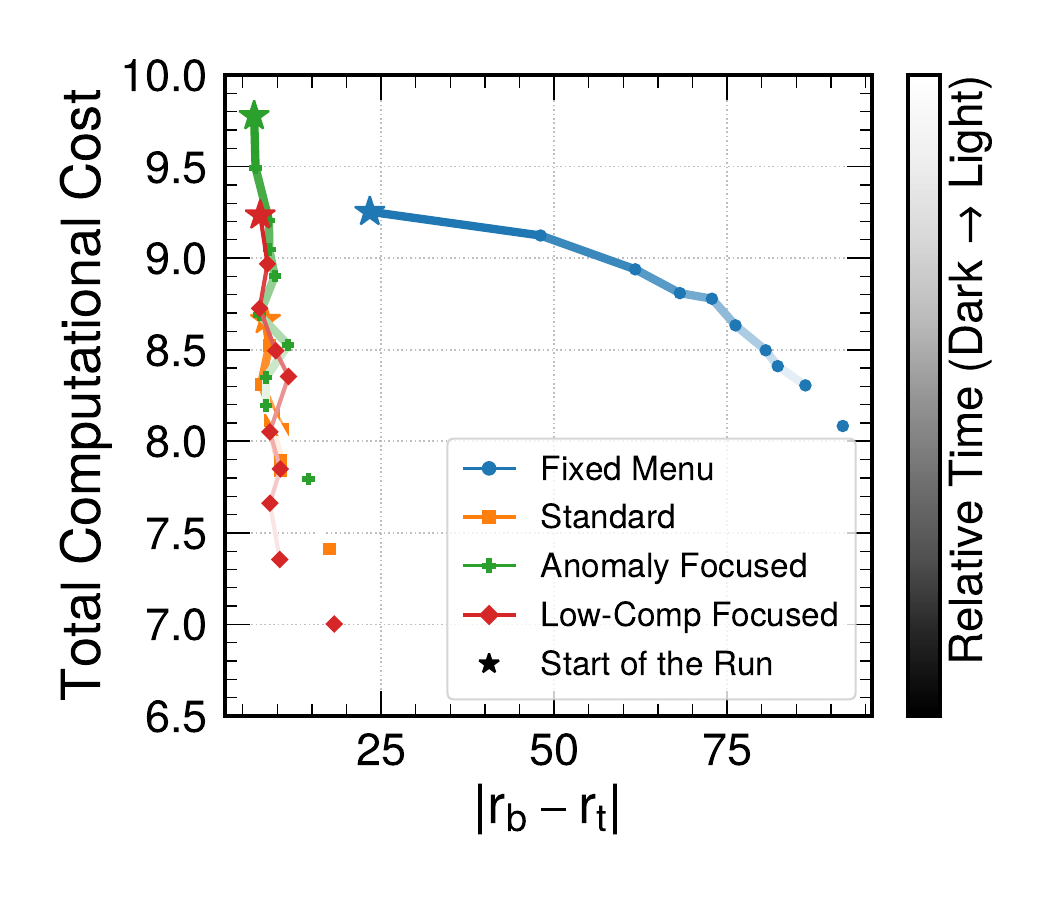}
      \caption{}
      \label{fig:summary_panel4-b}
    \end{subfigure} &
     \\[2mm]

    \begin{subfigure}{0.42\linewidth}
      \includegraphics[width=\linewidth]{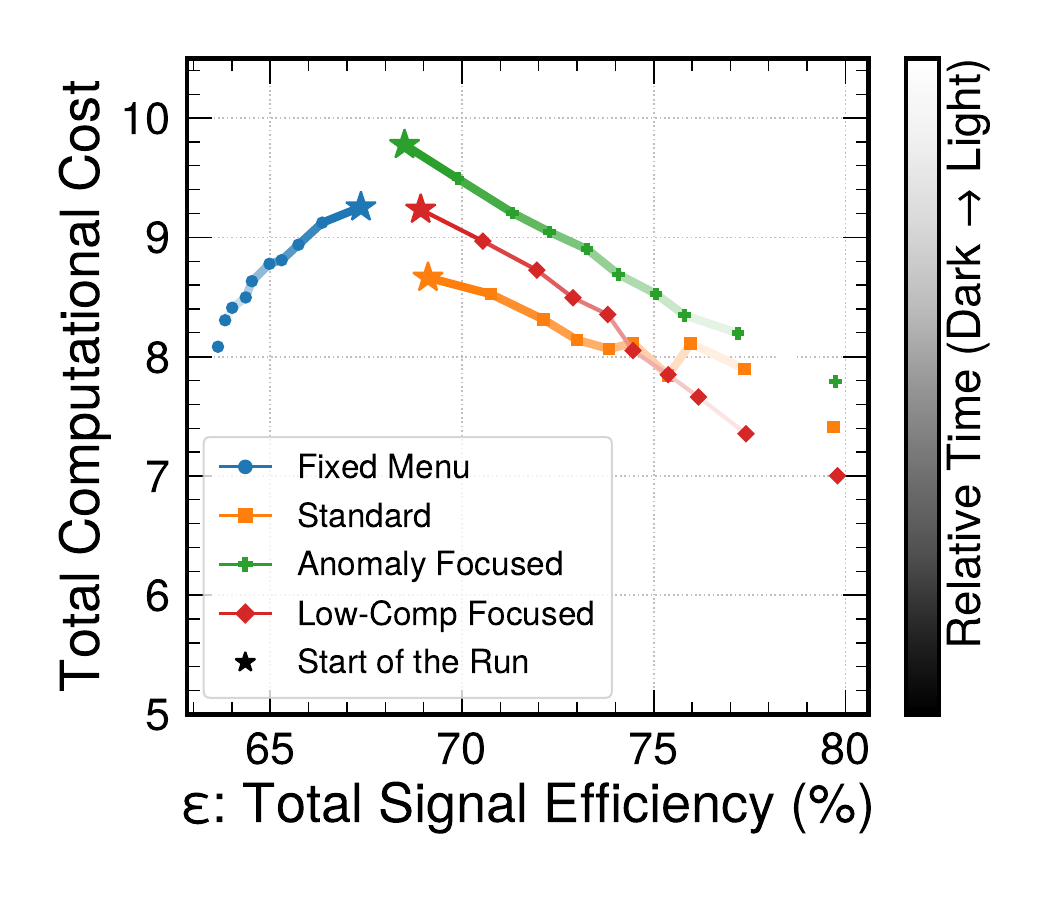}
      \caption{}
      \label{fig:summary_panel4-c}
    \end{subfigure} &
    \begin{subfigure}{0.42\linewidth}
      \includegraphics[width=\linewidth]{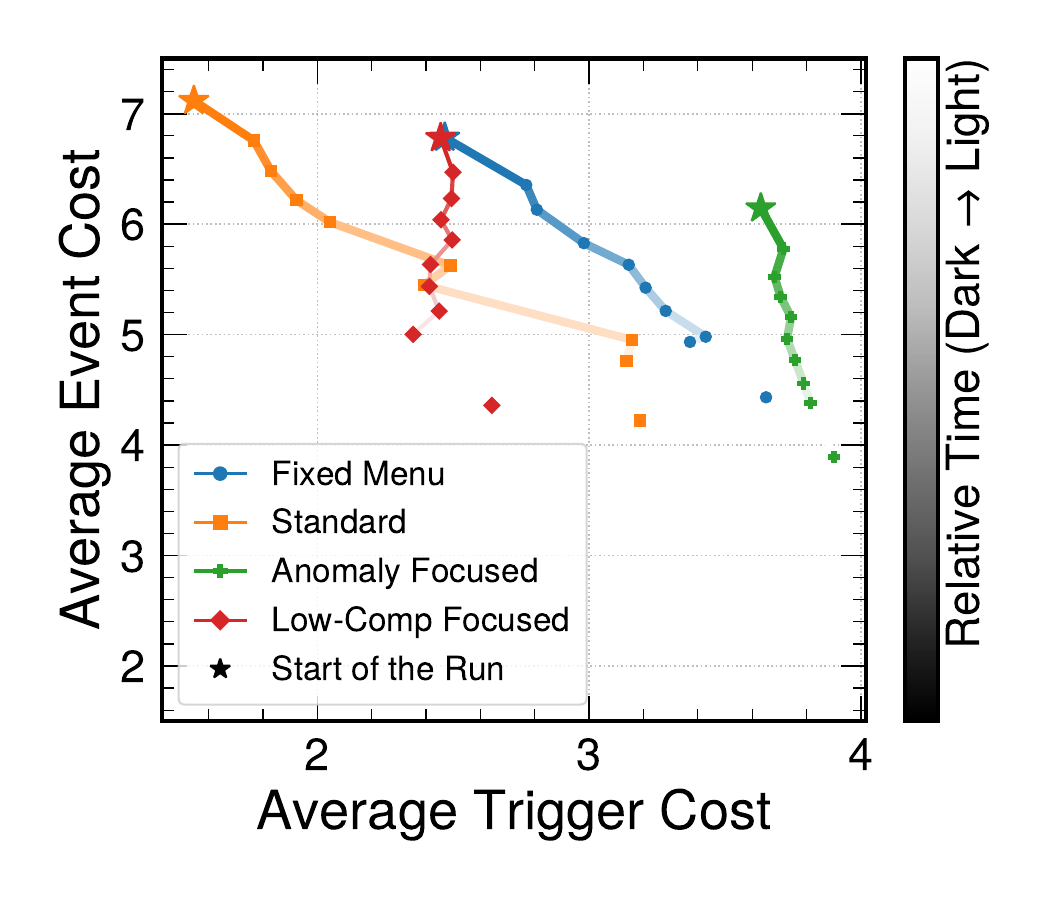}
      \caption{}
      \label{fig:summary_panel4-d}
    \end{subfigure} &
    
  \end{tabular}

  \caption{Summary of different agents’ performance on simulation samples.}
  \label{fig:summary_panel4}
\end{figure}

Panel~\ref{fig:summary_panel4-a} visualizes the joint evolution of rate control and signal performance: the horizontal axis quantifies the deviation from the target background rate, while the vertical axis captures signal inefficiency.
The Fixed Menu drifts away from the desired operating point as conditions change, exhibiting both increasing rate mismatch and degraded signal efficiency. 
By contrast, all adaptive agents remain confined to a region of small rate deviation while simultaneously improving signal efficiency, indicating that active control stabilizes the background response without sacrificing signal acceptance.
Panel~\ref{fig:summary_panel4-b} adds the resource dimension by showing total computational cost versus background-rate deviation.  The total computational cost is defined as the sum of trigger-path and event-level contributions.
The adaptive agents optimize costs over the course of the run while operating in a comparatively low-\(\lvert r_b - r_t \rvert\) regime. 
The Fixed Menu also exhibits a monotonic cost decrease, but along a trajectory in which \(\lvert r_b - r_t \rvert\) grows substantially.

Panel~\ref{fig:summary_panel4-c} makes the efficiency--cost trade-off explicit: total signal efficiency (over $t\bar t$ and \haaFourB\ samples) is plotted against total computational cost. 
As conditions evolve, the adaptive trajectories tend to move toward a more favorable region with higher efficiency at lower cost, showing that the controllers exploit the changing environment to reduce average complexity while maintaining or improving signal yield. By the end of the run, the Low-Computational-Cost agent achieves the most aggressive cost reduction at comparable efficiency.
Finally, panel~\ref{fig:summary_panel4-d} decomposes the computational cost by separating the trigger-level contribution (e.g.\ \HT versus AD composition) from the event-level contribution (proxied by jet multiplicity). All strategies follow the global trend of decreasing event complexity over the course of the run.

Overall, the simulation summary presents a coherent picture across all four projections: static menus drift and lose efficiency under time-dependent conditions, whereas adaptive control keeps the background stable and improves the position of the menu in the efficiency--cost trade space. Importantly, the differences among agents reflect their respective cost-function priorities, demonstrating how the framework can encode experiment-specific objectives.

\subsection{Deployment on Collision Data}
\label{subsec:summary_data}

We now apply the same control strategies using CMS Run~283408, and summarize the four agents' trajectories over the full run in Fig.~\ref{fig:summary_data}. 
The intent of this comparison is not to reproduce simulation point-by-point, but to verify whether the qualitative control behavior remains consistent under real operating conditions, including rate regulation, efficiency preservation, and cost adaptation.

\begin{figure}[htbp]
  \centering
  \setlength{\tabcolsep}{4pt}
  \renewcommand{\arraystretch}{1.0}
  \begin{tabular}{ccc}
    \begin{subfigure}{0.42\linewidth}
      \includegraphics[width=\linewidth]{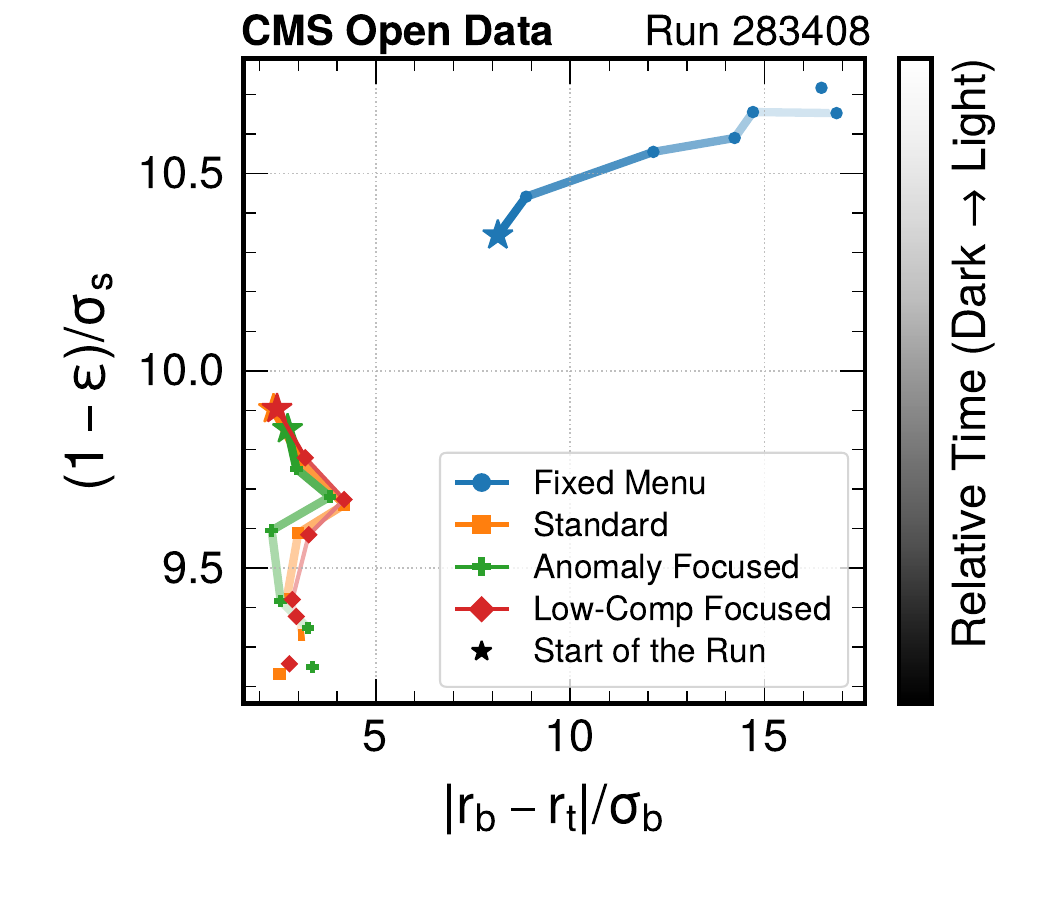}
      \caption{}
      \label{summ_data_a}
    \end{subfigure} &
    \begin{subfigure}{0.42\linewidth}
      \includegraphics[width=\linewidth]{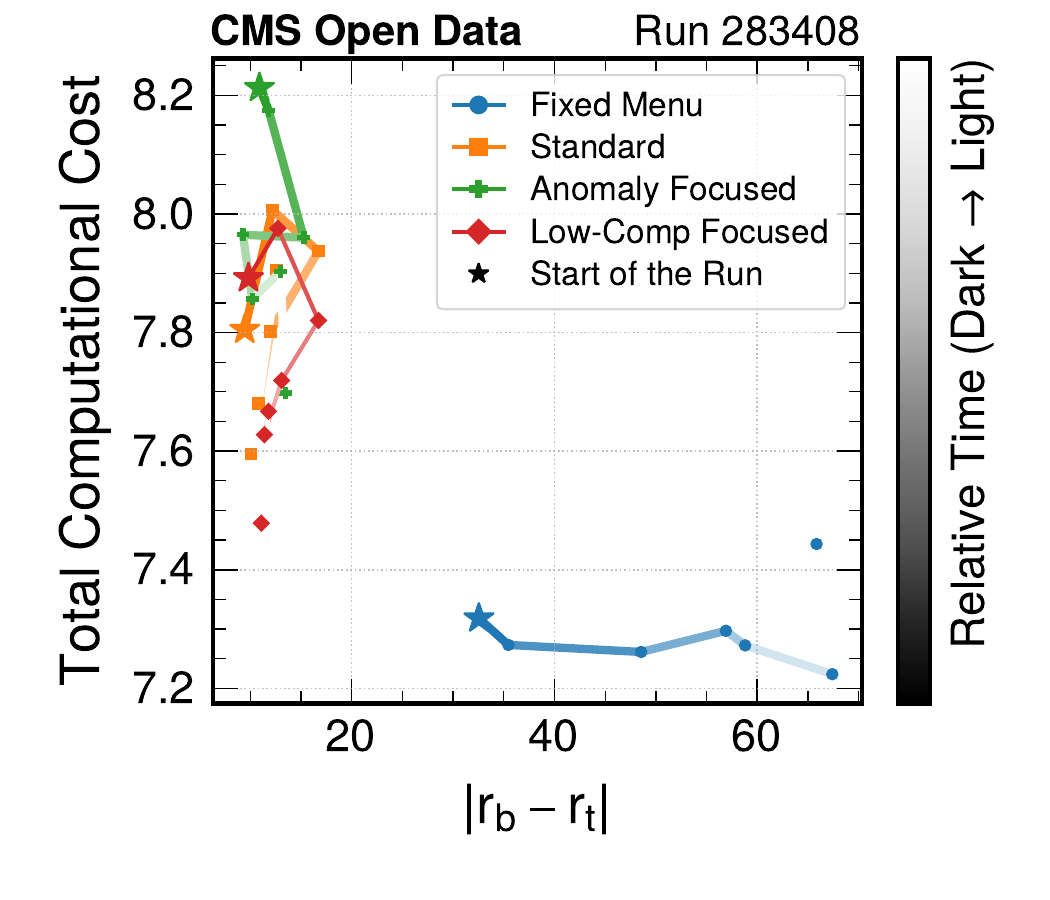}
      \caption{}
      \label{summ_data_b}
    \end{subfigure} &
     \\[2mm]

    \begin{subfigure}{0.42\linewidth}
      \includegraphics[width=\linewidth]{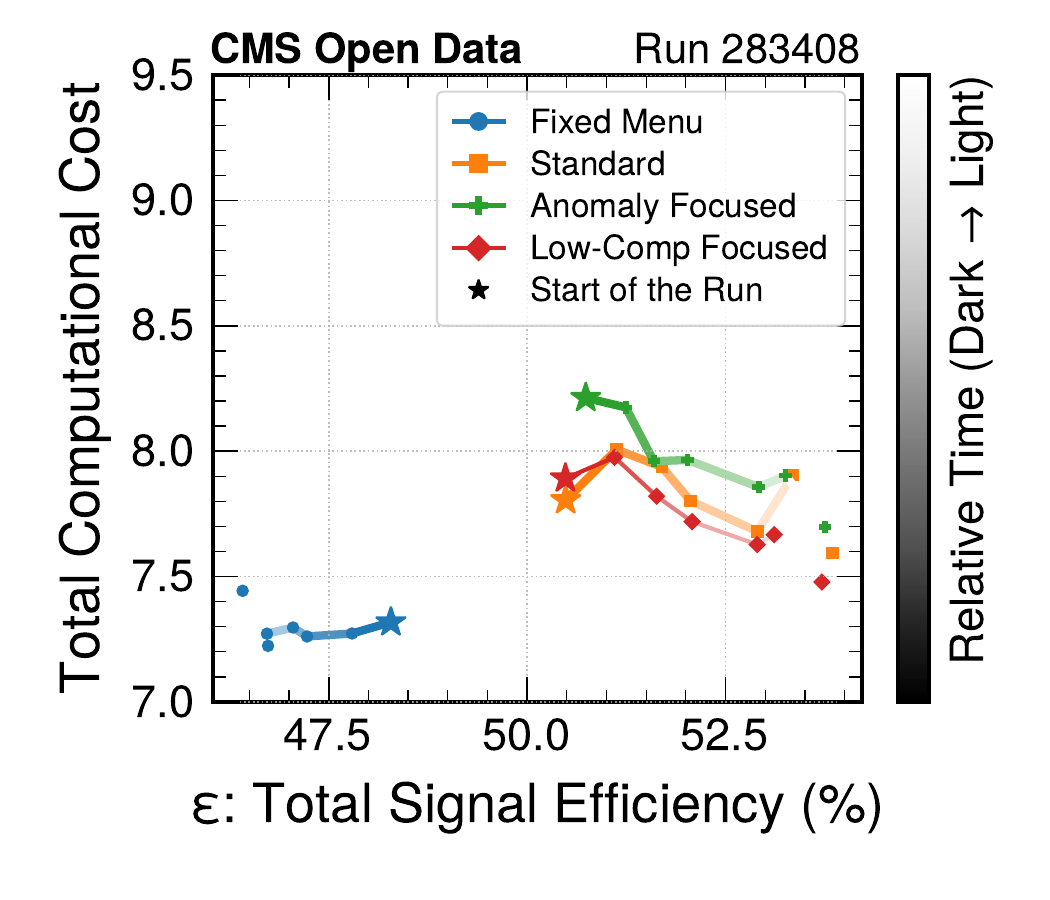}
      \caption{}
      \label{summ_data_c}
    \end{subfigure} &
    \begin{subfigure}{0.42\linewidth}
      \includegraphics[width=\linewidth]{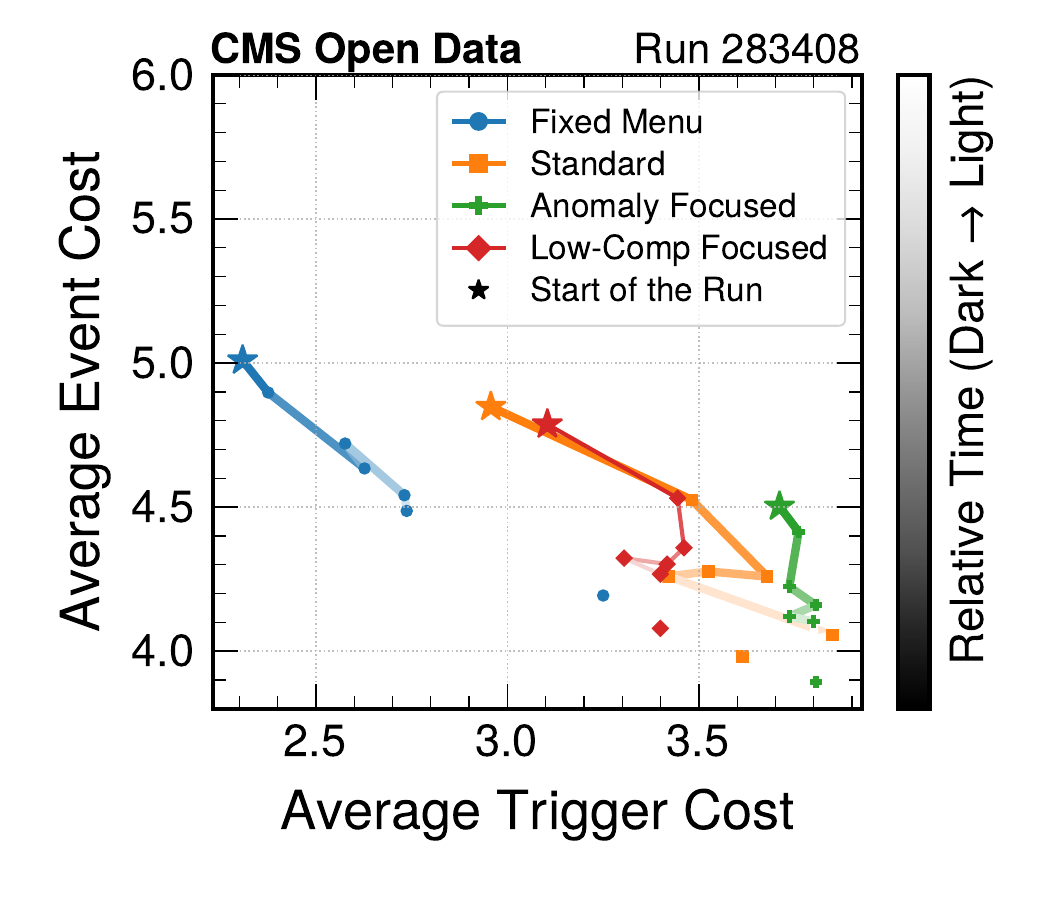}
      \caption{}
      \label{summ_data_d}
    \end{subfigure} &
  \end{tabular}
  \caption{A summary of each trigger menu control agents’ performance across various metrics over the course of CMS Run 283408.}
  \label{fig:summary_data}
\end{figure}

Panel~\ref{summ_data_a} shows the deviation of the background rate from the target versus signal inefficiency. 
The Fixed Menu shows the characteristic drift away from the target background rate, while maintaining systematically poorer signal performance than the adaptive approaches. By contrast, all three agents cluster at smaller, more stable rate deviations and exhibit a clear reduction in signal inefficiency as the run evolves.
Panel~\ref{summ_data_b} shows total computational cost versus background-rate deviation. 
The adaptive agents display a monotonic reduction in total cost over time while maintaining rate control, indicating that the controllers naturally migrate toward less expensive configurations as pileup decreases. 
The Low-Computational-Cost agent shows a lower computational cost value, especially at the end of the run, consistent with the trend observed in simulation.

Panel~\ref{summ_data_c} relates total signal efficiency to total computational cost. 
The adaptive strategies evolve toward a more favorable operating region, achieving higher average signal efficiency at reduced total cost as the run progresses.
This indicates that the efficiency--cost improvement is not an artifact of the simulated stream ordering, but persists in collision data.
Finally, panel~\ref{summ_data_d} separates event-level and trigger-level contributions. 
The average event-level cost decreases through the fill, reflecting the expected evolution of event activity, and all strategies follow this trend.

Overall, Fig.~\ref{fig:summary_data} confirms the core conclusion suggested by simulation: adaptive trigger control stabilizes the background response and preserves or improves signal acceptance, while automatically exploiting the time evolution of the run to reduce computational burden. 
The qualitative agreement between MC and data trajectories supports the robustness of the benchmark and motivates the applicability of feedback-based trigger-control strategies in realistic operational environments.

\section{Conclusions and Outlook}
\label{sec:outlook}

In this work, we have presented the framework of a dynamic, \textit{self-driving} trigger system, which incorporates real-time feedback to maximize scientific and operational objectives.
We set up a benchmark ecosystem that captures the tools, objectives, and constraints that are typical of LHC experiments: a combination of traditional and machine learning algorithms are employed to collect the most promising collision events while simultaneously controlling the rate of false positives and respecting limited computational resources.
All aspects have been tested with historical trigger data from collisions recorded by the CMS experiment, focusing on events with interesting hadronic jet activity that could indicate the presence of rare Standard Model processes or yet-to-be discovered phenomena.
To this end, we empowered an autonomous control agent to reconfigure the trigger selection criteria at regular time intervals, acting based on both external policies and in-situ performance estimates.

This benchmark began by studying simulated events that were given a temporal component, namely an arrow of time, characterized by a consistently decreasing instantaneous luminosity delivered to the experiment. First, we examined independent control of each trigger path based on a PD feedback loop. Then, we introduced and evaluated an ideal, generalized agent that could adjust all paths of the trigger menu at once, taking advantage of the time-varying correlations among all active algorithms.
Depending on the specified reward, the controller's policy prioritized actions to collect signal processes, explore novel event topologies with AD, or reduce computational costs associated with further trigger processing.
We showed that a simple control algorithm based on modest, adiabatic changes to the prescribed menu performed near the theoretical ideal in each of these cases.

After considering this synthetic dataset, the controller's performance was evaluated using historical CMS collision data, which can provide a unique window into time-varying factors that a real agent would encounter.
Here, stable collection rates and improved efficiencies to candidate signals were observed, as in the pure simulation setup.
Details of the agents' evolving actions over time were presented in Sec.~\ref{sec:summary_plots}, illustrating the tradeoff in priorities that were ultimately dictated by each control policy.
Despite having only an approximate emulation of time in our simulation samples, their observed trends largely coincided with those seen in data.

This study provides a proof-of-concept that an adaptive trigger system can be designed in a relatively simple manner while realizing dramatic improvements over the current experimental baseline of static menus.
We further hope that providing all details of this work as an open benchmark task will further stimulate the community to produce new ideas that extend the self-driving trigger concept to even more powerful control systems.
At the same time, additional considerations may need to be addressed before a fully autonomous system can be realized; the level of integration of the controller with the trigger and data acquisition system, the ideal methodology for the book-keeping of trigger configuration data, and how occasional human intervention might be incorporated into this system could all provide interesting questions for future work.
Extensions of this study could also naturally study more powerful decision-making agents, including reinforcement learning and active learning strategies that promise even more predictive and efficient means of control.

In conclusion, our results paint a clear picture of the novel sensitivity and numerous operational advantages over legacy systems, ideally 
opening the door to a new generation of autonomous triggers for high-throughput environments like the LHC.
Self-driving triggers are well within reach of modern experimental capabilities, and their implementation offers a uniquely focused strategy to navigate the landscape of complex and wide-ranging constraints while maximizing our ultimate goal of sensitivity to interesting physics.

\vspace{0.5cm}
\noindent \textbf{Code and Datasets}:
The workflow used for this study can publicly be accessed at Ref.~\cite{adaptive_particlephysics_triggers}, and the datasets generated are available on Zenodo at Ref.~\cite{Zenodo}.

\vspace{0.5cm}
\noindent \textbf{Acknowledgments}:
This work is supported by the University of Chicago Joint Task Force Initiative (JTFI) – Partnerships for Emerging Technologies Seed Funding. A.G, J.N, and N.T are supported by FermiForward Discovery Group, LLC under Contract No. 89243024CSC000002 with the U.S. Department of Energy, Office of Science, Office of High Energy Physics.

\FloatBarrier 

\bibliography{bibliography}

\end{document}  
\typeout{get arXiv to do 4 passes: Label(s) may have changed. Rerun}